\documentclass[10pt,a4paper,final]{iopart}

\usepackage{geometry}
\usepackage{amssymb}
\usepackage{mathrsfs}

\usepackage{multirow}
\usepackage{ulem}

\expandafter\let\csname equation*\endcsname\relax

\expandafter\let\csname endequation*\endcsname\relax

\usepackage{graphicx}
\usepackage[breaklinks=true,colorlinks=true,linkcolor=blue,urlcolor=blue,citecolor=blue]{hyperref}
\usepackage{footmisc}

\usepackage[utf8x]{inputenc}
\usepackage{color}
\usepackage[dvipsnames]{xcolor}
\usepackage{enumitem}
\usepackage{epsfig}
\usepackage{lscape}
\usepackage{hyperref}
\usepackage{mathtools}
\hypersetup{
    colorlinks=true,
    linkcolor=blue,
    filecolor=magenta,      
    urlcolor=blue,
}
\usepackage{verbatim}
\usepackage{soul}

 \usepackage{fancyhdr}
\usepackage{tikz} 
\usepackage{pgfplots}
 
 \usepackage{subcaption}

\newcommand{\BEQ}{\begin{equation}}     
\newcommand{\BEA}{\begin{eqnarray}}
\newcommand{\BD}{\begin{displaymath}}
\newcommand{\EEQ}{\end{equation}}       
\newcommand{\EEA}{\end{eqnarray}}
\newcommand{\ED}{\end{displaymath}}
\newcommand{\bb}{\begin{eqnarray}}
\newcommand{\ee}{\end{eqnarray}}

\newcommand{\lap}[1]{\overline{#1}}     
\newcommand{\dlap}[1]{\overline{\overline{#1}}}    
\newcommand{\mel}[1]{\breve{#1}}        

\newcommand{\D}{{\rm d}}                
\newcommand{\II}{{\rm i}}               

\newcommand{\demi}{\frac{1}{2}}         
\newcommand{\wit}[1]{\tilde{#1}}    
\renewcommand{\vec}[1]{\boldsymbol{#1}} 
\newcommand{\vekz}[2]
     {\mbox{${\begin{array}{c} #1  \\ #2 \end{array}}$}}

\usepackage{caption}

\newcommand{\tto}{\mathsf{t}_0}
\newcommand{\dast}{{\ast \ast}}

\definecolor{gruen}{HTML}{006800}

\newcommand{\sw}[1]{\textcolor{black}{#1}}

\definecolor{MyGray}{rgb}{0.95,0.95,0.95}
\definecolor{darkgreen}{HTML}{00BB00}
\definecolor{MyGray}{rgb}{0.90,0.90,0.90}
\definecolor{MyGray2}{rgb}{0.6,0.6,0.6}
\definecolor{MyRed}{rgb}{1.0,0.95,0.95}
\definecolor{MyRed2}{rgb}{0.9,0.4,0.4}
\definecolor{MyBlue}{rgb}{0.2,0.2,0.7}
\definecolor{MyGreen}{rgb}{0.2,0.7,0.2}

\usepackage{etoolbox}
\makeatletter
\def\@mkboth#1#2{}
\newlength\appendixwidth
\preto\appendix{\addtocontents{toc}{\protect\patchl@section}}
\newcommand{\patchl@section}{%
  \settowidth{\appendixwidth}{\textbf{Appendix }}%
  \addtolength{\appendixwidth}{1.5em}%
  \patchcmd{\l@section}{1.5em}{\appendixwidth}{}{\ddt}%
}

\newcommand{\mainmatter}{%
  \setcounter{footnote}{0}%
  \patchcmd{\@makefntext}{\fnsymbol}{\arabic}{}{}%
  \patchcmd{\@thefnmark}{\fnsymbol}{\arabic}{}{}%
  \def\@makefnmark{\textsuperscript{\arabic{footnote}}}%
}

\makeatother

\renewcommand{\emph}[1]{\textit{#1}}
\renewcommand{\em}{\it}

\begin{document}
\title{Non-equilibrium dynamics of the open quantum $O(n)$-model 
with non-Markovian noise: exact results}

\author{Sascha Wald$^{1}$, Malte Henkel$^{2,3}$ and Andrea Gambassi$^{4}$}

\address{$^1$Statistical Physics Group, Centre for Fluid and Complex Systems, Coventry University, Coventry, England}


\address{$^2$Laboratoire de Physique et Chimie Th\'eoriques (CNRS UMR 7019),
Universit\'e de Lorraine, B.P. 70239, \\F -- 54506 Vand{\oe}uvre-l\`es-Nancy Cedex, France\footnote{Permanent address.}}%

\address{$^3$Centro de F\'{i}sica Te\'{o}rica e Computacional, Universidade de Lisboa, P--1749-016 Lisboa, Portugal\looseness=-1}%

\address{$^4$SISSA - International School for Advanced Studies and INFN, via Bonomea 265,
I-34136 Trieste, Italy}

\ead{sascha.wald@coventry.ac.uk}

\begin{abstract}
The collective and  purely relaxational dynamics of quantum many-body systems 
after a quench at temperature $T=0$, from a disordered state to various phases 
is studied through the exact solution of the quantum Langevin equation of the 
spherical and the $O(n)$-model in the limit $n\to\infty$. The stationary state 
of the quantum dynamics is shown to be a non-equilibrium state. The quantum 
spherical and the quantum $O(n)$-model for $n\to\infty$ are in the same 
dynamical universality class. The long-time behaviour of single-time and 
two-time correlation and response functions is analysed and the universal 
exponents which characterise quantum coarsening and quantum ageing are derived. 
The importance of the non-Markovian long-time memory of the quantum noise is 
elucidated by comparing it with an effective Markovian noise having the same 
scaling behaviour and with the case of non-equilibrium classical 
dynamics. 
\end{abstract}

\vspace{2pc}

\noindent
\newpage


\setcounter{page}{1}
\mainmatter

\section{Introduction}
%

\setcounter{footnote}{0}   
%

\noindent
Investigating the non-equilibrium quantum dynamics of complex many-body systems 
is of fundamental importance for understanding the cooperative behaviour that 
may emerge from a large number of strongly interacting degrees of freedom. 
Experimentally accessible systems include cold atoms~\cite{Bloc08,Pol11,Mit18}, 
scaled-up quantum circuits \cite{Houc12,Bluh19}, ultrafast pump-probe 
measurements in correlated materials~\cite{Faus11,Smal12} and quark-gluon 
plasma~\cite{Arse05}. Among the central questions are those about the nature of 
the stationary states after the system has been forced out equilibrium. One 
needs to carefully verify whether the system ``thermalises'' towards an 
equilibrium state or not, and how to describe the relaxation towards 
stationarity~\cite{Kame11,Weis12,Cald14,Giam16}. A systematic approach to these 
issues is to prepare the system in some (in general non-equilibrium) initial 
state and to subsequently {\it quench} at least one macroscopic control 
parameter and to let the system relax~\cite{Bray94,Ham97,Cugl03,Henk10}.

In particular, the possible presence of criticality in the system is likely to 
affect the non-equilibrium dynamics and the relaxation after a quench onto the 
critical point, i.e. the so-called {\it non-equilibrium critical dynamics}, or 
into the ordered phase, corresponding to {\it coarsening}. In these scenarios, 
novel qualitative features, distinct from, e.g., those of the equilibrium 
dynamics may be observed~\cite{Giam16,Cugl03,Henk10,Stru78}. Quite generically, 
systems quenched onto or across a critical point, will fail to thermalise and 
rather undergo an ``{\it ageing dynamics}'' which never reaches a stationary 
state. This dynamics is characterised by the three properties~\cite{Henk10}: (i) 
slow dynamics, (ii) absence of time-translation-invariance and (iii) dynamical 
scaling. These features are mainly studied through the long-time behaviour of 
two-time correlation functions and responses [see Eqs.~(\ref{eq:scale2}), 
(\ref{gl:vieux}), (\ref{eq:lambdas})]. Ageing dynamics is usually characterised 
by a single, emergent and time-dependent length scale $L$ which generically 
grows as $L(t)\sim t^{1/z}$ at long times $t$, where $z$ is the so-called {\it 
dynamical exponent}. Quantities like correlators and response functions then 
display {\it dynamical scaling} in which the associated exponents and scaling 
functions are {\it universal}, i.e., largely independent of the microscopic 
details of the system. In contrast to equilibrium systems, dynamical scaling 
after a quench is observed in large portions of the parameter 
space~\cite{Stru78}.

Ageing effects and their reproducible and universal aspects were first studied 
in glassy systems~\cite{Stru78} in contact with a thermal bath at temperature 
$T$, before it became apparent that analogous phenomena also arise in much 
simpler systems without disorder or frustration~\cite{Ronc78,Cugl94,Godr00,Calabrese_2005}. The 
majority of systems studied in the literature have classical dynamics 
\cite{Henk10} with some notable exceptions concerning anomalous coarsening in 
pre-thermal phases \cite{Mara15,Lem16}. If these systems are  quenched into the 
ordered phase with $T<T_c$, where $T_c$ denotes the critical temperature of the 
system, the long-time behaviour is fully characterised by the gross features of 
the initial state, while the coupling to the external heat bath at temperature 
$T>0$ turns out to be irrelevant. Conversely, for a critical quench onto 
$T=T_c>0$, the leading behaviour is governed by the thermal noise and the 
initial state correlations are largely irrelevant as long as they are 
short-ranged~\cite{Henk10,Taeu14,Cala07}. One system used for the theoretical 
analysis of generic non-equilibrium dynamics and ageing is the {\it spherical 
model}~\cite{Berl52,Lew52}, first introduced as a simple exactly solvable model 
of a magnetic phase transition in $d$ spatial dimensions with a non-mean-field 
critical behaviour for $2<d<4$. Its classical, purely relaxational dynamics 
(model A), described by a Langevin equation with a Gaussian white noise, can be 
solved exactly~\cite{Ronc78,Godr00,Pico02} and does confirm the generic scaling 
behaviour expected as indicated above. The successful confirmation of classical, 
dynamical scaling makes this model a promising candidate for similar studies in 
the quantum realm. In particular, the noisy description of open quantum systems 
differs qualitatively from the classical case and one may ask whether it is 
possible to extend the classical characterisations towards quantum systems. We 
shall attempt to answer this question by analytically studying the long-time 
dynamics of the simplest open quantum model with non-trivial many-body 
interactions.

The non-equilibrium dynamics of isolated quantum systems has been analysed 
intensively, see, e.g., 
Refs.~\cite{Barb19,Cugl18,Heyl18,Mara15,Chan13,Pol11,Dzia10,Bloc08,Mit18} and 
references therein. Much less is known, in general, about non-equilibrium {\it 
open} quantum systems \cite{Cugl98,Aron09,Gagel14,Gagel15,Wald16,Wald18}. 
Partially, this might be due to the widespread expectation, summarised in 
Ref.~\cite{Aron09}, that \textit{``\ldots a large class of coarsening systems 
(classical, quantum, pure, and disordered) should be characterised by the same 
scaling functions.''}. While there are good reasons to accept this statement in 
the case of finite temperatures, this is not obvious in the limit $T\to 0$ where 
quantum fluctuations govern the bath structure. An important distinction is that 
zero-temperature quantum noise is necessarily {\it non-Markovian} 
\cite{Cal81,Ford65,Ford87,Ford88,Gard04,Hang05,Kame11,Weis12,Arau19} and the 
resulting memory effects might become important in the long-time quantum ageing 
behaviour. Comparative studies, see, e.g., Ref.~\cite{Weis12}, of the classical 
and quantum Brownian motion lead, respectively, to growth laws $L_{\rm 
cl}(t)\sim t^{1/2}$ and $L_{\rm qu}(t) \sim \ln t$ for the typical length scale 
$L$, with $L_{\rm cl}(t)\gg L_{\rm qu}(t)$ at long times. In a certain sense, 
this suggests that quantum noise can be considered ``weaker'' than the classical 
white noise. Accordingly, one might expect that the relative importance of the 
initial and bath correlations could be different when comparing quantum and 
classical dynamics.

Exactly solvable models are useful in this context, as they permit 
mathematically controlled statements on a well-defined physical system, see, 
e.g., Ref.~\cite{Bax16}. Here, we shall analyse the non-equilibrium quantum 
dynamics of two closely related models:
 
\begin{enumerate}[label=(\alph*)]
 
 \item The {\it quantum $O(n)$-model} in the large-$n$ 
limit~\cite{Moshe03,Grac04} which provides the simplest approximation of 
non-linear interactions on top of a free quantum field theory.
  
 \item The {\it quantum spherical model}~\cite{Ober72,Henk84,Voj96}, which is a 
mathematical extension of the quantum Ising model to obtain analytical insights 
beyond the latter.

\end{enumerate} 
These models have the appealing feature that the many-body dynamics for 
arbitrary spatial dimension $d$ can be reduced to the solution of a single 
integro-differential equation, from which all observables of physical interest 
can be determined. We shall describe the non-equilibrium dynamics of these 
models by a quantum Langevin equation, which is known to guarantee physically 
desirable properties for a relaxation process, including the validity of the 
quantum fluctuation-dissipation 
theorem~\cite{Cal81,Ford65,Ford87,Ford88,Gard04,Hang05,Arau19,Oliv20}. Since the 
emerging equations are linear and we focus on observables which are at most 
quadratic in the fluctuating fields, this scheme is self-consistent and more 
advanced field-theoretical treatments, that are usually needed in order to 
describe interacting models~\cite{Kame11,Giam16}, are not required. We study a 
quantum bath at temperature\footnote{It is conceivable that the long-time limit 
and the limit $T\to0$ may not commute.} $T=0$ and shall address the following 
questions: 
\begin{enumerate}

\item Are the leading long-time dynamics of the two models mentioned above 
equivalent?

\item Do these systems eventually relax to an equilibrium state? 

\item What are the (quantum) phase transitions in these systems? 

\item What is the relative importance of the spatial correlations existing in 
the initial state and the  bath noise correlators? 

\item What are the differences between the actual quantum noise and a suitable 
effective Markovian noise? In which observables could such differences be seen? 

\item Is there a ``quantum ageing'' distinct from ageing in classical dynamics? 
For isolated systems quantum ageing after a quench has been found in the 
pre-thermal phase \cite{Mara15,Chio17}. Despite being distinct from classical 
ageing, its actual quantum character can be debated as the ageing occurs in 
highly excited states.

\end{enumerate} 
As an intermediate step, it will be useful to study a model with an effective 
Markovian noise, introduced artificially and tailored such that the leading 
scaling behaviour in the presence of the actual quantum noise is reproduced. 
However, the treatment of the non-Markovian noise requires the introduction of 
suitable mathematical tools which are discussed below, see 
also~\ref{app:prop_lap}, \ref{app:F} and~\ref{app:Lap}.
\textcolor{black}{We find that the overdamped quantum Langevin dynamics at 
zero temperature shows several qualitative differences from classical 
dynamics. These concern the non-equilibrium nature of the stationary state
(even for relaxations occuring in the disordered phase), the inequivalence of regimes
of non-equilibrium quantum dynamics and those of the classical dynamics and the 
relevance of the non-Markovian quantum noise for the scaling of the 
single-time correlators. For clarity, we summarise these findings
in Sec.~\ref{sec:results} without focussing on technical details.}

This work is organised as follows. In Sec.~\ref{sec:equilibrium} we introduce 
the quantum spherical and the quantum $O(n)$-model, in the limit $n\to\infty$ at 
thermal equilibrium and we recall the main features of their quantum phase 
diagrams. In Sec.~\ref{sec:dynamics} we formulate the quantum non-equilibrium 
dynamics and review the scaling argument by which these models can be reduced to 
a single over-damped quantum Langevin equation, in which the different types of 
dynamics (classical, quantum, etc.) are solely distinguished by the specific 
expression of the noise correlation functions. In Sec.~\ref{sec:results} we 
summarise our predictions for one- and two-time correlation and response 
functions of the fluctuating fields, obtained from the exact solution of the 
non-equilibrium dynamics and we discuss their physical interpretation. This is 
followed in Sec.~\ref{sec:solution} by the detailed solution of the spherical 
constraint, for the non-Markovian quantum noise. Finally, 
Sec.~\ref{sec:observables} discusses the derivation of the time-dependent 
physical observables from the formal solution of the dynamical constraints of 
the models while we present our conclusions in 
Sec.~\ref{sec:conclusion}\textcolor{black}{, notably via a detailed comparison 
with classical dynamics.} 
Several appendices discuss the technical details of our analysis.

%
\section{Equilibrium behaviour of the spherical and $O(n)$ model for $n\to\infty$}
\label{sec:equilibrium}
%

The spherical model and the $O(n)$-model with $n\to\infty$, are introduced as 
two exactly solvable quantum statistical systems that show non-mean-field phase 
transitions:
\begin{enumerate}[label=(\alph*)]

\item
The {\it  $O(n)$-model} is described by the quantum $\phi^4$ field 
theory \cite{Moshe03,Grac04}
\begin{equation}
 H_n = \frac{1}{2}\int_{\vec{x}}\left[ \vec{\pi}^2+ (\nabla_{\vec{x}} \vec{\phi})^2 
 +r_0 \vec{\phi}^2 +\frac{u}{12 n }\left(\vec{\phi}^2\right)^2 \right],
\end{equation}
with the bosonic $n$-component vector field $\vec{\phi} = 
(\phi_1,\ldots,\phi_n)$. The canonically conjugate momentum $\vec{\pi} = 
(\pi_1,\ldots,\pi_n)$ satisfies $[\phi_a(\vec{x}), \pi_b(\vec{x}')] = \II \hbar 
\delta(\vec{x}-\vec{x}')\delta_{ab}$. The integral notation is to be understood 
as $\int_{\vec{x}} = \int_{\mathbb{R}^d} \D^dx$. The parameter $u$ controls the 
strength of the anharmonic coupling, with $u=0$ corresponding to the Gaussian 
model, and $r_0$ is the bare square mass of the theory. In the limit 
$n\to\infty$ of the number of components of the field, the anharmonic 
interaction can be decoupled and accounted for by adding fluctuations to $r_0$. 
The effective Hamiltonian then describes the scalar field theory \cite{Moshe03} 
\begin{align}
 H_\infty = \frac{1}{2}\int_{\vec{x}}\left[ \pi^2+ (\nabla_{\vec{x}} \phi)^2 
 +r \phi^2 \right], \quad   \mbox{with}\quad 
r= r_0 +\frac{u}{6} \left\langle\phi^2\right\rangle ,
\label{eq:Hon}
\end{align}
where $\langle \cdots\rangle$ indicates the expectation value with respect to 
the system density matrix. In this limit, the equilibrium critical properties 
can be determined analytically by formally solving the external constraint on 
the effective parameter $r$. 

\item 
The \textit{quantum spherical model} \cite{Ober72,Henk84,Voj96} is described by 
the lattice Hamiltonian 
\begin{align}
 H_{\text{sm}} &= \sum_{n\in\mathscr{L}}\bigg[ 
                  \frac{\lambda}{2}p_n^2+\frac{\sigma}{2}s_n^2
                  -J\sum_{\left<n,m\right>}s_ns_m \bigg],
\label{eq:sm}
\end{align}
where the ``spin'' operator\footnote{The operator $s_n$ is referred to as spin 
operator, motivated by the analogy to the classical spin model in terms of which 
the spherical model was defined. In the quantum model, $s_n$ are position 
operators.} $s_n$ is located at the site $n\in\mathscr{L}$ of the hypercubic 
lattice $\mathscr{L}\subset \mathbb{Z}^d$ and $p_n$ is its canonically conjugate 
momentum operator, i.e., $[s_n,p_m] = \II \hbar \delta_{nm}$. The exchange 
coupling is $J>0$ and the parameter $\lambda$ quantifies the strength of quantum 
fluctuations in the system with $\lambda=0$ corresponding to the 
\textit{classical} spherical model \cite{Berl52,Lew52}. The parameter $\sigma$ 
is a Lagrange multiplier imposing the \textit{spherical constraint}
\begin{align}
 \sum_{n\in\mathscr{L}} \left\langle s_n^2\right\rangle &= \mathscr{N},
 \label{eq:sc}
\end{align}
where $\mathscr{N}=|\mathscr{L}|$ is the number of sites  of the lattice. This 
constraint distinguishes the spherical model in Eq.~\eqref{eq:sm} from a set of 
non-interacting quantum harmonic oscillators. We rescale this ``standard'' 
formulation of the spherical model as  $s_n/\sqrt{\lambda}\to s_n$, 
$\sqrt{\lambda}p_n \to p_n$ in such a way that the canonical commutation
relation is preserved. The rescaled interaction constant 
reads $2J \lambda$ and is set to $1$. With the substitution 
$r:=\sigma \lambda-d$ we obtain 
\begin{align}\label{eq:Hsm}
 H_{\text{sm}} = \frac{1}{2}\sum_{n\in\mathscr{L}}\bigg[ p_n^2+ (r+d) s_n^2-\sum_{\left< n,m\right>}s_ns_m\bigg] \quad\mbox{with}\quad
 \quad  \sum_{n\in\mathscr{L}} \left\langle s_n^2\right\rangle = \mathscr{N}/\lambda.
\end{align}
\end{enumerate}

\begin{figure}[t]
 \centering
 \includegraphics[width=0.8\textwidth]{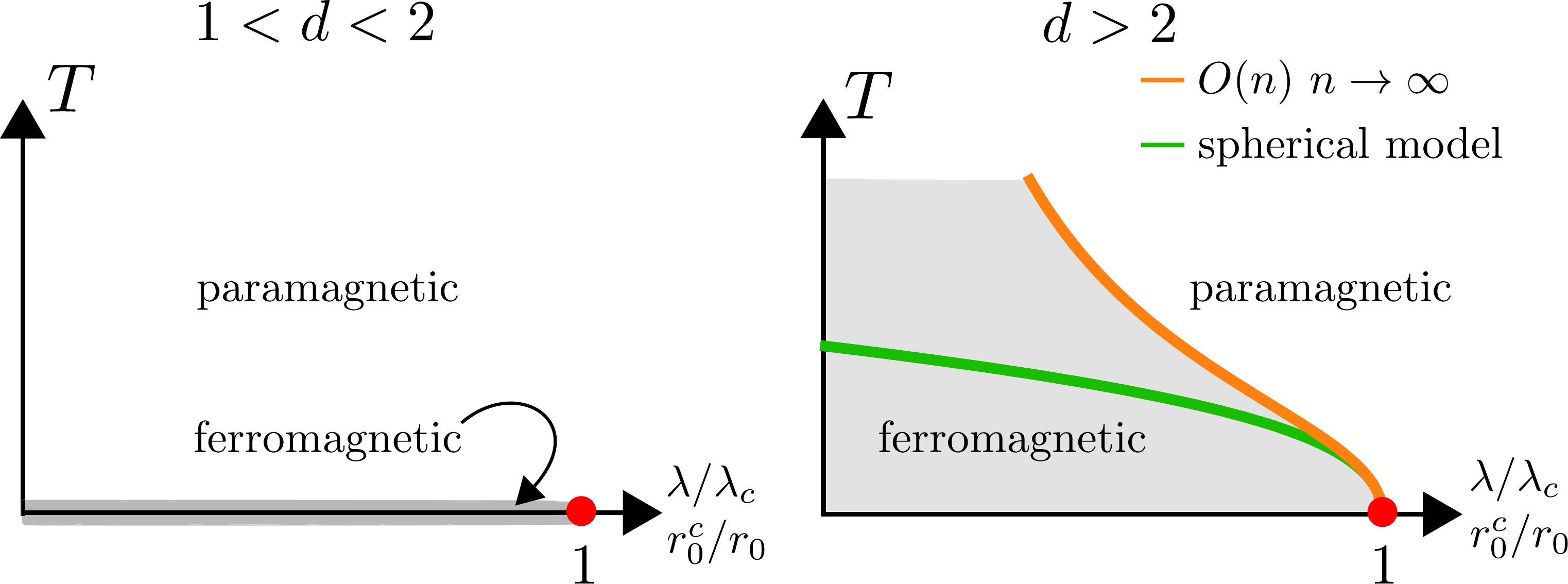}
 \caption[Schematic phase diagram]{
 Schematic equilibrium phase diagram of the quantum spherical model and the 
$O(n)$-model at large $n$, for various spatial dimensions $d$, \sw{scaled such 
that $\lambda_c = 1/(r_0)_c$ at zero temperature.} For $1<d<2$ the models 
undergo a quantum phase transition at zero temperature. For $d>2$, a thermal 
critical  line appears \cite{Henk84,Voj96,Oliv06,Bien12,Wald15}.}
 \label{fig:SM-pd}
\end{figure}

\noindent
In the thermodynamic limit $\mathscr{N}\to\infty$ the quantum spherical model 
and the $O(n)$ model for $n\to\infty$, are characterised  by a non-trivial 
equilibrium phase diagram~\cite{Henk84,Nieu95,Voj96,Oliv06,Wald15}. In 
particular, for spatial dimensions $d>1$, a quantum critical point $r_0^c$ 
(respectively $\lambda_c$) is present at $T=0$, separating a ferromagnetic and a 
paramagnetic phase. For $d>2$ such a phase transition occurs also at $T>0$ along 
the line of critical points $r_0^c(T)$  (respectively $\lambda_c(T)$). The 
qualitative phase diagram is shown in Fig.~\ref{fig:SM-pd}~\cite{Wald15}.  The 
critical behaviour at these equilibrium transitions is exactly solvable since 
the complex many-body problem is reduced to the solution of a single 
transcendental equation. The phase transition at $T\neq 0$ belongs to the same 
universality class as the classical finite-temperature phase transition, while 
the phase transition occurring at $T=0$ in $d$ spatial dimensions belongs to the 
same universality class as the classical thermal transition in $d+1$ spatial 
dimensions~\sw{\cite{Henk84, Voj96, Stan68}}. The close relationship between 
these models is apparent from the comparison of the Hamiltonians in 
Eqs.~\eqref{eq:Hon} and \eqref{eq:Hsm}, each of which is subject to an external 
constraint. The universality classes of the corresponding transitions in the 
bulk are the same \cite{Stan68,Grac04} and the phase diagrams of the models look 
qualitatively similar, even though microscopic details, such as the exact 
critical values of the relevant parameters may vary. We shall show below that 
this analogy carries over to the leading relaxation behaviour out of 
equilibrium.

%
\section{Non-Equilibrium Dynamics}
\label{sec:dynamics}
%
Here, we discuss the effects of a coupling to a dissipative environment and 
formulate the non-equilibrium quantum dynamics of the statistical systems 
introduced in Sec.~\ref{sec:equilibrium}. This dynamics is governed by 
stochastic {\it quantum Langevin equations} which describe the dissipative 
aspects of an environment via a viscous and a random force. 

Since the models presented in Sec.~\ref{sec:equilibrium} are translational 
invariant, it is convenient to consider the Fourier components of the 
time-dependent fields $\phi(t,\vec{r})$, i.e.,
\begin{equation}
 \phi_{\vec{k}}(t) = \int_{\vec{x}} \phi(t,\vec{x})\, e^{-\II \vec{k}\vec{x}}.
\end{equation}
Hereafter, we focus specifically on the $O(n)$-model with $n\to \infty$, but the 
discussion can be repeated for the spherical model by simply replacing 
$\phi_{\vec{k}}\mapsto s_{\vec{k}}$ and $\pi_{\vec{k}}\mapsto p_{\vec{k}}$. For 
the model Hamiltonians \eqref{eq:Hon} and \eqref{eq:Hsm}, generically indicated 
below by $H$, the dynamics of the Fourier components reads~\cite{Arau19}
\begin{subequations} 
\label{eq:langevin}
\begin{align}
 \partial_t \phi_{\vec{k}}  &= \frac{\II}{\hbar}\left[ H,\phi_{\vec{k}} \right] 
+ \eta_{\vec{k}}^{(\phi)},\\[.25cm]
 \partial_t \pi_{\vec{k}}  &= \frac{\II}{\hbar}\left[ H,\pi_{\vec{k}} \right] 
-\gamma \pi_{\vec{k}}+ \eta_{\vec{k}}^{(\pi)},
\end{align}
\end{subequations}
with two distinct noise operators $\eta_{\vec{k}}^{(\phi)}$ and 
$\eta_{\vec{k}}^{(\pi)}$ whose properties will be specified further below. The 
damping parameter $\gamma$ is positive and we restrict our analysis to Ohmic 
damping, i.e. we assume that the damping is frequency-independent. All the 
information on the environment and its coupling to the system is contained in 
the correlation functions of the noises. There are two mathematically equivalent 
ways to specify these correlation functions. First, following 
Refs.~\cite{Gard04,Weis12,Ford65,Ford87,Cal81}, one may take the environment 
into account explicitly and indicate its quantum Hamiltonian $H_{\rm env}$ and 
its interaction $H_{\rm int}$ with the system. The composite system $H_{\rm tot} 
= H +H_{\rm env}+ H_{\rm int}$ then evolves unitarily and by admitting the 
environment to be a bath of thermal harmonic oscillators much larger than the 
system, the equation of motion for the degrees of freedom of the environment can 
be explicitly solved. The average over the distribution of the environment can 
then be carried out, provided one specifies an adequate spectral function for 
the bath \cite{Gard04}. Alternatively, one may consider the system $H$ to be a 
set of harmonic oscillators. Rather than specifying the properties of the 
environment explicitly, one may model the dissipative aspects by an Ohmic 
damping and two noises (one for each conjugate bosonic variable). The 
correlations of these {\it ad hoc} noises cannot explicitly contain system 
parameters and should be determined in such that the resulting evolution of the 
system satisfies the following fundamental properties \cite{Arau19}: (i) 
canonical equal-time commutation relation, (ii) Kubo formula, (iii) the virial 
theorem and (iv) the quantum fluctuation-dissipation theorem for any $T>0$. Both 
procedures lead exactly to the {\it same} noise specifications. For times $t\neq 
t'$ the non-vanishing two-time noise correlators are given by 
\begin{subequations}  \label{ara3.29} 
\begin{align}
\left\langle\bigl\{ \eta_{\vec{k}}^{(\phi)}(t), {\eta}_{\vec{k}'}^{(\pi)}(t') 
\bigr\}\right\rangle &= \gamma T \coth\left(\frac{\pi}{\hbar} T(t-t')\right) 
\delta(\vec{k}+\vec{k}'),\\[.25cm]
\left\langle\bigl[ {\eta}_{\vec{k}}^{(\phi)}(t), {\eta}_{\vec{k}'}^{(\pi)}(t') 
\bigr]\right\rangle &=\II\hbar\gamma \, \delta(t-t')\delta(\vec{k}+\vec{k}').
\end{align} 
\end{subequations}
Here, the average is done on these noises, and all other averages of the noise 
(anti-)commutators vanish.\footnote{ The construction of the noise 
anti-commutators is done in frequency space. Thus, Eq.~(\ref{ara3.29}) is 
correct up to a set of time differences of measure zero \cite{Arau19}. In 
particular, this implies that the equal-time anticommutators may remain finite. 
The noise correlators have to be understood as distributions.} We shall be 
mainly interested in the long-time behaviour of the models after a quench. The 
analysis of the ensuing dynamics will be greatly simplified if one eliminates 
the equation of motion for the momentum $\pi_{\vec{k}}$, which can be done by 
taking a formal scaling limit~\cite{Arau19}
\begin{equation}\label{eq:over}
 \lambda\to 0,\ \  t\to\infty, \ \ 
\quad  \text{with} \quad \textcolor{black}{\tilde{t}=}\lambda t = \text{cst.},\ \
\end{equation}
\textcolor{black}{with the fixed and finite damping constant $\wit{\gamma} = \lambda \gamma$}
\textcolor{black}{and the rescaled temperature $\tilde{T} = T/\lambda$}.
Although the limit $\lambda\to0$ would 
\textcolor{black}{correspond to a classical dynamics}
for $t={\rm cst.}$ we 
\textcolor{black}{emphasise here}
that this scaling limit does not reduce to
\textcolor{black}{that case}
since time is 
\textcolor{black}{also}
scaled appropriately.
\textcolor{black}{We now relabel the rescaled variables by dropping all tildes from 
the variables $\tilde{t}$, $\tilde{\gamma}$ and $\tilde{T}$ and focusing on the 
zero-temperature case $T=0$.}
The Langevin equations~\eqref{eq:langevin} then reduce to a single over-damped 
Langevin equation with a ``composite'' noise $\xi_{\vec{k}}$ 
\cite{Gard04,Arau19,Weis12,Ford65,Ford87,Cal81}, i.e.,
\begin{equation}\label{eq:qle}
 \gamma \partial_t \phi_{\vec{k}}(t) + \left(r(t)+k^2\right) \phi_{\vec{k}}(t) = 
\xi_{\vec{k}}(t),
\end{equation}
where $r=r(t)$ is to be found self-consistently from the constraint 
[cf.~Eqs.~\eqref{eq:Hsm} and \eqref{eq:Hon}] \textcolor{black}{and $\gamma$
is the rescaled damping parameter. Additional details on this 
long-time limit are provided in~\ref{app:over}.} 
The properties of the noise 
$\xi_{\vec{k}}$ will largely determine the (quantum) character of the resulting 
dynamics and since we shall consider various cases, we shall specify its 
properties below in more detail.

\begin{figure}[tb]
 \centering
 \centering
\begin{tikzpicture}
  \begin{axis}[xmin=-4,xmax=4,ymin=-0.2,ymax=1,
    xlabel={$\tau$},
    ylabel={normalised noise correlation}]

    \addplot[domain = -4:4,samples = 100, smooth, line width = 2pt, blue] {
     exp(-abs(x))};
    \addlegendentry{WP}
    
    \addplot[domain = -4:4,samples = 100, smooth, line width = 2pt, red] {
     (1-x^2)/(1+x^2)^2};
    \addlegendentry{QN}

        \addplot[domain = -4:4,samples = 100, dashed, gray] {
     0};
    \end{axis}
\end{tikzpicture}

 \caption[Normalised noise]{Dependence of the  noise correlations $\langle 
\{\xi_k(t),\xi_{k'}(t') \}\rangle$ in Eq.~\eqref{eq:qn_anticommu_T}, with an
exponential regularisation parameter $\tto$, on the dimensionless time 
difference $\tau \equiv (t-t')/\tto$, normalised by its value at $t'=t$. We 
compare the case of the quantum noise (QN) at zero temperature, corresponding to 
 $T\to 0$ in Eq.~\eqref{eq:qn_anticommu_T} with the case of the \sw{Wiener} 
process (WP)~\cite{Uhl30} with the same width, corresponding to  $\hbar\to 
0$ in Eq.~\eqref{eq:qn_anticommu_T}, which is equivalent to a classical white 
noise. }
\label{fig:QNOU}
\end{figure}
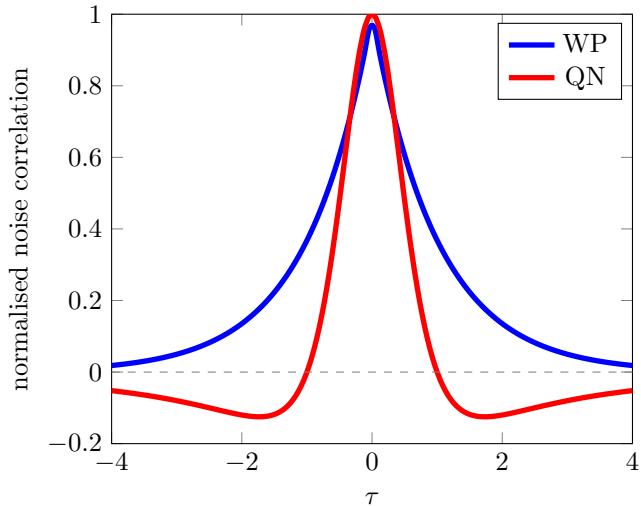

The non-equilibrium dynamics has been reduced to an effective over-damped 
Langevin equation, which is formally identical to the classical Langevin 
equation.\footnote{Despite the formal similarity of the classical and the 
quantum equation, the latter is an operator equation. To simplify the notation 
we shall not emphasise this distinction.} Any distinction between classical and 
quantum dynamics will now solely emerge from the form of the noise correlation 
functions. In this work, we shall distinguish the following three cases:
\begin{enumerate}
\item 
{\it Quantum dynamics}, derived from Eqs.~(\ref{ara3.29}), and described by  
\cite{Cal81,Ford65,Ford88,Gard04,Hang05,Weis12,Arau19} 
\begin{subequations} \label{eq:qudyn}
 \begin{align}
\left\langle \left\{ \xi_{\vec{k}}(t), \xi_{\vec{k}'}(t') \right\} 
\right\rangle &= \frac{2\gamma\hbar}{\pi} \int_0^\infty \!\D\omega\: 
\omega\operatorname{coth} \left( \frac{\hbar \omega}{T} \right) 
\cos(\omega(t-t')) \: \delta(\vec{k}+\vec{k}'), \label{eq:qn_anticommu_T} 
\\[.25cm]
\left\langle \left[ \xi_{\vec{k}}(t), \xi_{\vec{k}'}(t') \right] \right\rangle 
&= 2\II\hbar \gamma\left(  \frac{\D}{\D t} \delta(t-t')\right) \: 
\delta(\vec{k}+\vec{k}') ,
\label{eq:qn_commu}
 \end{align}
\end{subequations}
where $T$ is the bath temperature and the noise correlation function is non-Markovian. 
\item 
{\it Classical dynamics}, obtained from Eq.~\eqref{eq:qudyn} in the  limit 
$\hbar\to 0$, leading to
\begin{align} \label{eq:classical}
    \left\langle \left\{ \xi_{\vec{k}}(t), \xi_{\vec{k}'}(t') \right\} \right\rangle 
    = 4 T \gamma \delta(t-t')\delta(\vec{k}+\vec{k}'),\quad \mbox{with}   
    \quad
    \left\langle \left[ \xi_{\vec{k}}(t), \xi_{\vec{k}'}(t') \right] \right\rangle = 0.
\end{align}
This is the well-studied Markovian  white noise, see, e.g., 
Refs.~\cite{Ronc78,Godr00,Pico02}.  The central question in this work 
essentially concerns the consequences of the non-Markovian quantum noise in 
Eq.~\eqref{eq:qudyn} in comparison with the Markovian classical white noise in 
Eq.~\eqref{eq:classical}. 
\item
{\it Effective dynamics} \cite{Mari16}, inspired by a simple scaling argument of 
the zero-temperature limit of Eqs.~\eqref{eq:qudyn}, i.e.,
\begin{equation}  \label{eq:eff_qn_anticommu}
  \left\langle \left\{ \xi_{\vec{k}}(t), \xi_{\vec{k}'}(t') \right\} \right\rangle 
  = \mu |\vec{k}|^2 \delta(t-t') \delta(\vec{k}+\vec{k}'),\quad \mbox{with}  
  \quad
    \left\langle \left[ \xi_{\vec{k}}(t), \xi_{\vec{k}'}(t') \right] \right\rangle = 0,  
\end{equation}
with a dimensionless control parameter $\mu$. This is a classical noise 
with a momentum-dependent effective temperature $T_{\rm eff}=\mu |\vec{k}|^2/2$. 
As we shall see below, the analysis of the simplified noise correlators in 
Eq.~\eqref{eq:eff_qn_anticommu} is an efficient short-cut for studying ageing, 
since it readily reproduces the ageing behaviour which usually follows from a 
technically demanding analysis of the actual quantum noise in 
Eq.~\eqref{eq:qudyn}.
In particular, the effective description of the noise in 
Eq.~\eqref{eq:eff_qn_anticommu} circumvents the difficulties due to the 
non-locality in time of the actual correlator by introducing a more complicated 
spatial structure of the noise, which is however, amenable to analytical 
calculations. In this spirit, the factor $|\vec{k}|^2$ in 
Eq.~\eqref{eq:eff_qn_anticommu} is the result of the underlying spatio-temporal 
scaling of these models described by the dynamical exponent $z = 2$. 
Heuristically, the effective scaling dependence $\sim (t-t')^{-2}$ of the 
r.h.s.~of Eq.~\eqref{eq:qn_anticommu_T} is replaced in 
Eq.~\eqref{eq:eff_qn_anticommu} by $|\vec{k}|^2 \times \delta(t-t')$ where each 
of the two factors brings in a scaling dependence $\sim (t-t')^{-1}$.
\end{enumerate} 
Non-Markovian effects are most prominent at zero temperature and we shall 
therefore focus our analysis on this case.  The limit $T\to 0$ in 
Eqs.~\eqref{eq:qudyn} does not affect the noise commutator, while the anticommutator in Eq.~\eqref{eq:qn_anticommu_T} 
\textcolor{black}{is given by a  
singular integral. Following standard procedures~\cite{Gard04} this singular
integral is regularised by introducing an additional microscopic time-scale $\tto>0$ such 
that the noise anticommutator reads}
\begin{equation}
 \left\langle \left\{ \xi_{\vec{k}}(t), \xi_{\vec{k}'}(t') \right\} \right\rangle 
 = \frac{\gamma\hbar}{\pi} \int_{-\infty}^\infty  \!\D\omega\: |\omega|\, 
 e^{\II \omega (t-t')} e^{-\tto |\omega|}\: \delta(\vec{k}+\vec{k}').
       \label{eq:qn_anticommu}
\end{equation}
While for 
$\tto=0$, the integral formally diverges, it is possible to interpret it as a 
distribution \cite{Gelf64}. \textcolor{black}{ We prefer to avoid 
the explicit use of distributions and consider below the noise correlator Eq.~(\ref{eq:qn_anticommu}) in its regularised 
form.} Equation~\eqref{eq:qn_anticommu} then gives explicitly~\cite{Gard04}
\begin{equation}
 \left\langle \left\{ \xi_{\vec{k}}(t), \xi_{\vec{k}'}(t') \right\} 
\right\rangle = 2\frac{\gamma\hbar}{\pi} \frac{\tto^2 - 
(t-t')^2}{\left[\tto^2+(t-t')^2\right]^2}\, \delta(\vec{k}+\vec{k}'). 
\label{eq:qn_anticommu_cut}
\end{equation}
In Fig.~\ref{fig:QNOU} this regularised form is compared with that of 
%
%
%
%
equal width of the corresponding regularised classical 
Wiener process obtained by formally taking the limit $\hbar \to 0$ of 
Eq.~\eqref{eq:qudyn} after introducing the factor $e^{-\tto|\omega|}$ in 
Eq.~\eqref{eq:qn_anticommu_T}, as done above for the case $T\to 0$. This process 
is known to describe Brownian motion in the long-time limit, as the noise 
correlator is $ \propto \delta(t-t')$ as $\tto \to 0$. While the central peaks 
in Fig.~\ref{fig:QNOU} look quite similar, the quantum noise decays with a 
power-law tail for large time differences $\tau=(t-t')/\tto$ and, furthermore, 
it is anti-correlated (but within a narrow region, of width $\sim \tto$ around 
$\tau=0$), which is distinct from the exponential decay of the positively 
correlated Wiener process. The present work investigates the consequences of the 
differences in the noise anti-commutators  illustrated in Fig.~\ref{fig:QNOU} on 
the long-time dynamics of the two models described in 
Sec.~\ref{sec:equilibrium}. In particular, we shall address the extent to which 
the non-Markovian character of the quantum noise is important and if the 
effective noise is successful in reproducing quantum properties. 

The role of the cut-off $\tto$, which introduces a new time scale into the 
dynamics, can be better understood as follows. In~\ref{app:commu} we analyse the 
case of a single quantum harmonic oscillator, focusing on the equal-time 
commutation relation and on the virial theorem that are both known to hold true 
for the quantum Langevin dynamics. In the limit $T\to 0$, the inertial term is 
responsible for these quantities to be well-defined by regularising the 
integrals over the bath degrees of freedom. We show that, for the physical 
quantities we focus on, the inertial term can be effectively substituted by the 
regularised noise correlator in the equations of motion. Although the noise 
structure is inherently responsible for conserving these properties, the 
elimination of the inertial terms substantially weakens this effect and a 
cut-off is needed to ensure a sensible quantum dynamics. In this way, the 
damping rate $\gamma$ also sets the cut-off scale as $\tto\sim 1/\gamma$.

Finally, we notice that the over-damped Langevin equation will lead to a 
dynamical exponent $z=2$, in contrast to closed quantum systems where the 
unitary evolution generically leads to $z=1$ \cite{Cala16,Delf18}. 
As stated in point (iii) above a simple dimensional analysis with $z=2$ naturally leads to the 
effective noise correlator~(\ref{eq:eff_qn_anticommu}).

In summary,  we have seen that the long-time behaviour of the spherical model 
and $O(n)$ model for $n\to\infty$ can always be described by the over-damped 
Langevin equation \eqref{eq:qle}, in which the actual physical nature of the 
dynamics at temperature $T=0$ only enters via the noise correlators. In 
particular: (i) for quantum dynamics these correlators are given by 
Eqs.~\eqref{eq:qn_commu} and \eqref{eq:qn_anticommu_cut}, (ii) for classical 
white noise by Eq.~\eqref{eq:classical}, and (iii) for the effective noise, by 
Eq.~\eqref{eq:eff_qn_anticommu}.

\section{Relaxation and Ageing - Analytical Predictions}
\label{sec:results}

In this section, we summarise and interpret our main results. The detailed 
analysis is presented in Secs.~\ref{sec:solution} and~\ref{sec:observables}.  

The formal solution of Eq.~\eqref{eq:qle}, which provides the basis of all 
further analyses, is
\begin{equation}
 \phi_{\vec{k}}(t) =\frac{\exp(-k^2 t/\gamma)}{\sqrt{g(t)\,}\,} \left[ 
\phi_{\vec{k}}(0) +\frac{1}{\gamma}\int_0^t \!\D s\: 
\sqrt{g(s)\,}\,\exp\left(k^2 s/\gamma\right)\, \xi_{\vec{k}}(s) \right],
\label{eq:sol}
\end{equation}
where we introduced
\begin{equation}  \label{gl:g}
g(t) := \exp\left(\frac{2}{\gamma} \int_0^t \!\D s\:  r(s)  \right),
\end{equation}
in analogy with the treatment of the classical non-equilibrium spherical and 
$O(n)$-models for $n\to\infty$ \cite{Godr00,Pico02,Wald18}. Once the function 
$g(t)$ is determined from the consistency conditions in Eqs.~\eqref{eq:Hon} and 
\eqref{eq:Hsm}, the non-equilibrium dynamics of these models is solved.

First, we focus on the upper and lower critical dimensions of these 
models, denoted by $d_u$ and $d_l$, respectively, in comparison with the
classical critical behaviour. At equilibrium, the $T=0$ quantum phase transition 
in $d$ spatial dimensions belongs to the same universality class as the thermal 
phase transition of the corresponding $(d+1)$-dimensional classical 
system~\cite{Hertz76,Henk84,Voj96}. The upper and lower critical dimensions
for the classical system are $d_l^{({\rm cl, eq})}=2$ and $d_u^{({\rm 
cl, eq})}=4$~\cite{Berl52} and thus their quantum analogues are
$d_l^{({\rm qu, eq})}=1$ and $d_u^{({\rm qu, eq})}=3$.
As we shall see, the coupling to an external reservoir does affect this 
behaviour. However, in the diffusive scaling limit $k^2 t={\rm cst.}$ which is 
most natural for the over-damped Langevin equation,
%
%
the lower and upper critical dimensions are shifted as $d_l^{({\rm qu})}=0$ and 
$d_u^{({\rm qu})}=2$ for quantum noise. This differs from both the classical 
dynamics --- for which the lower and upper critical dimensions are the same as 
at equilibrium --- and from the quantum behaviour at equilibrium. It follows that the 
stationary state of the dynamics driven by the quantum Langevin equation 
\eqref{eq:qle} with noises \eqref{eq:qn_anticommu_cut}, and \eqref{eq:qn_commu} 
is neither a classical nor a quantum equilibrium state at the specific time- and 
length-scales dictated by the diffusive scaling limit. This statement is 
supported by the analyses presented below, from which it turns out that it is not 
possible to satisfy a fluctuation-dissipation relation, even for a quench to the 
disordered phase. Accordingly, any phase transitions eventually found in the 
dynamics should be interpreted as a sort of kinetic phase transition. In 
particular, for $0<d\leq1$, there is no equilibrium analogue of an ordered 
phase. 

Our analysis of the dynamics consists in the calculation of the following 
quantities. First, we study the {\it equal-time correlation function} 
$C_{\vec{k}}(t)$ defined by
\begin{equation}
 \delta(\vec{k}+\vec{k}') C_{\vec{k}}(t) :=
 \left\langle \left\{ \phi_{\vec{k}}(t),\phi_{\vec{k}'}(t) \right\} \right\rangle.
\end{equation}
In order to obtain spatio-temporal information about the non-equilibrium state 
at different time and length scales we shall study $C_{\vec{k}}(t)$ in the 
scaling limit
\begin{equation}\label{eq:spatiotemp} t\to\infty ,  \quad  k\to 0 , \quad  \rho 
:= \frac{k^2 t}{\gamma} = {\rm cst.}
\end{equation}
Next, we investigate {\it the two-time linear response functions}
\begin{equation}\label{eq:R}
 R_{\vec{k}}(t,s):=\frac{\delta 
\left\langle\phi_{\vec{k}}(t)\right\rangle}{\delta 
h_{\vec{k}}(s)}\bigg|_{h=0}\quad \mbox{and} \quad R(t,s) := 
\int_{\vec{k},(\Lambda)} \frac{\delta 
\left\langle\phi_{\vec{k}}(t)\right\rangle}{\delta h_{\vec{k}}(s)}\bigg|_{h=0},
\end{equation}
where, assuming spatial rotational invariance, we introduce the short-hand  
$\int_{\vec{k},(\Lambda)} = \int_0^{\Lambda}\int_{S^d} \D\vec{k}/(2\pi)^d$ for 
the momentum integration, over a hyper-sphere $S^d$ with radius up to  
$\Lambda$. $ R_{\vec{k}}(t,s)$ and $R(t,s)$, respectively, indicate the response 
of the order parameter $\left\langle \phi_{\vec{k}}(t)\right\rangle$ at time $t$ 
to a perturbation of its conjugate field $h_{\vec{k}}(s)$ at time $s$ and the 
(auto-) response of the order parameter at a certain point in space to an 
earlier perturbation applied at the same point. The auto-response function 
$R(t,s)$ is particularly useful for  studying ageing behaviour \sw{\cite{Bray94, 
Henk10,Cugl03,Godr00}}. Usually, one refers to $s$ as the {\it waiting time} and 
to $t$ as the {\it observation time}. Finally, we consider the two-time 
correlation functions 
\begin{equation} \label{gl:initial}
\delta(\vec{k}+\vec{k}') C_{\vec{k}}(t,s) := \left\langle \left\{ 
\phi_{\vec{k}}(t),\phi_{\vec{k}'}(s) \right\} \right\rangle \quad\mbox{and}\quad 
C(t,s) := \int_{\vec{k},(\Lambda)}C_{\vec{k}}(t,s) ,
\end{equation}
which describe, respectively, how correlations propagate across the system and 
how the auto-correlation evolves.\footnote{Since the equation of motion 
\eqref{eq:qle} is linear and we consider quantities at most quadratic in the 
order parameter $\phi_{\vec{k}}$, only the second moments of the noises are 
required. } 
\begin{figure}[t]
 \centering
 \includegraphics[width=.5\textwidth]{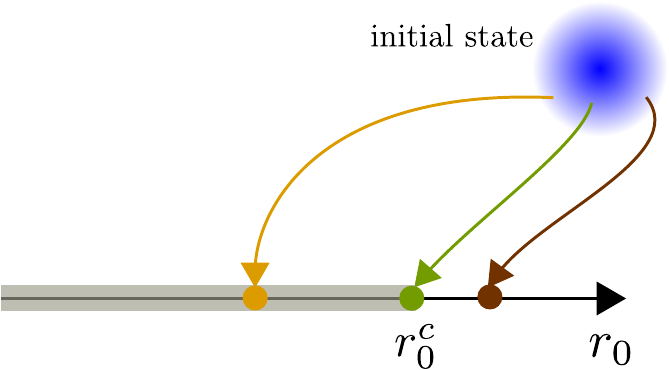}
 \caption[Quench protocols]{Schematic illustration of the three distinct quench 
protocols considered in the present work. The gray shaded area indicates 
qualitatively the stationary ordered phase of the models under investigation. 
The system is prepared in a certain excited state with vanishing order parameter 
and subsequently quenched either to the ordered regime, the critical point or 
the disordered phase. All quenches lead to a formal zero-temperature state 
which, however, is not an equilibrium state.  }
 \label{fig:quenches}
\end{figure}

In Fig.~\ref{fig:quenches}, we illustrate the three different quench protocols 
that we consider below. First, for quenches which remain within the one-phase 
region, i.e., for $r_0>r_0^c$, we find a rapid relaxation towards a stationary 
state and we analyse the influence that the quantum noise statistics has on its 
properties, in particular how they differ from the equilibrium case. The ageing 
behaviour which is studied via the scaling of $R_{\vec{k}}(t,s)$ and 
$C_{\vec{k}} (t,s)$, defined in Eqs.~(\ref{eq:R}) and (\ref{gl:initial}) arises 
for quenches onto or below the critical point, i.e., for $r_0\leq r_0^c$. In 
this case, we analyse the dynamics in the long-time scaling limit \cite{Henk10} 
in which both times $t$ and $s$ are simultaneously large such that 
\begin{equation}\label{eq:scale2}
s\to\infty, \quad t\to \infty \quad \mbox{with fixed} \quad y:=t/s  >1.  
\end{equation}
Then the two-time auto-response and auto-correlation functions are expected to 
scale as
\BEQ \label{gl:vieux}
R(t,s) = s^{-1-a} f_R\left(t/s\right) \quad\mbox{and}\quad C(t,s) = s^{-b} 
f_C\left(t/s\right). 
\EEQ
In the limit $y\to\infty$ one expects for the asymptotics of the scaling 
functions $f_{R,C}$
\begin{equation}\label{eq:lambdas}
f_R(y) \simeq f_{R,\infty}\: y^{-\lambda_R/z} \quad\mbox{and}\quad f_C(y) \simeq 
f_{C,\infty}\: y^{-\lambda_C/z},
\end{equation}
which define the auto-response exponent $\lambda_R$, the auto-correlation 
exponent $\lambda_C$ and the ageing exponents $a$ and $b$. Their values depend 
on whether $r_0<r_0^c$ or $r_0=r_0^c$. We shall also see that $z=2$ throughout.

We shall study the coarsening 
as well as the ageing behaviour by preparing the system in an initial state with
\begin{equation}\label{eq:C0}
\left\langle \phi_{\vec{k}}(0)\right\rangle = 0 \quad
\text{and} \quad C_{\vec{k}}(0) 
\stackrel{k\searrow 0}{=} c_\alpha k^\alpha,
\end{equation}
where $C_{\vec{k}}(t)$ is defined in Eq.~\eqref{gl:initial} such that the 
initial order parameter vanishes but yet we admit long-range initial 
correlations $C(0,\vec{r})\sim |\vec{r}|^{-d-\alpha}$ as $|\vec{r}|\to\infty$ 
with $d+\alpha>0$ such that the initial correlations decay upon increasing the 
distance. While the case $\alpha=0$ describes short-ranged initial correlations 
\cite{Godr00}, long-ranged initial correlations are obtained for $\alpha<0$ 
\cite{Pico02}. Because of the vanishing initial order parameter, we interpret 
the initial state (\ref{eq:C0}) as being disordered, see 
Fig.~\ref{fig:quenches}.

With the initial conditions \eqref{eq:C0} and a generic bath correlator 
\begin{equation}
\label{eq:def-k-m}
\left\langle \{ \xi_{\vec{k}}(t'), \xi_{\vec{k}'}(s') \}\right\rangle = 
m_{\vec{k}}(t'-s')\delta(\vec{k}+\vec{k}'),
\end{equation}
the two-time response and correlation functions, defined according to 
Eqs.~\eqref{eq:R} and \eqref{gl:initial},  can be expressed as
\begin{subequations} \label{gl:RC-formels}
\begin{align}
R_{\vec{k}}(t,s) &= \frac{1}{\gamma} \sqrt{\frac{g(s)}{g(t)}\,}\, 
e^{-k^2(t-s)/\gamma} \Theta(t-s),\label{gl:R-formel} \\ 
C_{\vec{k}}(t,s) &= \gamma^2 R_{\vec{k}}(t,0) R_{\vec{k}}(s,0) C_{\vec{k}}(0)  + 
\int_0^t\!\D t' \int_0^{s}\!\D s'\: R_{\vec{k}}(t,t')R_{\vec{k}}(s,s') 
m_k(t',s'), \label{gl:C-formel}
\end{align}
\end{subequations}
where the Heaviside function $\Theta(t-s)$ imposes the causality condition 
$t>s$ for the response function.  By setting $t=s$ in (\ref{gl:C-formel}), one 
obtains the equal-time (or one-time) correlator $C_{\vec{k}}(t)=C_{\vec{k}}(t,t)$.

\subsection{Effective Dynamics}
\label{ssec:eff}

Compared to solving the dynamics of the model in the presence of the actual 
quantum noise, it turns out that it is considerably easier to determine the 
dynamical behaviour driven by the Markovian effective 
noise~(\ref{eq:eff_qn_anticommu}), starting from the initial 
conditions~(\ref{eq:C0}). Since the effective noise is constructed such that its 
scaling dimension is the same as the one of the actual quantum noise 
(\ref{eq:qn_anticommu_cut}), the leading scaling behaviour and exponents which 
characterise the emerging ageing behaviour in the presence of the effective 
Markovian dynamics will also hold for the actual quantum dynamics, as we shall 
see below. For concreteness, we use here the language of the $O(n)$-model in the 
$n\to\infty$ limit, but all universal quantities concerning the long-time 
behaviour will be the same as for the spherical model. The  constraint in 
Eq.~(\ref{eq:Hon}) reduces to a linear integro-differential equation for the 
function $g(t)$,  defined in Eq.~(\ref{gl:g}). Standard techniques for the 
solution are available \cite{Henk10,Godr00,Pico02}, the main steps of which we 
recall in~\ref{app:eff}. We obtain the (non-universal) critical point of the 
dynamics 
\begin{equation}
   {r}_0^c = - \frac{u}{12}\frac{\mu }{\gamma} \frac{\Omega_d}{(2\pi)^d} \frac{\Lambda^{d}}{d},
\end{equation}
where $\Omega_d$ is the $d$-dimensional solid angle. For a fixed 
{\sc uv} cutoff $\Lambda<\infty$, ${r}_0^c$ is finite for all spatial 
dimensions $d>0$, which confirms the above argument on the lower critical 
dimension $d_l=0$. Depending on the sign of the difference $r_0-r_0^c$ we can 
distinguish the various cases schematically represented in 
Fig.~\ref{fig:quenches}, which we discuss below. 
\begin{enumerate}
 \item[{\bf (1)}]
 For a quench remaining in the disordered phase ${r}_0 > {r}_0^c$, we find an 
exponential long-time growth 
\begin{equation}
\label{eq:g-exp}
g(t)\sim \exp(t/\tau_{\text{r}})
\end{equation}
of the function $g(t)$, with a characteristic time scale $\tau_{\text{r}}$. This 
time scale diverges as the quench parameter $r_0$ approaches criticality from 
above, i.e., as $r_0-r_0^c\to 0^+$. At long times after the quench, i.e., in the 
stationary limit, the equal-time correlation function $C_ {\vec{k}}(\infty)$ 
becomes 
\begin{equation} \label{eq:eff-C-single} 
 C_ {\vec{k}}(\infty) \simeq \frac{\mu}{\gamma^2}\frac{k^2}{1/\tau_{\text{r}} + 
2 k^2/\gamma}.
\end{equation}
This reproduces the standard classical Ornstein-Zernicke form (see, e.g., 
Ref.~\cite{Godr00}), up to the momentum-dependent effective temperature $T_{\rm 
eff}=T_{\rm eff}(k) =\mu k^2/2$ which comes from the noise correlator in 
Eq.~(\ref{eq:qn_anticommu}). In the same stationary limit, the two-time response 
and correlation functions rapidly converge to 
\begin{subequations}
\label{eq:effR}
\begin{align}
 R_{\vec{k}} (t,s) &\simeq \frac{1}{\gamma} 
\exp\left(-\left(\frac{1}{2\tau_{\text{r}}}+\frac{k^2}{\gamma}
\right)(t-s)\right),\\[.25cm]
C_{\vec{k}}(t,s) &\simeq  \frac{\mu k^2}{\gamma^2} 
\frac{1}{\frac{1}{\tau_{\text{r}}}+2\frac{k^2}{\gamma}} \exp 
\left(-\left(\frac{1}{2\tau_{\text{r}}} +\frac{k^2}{\gamma}\right)(t-s)\right),
\end{align}
\end{subequations}
which only depend on the time difference $\tau=t-s$ (with exponentially small 
corrections in $t,s$ and $\tau$).  Note that all these expressions do not 
contain any reference to the parameter $\alpha$ which describes the initial 
correlations, indicating that the memory of the initial state is lost in the 
long-time limit. In addition, the expressions \eqref{eq:effR} satisfy an 
effective version of the classical fluctuation-dissipation theorem for every 
mode, i.e., for $\tau>0$,\footnote{One should not confuse the equal-time correlator $C_{\vec{k}}(t)$ with the time-translation-invariant two-time correlator $C_{\vec{k}}(\tau) = C_{\vec{k}}(\tau+s,s)$ in those cases where it is independent of the waiting time $s$.}
\begin{equation}\label{eq:fdt_eff}
\frac{\partial C_{\vec{k}}(\tau)}{\partial \tau}  =  
-\frac{\mu}{2\gamma}k^2R_{\vec{k}}(\tau)  = -\frac{T_{\rm eff}(k)}{\gamma} 
R_{\vec{k}}(\tau), 
\end{equation}
with the same mode-dependent effective temperature as the one determined above. 
This classical behaviour controlled by $T_{\rm eff}(k)$ is a direct consequence 
of the form of the noise (\ref{eq:eff_qn_anticommu}). 
\begin{table}[b]
\captionsetup{width=.8\textwidth}
\setlength{\tabcolsep}{10pt} 
\renewcommand{\arraystretch}{1.5} 
\centering
\caption{Non-equilibrium exponents for critical and subcritical quenches. In the 
former case, one distinguishes three possible regions I -- III (see 
Fig.~\ref{fig:dalpha}) depending on the dimension $d$.
The exponent $\digamma$ is defined in Eq.~(\ref{gl:g-crit}).
} 
\begin{tabular}{|c|lc||c|c|c|c|c|}\hline
\multicolumn{2}{|c||}{quench}  & $\digamma$ & $\lambda_C$ & $\lambda_R$ & $a$ & 
$b$  \\ \hline \hline
\multirow{3}{*}{$r_0=r_0^c$} &\multicolumn{1}{|l||}{I. $0<d<2$} & 
$-\frac{\alpha}{2}$   & $d+\frac{\alpha}{2}$ & $d-\frac{\alpha}{2}$ & 
$\frac{d}{2}-1$ & $\frac{d}{2}$ \\\cline{2-7}
&\multicolumn{1}{|l||}{II. $2<d$, $d+\alpha<2$} & $1-\frac{d+\alpha}{2}$ & 
$1+\frac{d+\alpha}{2}$ & $\frac{d-\alpha}{2}+1$ & $\frac{d}{2}-1$ & $1$ \\ 
\cline{2-7}
&\multicolumn{1}{|l||}{III. $2<d$, $d+\alpha>2$} & 0         & $d+\alpha$       	
 & $d$              	& $\frac{d}{2}-1$ & $\frac{d+\alpha}{2}$ \\ \hline 
\multicolumn{2}{|l||}{$r_0<r_0^c$}& $-\frac{d+\alpha}{2}$ & $\frac{d+\alpha}{2}$ 
& $\frac{d-\alpha}{2}$ & $\frac{d}{2}-1$ & $ 0$		 	  \\\hline
\end{tabular}
\label{tab:eff}
\end{table}
\setlength{\tabcolsep}{6pt}
\renewcommand{\arraystretch}{1.0}

\item[{\bf (2)}]
For a critical quench, $g(t)$  displays a very different long-time behaviour, 
being algebraic rather than exponential. This is expected since the relaxation 
time-scale $\tau_{\text{r}}$ diverges as $r_0\to r_0^c$ approaches criticality. 
Depending on the spatial dimension $d$, it turns out that three different 
cases must be distinguished, as schematically shown in Fig.~\ref{fig:dalpha}:
\begin{figure}[t]
 \centering
 \includegraphics[width=.5\textwidth]{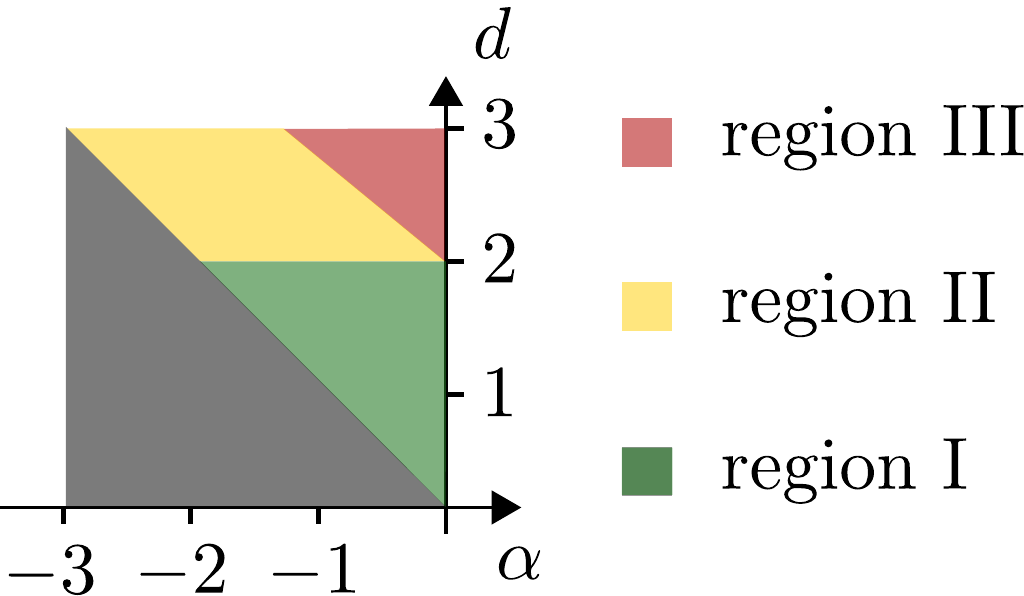}
 \caption{Illustration of the different regions as a function of the 
dimension $d$ and the exponent $\alpha$, corresponding to the initial 
correlations different asymptotic dynamics for a quench to the critical point as 
summarised in Table~\ref{tab:eff}.}
\label{fig:dalpha}
\end{figure}
\begin{itemize}
 \item region I: $0<d<2$ and $0<d+\alpha $. The fluctuations introduced in the 
dynamics by the quantum reservoir are relevant.
 
 \item region II: $2<d$ and $0<d+\alpha <2$. The quantum fluctuations due to the 
reservoir are irrelevant but the initial fluctuations are relevant. 
 
 \item region III: $2<d$ and $2<d+\alpha$. The quantum fluctuations due to the 
reservoir and those due to the initial state are both irrelevant and the scaling 
behaviour of the system is accurately described by the mean-field theory. 
\end{itemize}
In all three critical cases we find
\begin{equation} \label{gl:g-crit}
 g(t) \sim t^{\digamma}.
\end{equation}
The values of the exponent $\digamma$ are listed in Table~\ref{tab:eff}. Region 
III corresponds to having $\digamma=0$, i.e., the spherical constraint does not 
affect the dynamics, which is therefore the same as that of a set of 
uncorrelated free bosonic modes. 

Using the formal solution (\ref{eq:sol}), we obtain the equal-time correlation 
function, directly in the scaling limit~(\ref{eq:spatiotemp})
\begin{subequations} \label{gl:effektiv_CRC}
\begin{equation} \label{gl:effektiv_Ct}
 C_{\vec{k}} (t) \simeq \rho^{-\digamma} e^{-2\rho}\left[ \mathscr{C}_{(1)} 
c_\alpha k^{\alpha+2\digamma} + \frac{\mu}{\gamma} \int_0^\rho \!\D x\: e^{2x} 
x^{\digamma} \right],
\end{equation}
where $\mathscr{C}_{(1)}$ is a known constant and $\rho$ is defined in 
Eq.~\eqref{eq:spatiotemp}. Depending on the sign of $\alpha+2\digamma$, either 
the initial correlations (first term) or the reservoir fluctuations (second 
term) dominate Eq.~(\ref{gl:effektiv_Ct}) in the scaling limit. The exponents in 
table~\ref{tab:eff} show that in region I, both terms contribute while in 
regions II and III, the initial correlations are dominant.  For the two-time 
auto-response and autocorrelation functions we find, in the scaling regime 
(\ref{eq:scale2}), with $y=t/s>1$,
\begin{align}
 R(ys,s) &= R_{(0)} s^{-d/2} y^{-\digamma/2} (y-1)^{-d/2} ,  \\[.2cm]
 C(ys,s) &\simeq s^{-d/2}\left[ \mathscr{C}_{(2),1} c_\alpha 
s^{-\digamma-\alpha/2} \frac{y^{-\digamma/2}}{(1+y)^{(d+\alpha)/2}} +  
\mathscr{C}_{(2),2} \frac{\mu}{\gamma} \int_{-1}^1 \D x  \, 
\frac{(1-x)^{\digamma}}{(y+x)^{d/2 + 1}} \right],
\end{align}
\end{subequations}

where $R_{(0)}$, $\mathscr{C}_{(2),1}$ and $\mathscr{C}_{(2),2}$ are known 
constants whose values will not be needed here. Up to an overall normalisation, 
the auto-response function is universal and is independent of both the initial 
and the noise correlations. For the autocorrelator, instead, we find that both 
sources of fluctuations contribute to the scaling function in region I,  
although the leading asymptotic behaviour of $f_C(y)$ (see Eq.~(\ref{gl:vieux})) 
for $y\to\infty$ is dominated by the initial noise correlations. On the other 
hand, in  regions II and III, the initial correlations dominate for $\alpha<0$. 
Based on their definitions in Eqs.~(\ref{gl:vieux}) and (\ref{eq:lambdas}), the 
auto-response, autocorrelation and ageing exponents can now be easily determined 
from the asymptotic behaviours of  Eq.~\eqref{gl:effektiv_CRC}, resulting in the 
values listed in Table~\ref{tab:eff}.

\item[{\bf (3)}] 
For a quench across the critical point and into the ordered phase, i.e., with 
$r_0<r_0^c$, the long-time behaviour is again algebraic as in 
Eq.~\eqref{gl:g-crit} with $\digamma = -(d+\alpha)/2$. In terms of this 
$\digamma$, the relevant correlation and response functions are still given by 
Eq.~(\ref{gl:effektiv_CRC}), provided that $d+\alpha <2$. In this case, 
Eqs.~(\ref{gl:effektiv_CRC}) hold exactly as the asymptotic form of $g(t)$ is 
integrable in the origin. For $d+\alpha>2$ the noise contributions in 
Eqs.~(\ref{gl:effektiv_CRC}) are formally not defined, which is merely an 
artifact of substituting the asymptotic limit too early. However, this does not 
actually matter as the $k$-exponent in the initial term remains negative and 
thus the initial contributions keep on dominating (as opposed to the critical 
case I). Critical and subcritical quenches are distinct in one important aspect. 
Namely, for subcritical quenches, the initial correlations always dominate the 
long-time asymptotic behaviour, which is not always the case for critical 
quenches. Comparing the auto-response and auto-correlation exponents $\lambda_R$ 
and $\lambda_C$, defined in Eq.~(\ref{eq:lambdas}) respectively, for quenches 
with $r_0\leq r_0^c$, we see that one always has
\BEQ
\lambda_C = \lambda_R + \alpha.
\EEQ
This relationship is known to hold also for classical dynamics, and it has been 
derived both from the analysis of explicit models and from general arguments, 
although within a different range of dimensions 
\cite{Bray94,Pico02,Pico04,Henk10}. 
\end{enumerate}
Summarising, is there any evidence for a clear quantum effect on the ageing 
behaviour? The answer is certainly affirmative because of the dimensional shift 
$d\mapsto d-2$ when going from classical dynamics at $T>0$ to \emph{quantum} 
dynamics at $T=0$. But is there any additional contribution coming from the 
noise correlations of the quantum bath? Our results for the effective quantum 
noise suggest that this can happen only in case I of the critical quench, i.e., 
for dimensions $d<2$. In addition, in the presence of long-range initial 
correlations with $\alpha<0$, the leading asymptotics of the scaling function 
$f_C(y)$  for $y\to\infty$ depends only on the initial noise correlator, see 
Eq.~(\ref{gl:effektiv_CRC}). However, for fully disordered initial conditions 
with $\alpha=0$, both quantum and initial noise correlators do contribute to the 
scaling function $f_C(y)$ at a generic value of $y$. Accordingly, also the 
asymptotic amplitude $f_{C,\infty}$ contains a non-vanishing contribution from 
the bath noise correlations and therefore a signature of the original quantum 
nature of the system.

\subsection{Quantum Noise}
\label{ssec:qresults}

The quantum noise in Eqs.~(\ref{eq:qn_anticommu_T}) and 
(\ref{eq:qn_anticommu_cut}) is not Markovian. This implies that the techniques 
used above and in~\ref{app:eff} for solving the spherical constraint in 
Eq.~(\ref{eq:Hon}) in the case of the effective dynamics are no longer 
applicable.  In fact, the non-Markovianity leads to a {\it non}-linear 
integro-differential equation for $g(t)$, rather than a linear one, as we shall 
see in Eqs.~(\ref{eq:selfconvoSM}) and~(\ref{eq:selfconvo}).

The approach we used to solve it is explained in Secs.~\ref{sec:solution} 
and~\ref{sec:observables}. Here we provide an overview of the conclusions of 
this analysis.  Various aspects of the results we present below for the emerging 
scaling behaviour of the dynamics can be expressed in terms of the asymptotics 
of the auxiliary function (see Eqs.~(5.2.13), (5.2.16), and (5.2.35) in 
Ref.~\cite{Abra65}) with $x>0$,
\begin{equation} \label{eq:aux}
 g_{\text{AS}}(x) := \int_0^\infty  \!\!\D t\, \frac{\cos t}{t+x}\simeq 
\begin{dcases} -\left({\rm C}_E +\ln x\right)&\mbox{for} \quad x\ll 1,\\ 
1/x^2\hspace{1.84truecm}&\mbox{for} \quad x\gg1,
 \end{dcases} 
\end{equation}
where ${\rm C}_E=0.5772\ldots$ is Euler's constant.  Using 
the formulation in terms of the 
$O(n)$-model with $n\to \infty$, the critical point at $T=0$ 
and in the limits $\gamma\to\infty$, $\Lambda\to\infty$, $\tto \to 0$ 
with $\Lambda^2 \tto/\gamma ={\rm cst.}$ can be evaluated. For any $d>0$ we find
\begin{equation} \label{gl:r0c-qu}
 {r}_0^c =\frac{u}{12} \frac{4\hbar}{\pi\gamma}\frac{\Omega_d}{(2\pi)^d} 
  \frac{\Lambda ^d}{d} \left[\ln \left(\Lambda^2\frac{\tto }{\gamma }\right)+
  {\rm C}_E  -\frac{2}{d}\right].
\end{equation}
Note that ${r}_0^c$ is finite for any $d$ and therefore we conclude that 
$d_l=0$, while the actual value of $r_0^c$ now  depends on both cut-off 
parameters $\Lambda$ and $\tto$.

As in Sec.~\ref{ssec:eff}, we discuss separately the different possible 
quenches.
\begin{enumerate}
 \item[\bf (1)] 
For a quench to the disordered phase $r_0<r_0^c$, we  still find the exponential 
long-time behaviour of $g(t)$ as in Eq.~\eqref{eq:g-exp}. Hence, we can identify 
a finite time-scale $\tau_{\rm r}$ which is distinct from that of the effective 
quantum noise but still diverges as $r_0 \to r_0^{c,+}$.  The equal-time 
correlation function then reads, in the stationary limit,
\begin{align} \label{gl:correl-stat-disor}
 C_{\vec{k}}(\infty) \simeq \frac{\hbar}{\pi \gamma} g_{\rm AS}\left( \tto 
\left( k^2/\gamma + (2\tau_{\text{r}})^{-1} \right) \right),
\end{align}
instead of Eq.~\eqref{eq:eff-C-single}. In Fig.~\ref{fig:1phase_corr} (left 
panel), this stationary correlator is shown as a function of the momentum $k$. 
For small values of $k$, its shape is almost identical to the one of the 
classical Ornstein-Zernicke form. However, at larger momenta a crossover towards 
a different behaviour occurs. The expression for the stationary correlations in 
(\ref{eq:eff-C-single}), corresponding to the effective dynamics is, however, 
completely different, as shown in the figure. If the Ornstein-Zernicke form 
were exact, the stationary spatial correlator 
$C(\infty,R) = \int_{\vec{k}} e^{\II \vec{k}\cdot \vec{R}} C_{\vec{k}}(\infty)$
in $d=1$ would become 
exponential, i.e., $C(\infty,R)= e^{-(\gamma/\tau_{r})^{1/2} |R|}$ as a function 
of the distance $R$. The right panel in Fig.~\ref{fig:1phase_corr} shows the 
stationary correlator $C(\infty,R)$, obtained from 
Eq.~(\ref{gl:correl-stat-disor}), as a function of the distance $R$, for several 
values of the relaxation time $\tau_{\rm r}$. Indeed, although there is an 
exponential decay for sufficiently large values of $|R|$, the correlator 
$C(\infty,R)$ is rounded, compared to the exponential, for $|R|\to 0$ which 
comes about since the spherical spins are softer than, e.g., those of the $d=1$ 
Glauber-Ising model~\cite{Henk10}. 
%
%
\begin{figure}[t]

\begin{subfigure}{.25\textwidth}
\begin{tikzpicture}
\begin{axis}[xmin=0,xmax=3,ymin=0,ymax=1,
    xlabel={$k$},
    ylabel={$C_k(\infty)/A$}]

        \node[draw,align=left] at (40,70) {quantum noise};
     \addplot[smooth, line width = 2pt, red]
     table [y=C, x=k]{quantum.dat};
     \addlegendentry{QN};
     
          \addplot[ smooth, samples = 100, line width = 2pt, blue,dashed]{x^2/(0.1+x^2)};
     \addlegendentry{eff}

               \addplot[ smooth, samples = 100, line width = 2pt, black]{
               .2/(.2+x^2)};
     \addlegendentry{OZ}

    \end{axis}
      
\end{tikzpicture}

\end{subfigure}
\hspace{4cm}
\begin{subfigure}{.25\textwidth}
\begin{tikzpicture}
\begin{axis}[xmin=0,xmax=10,ymin=0,ymax=1,
    xlabel={$R$},
    ylabel={$C(\infty,R)$}]

     \addplot[smooth, line width = 2pt, black]
     table [y=t10, x=R]{datslowk.dat};
     \addlegendentry{$\tau_r = 10^1$};
     
          \addplot[ smooth, line width = 2pt, blue,dashed] table [y=t100, x=R]{datslowk.dat};
     \addlegendentry{$\tau_r = 10^2$}

    \addplot[smooth, line width = 2pt, red] table [y={t1000}, x=R]{datslowk.dat};
     \addlegendentry{$\tau_r = 10^3$}
     


    \end{axis}
\end{tikzpicture}

\end{subfigure}
 

 \caption[Equal-time correlators]{
 Equal-time correlation function $C_k(\infty)$ for a quench 
to the disordered phase in $d=1$ spatial dimension and with $\gamma=1$. 
 Left panel: stationary correlation function in momentum space for the effective 
dynamics (eff) and the quantum noise (QN), compared to the Ornstein-Zernicke 
form (OZ). For illustration purposes we introduced a normalisation parameter $A$ 
and choose $A = C_{\infty}(\infty)$ for the effective 
noise and $A = C_{0}(\infty)$ for the OZ form and the quantum 
noise.
Right panel: stationary real-space correlator $C(\infty,{R})$, derived from 
Eq.~(\ref{gl:correl-stat-disor}), as a function of the distance $R$ and normalised 
such that $C(\infty, 0) = 1$.
}
\label{fig:1phase_corr}
\end{figure}
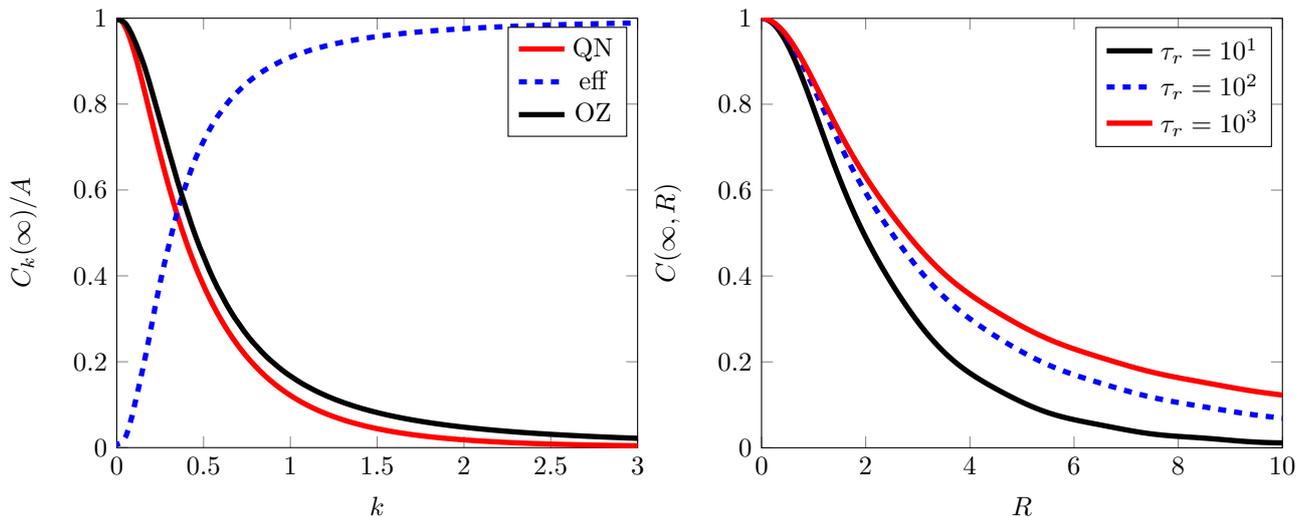
%
Since the Langevin equation (\ref{eq:qle}) is linear, the two-time response 
function is given by the same expression as in Eq.~\eqref{eq:effR}, found for 
the effective noise. However, the two-time correlation function shows a 
different behaviour, which  depends on the separation $\tau =t-s$  of the 
involved times $t$ and $s$, with $t>s$. We focus here on the case $t\gg 
s\gg 1$ with $\tau$ being kept fixed and large. Then 
\begin{subequations}
 \begin{align}
 C_{\vec{k}}(s+\tau,s) &\stackrel{s\gg\tau}{\simeq} -\frac{2\hbar}{\pi \gamma} 
\frac{1}{\left[(2\tau_{\text{r}})^{-1} + k^2/\gamma \right]^2} 
\frac{1}{\tau^2},\\[.25cm]
 R_{\vec{k}} (\tau) &\simeq \frac{1}{\gamma} 
\exp\left(-\left(\frac{1}{2\tau_{\text{r}}}+\frac{k^2}{\gamma}\right)
\tau\right).
\end{align}
\end{subequations}
Although these expressions are stationary, being dependent only on $\tau$, the 
correlator features an algebraic rather than an exponential decay for $\tau\gg 
1$. Thus, no obvious fluctuation-dissipation ratio emerges. This shows that {\it 
the steady state of the zero-temperature quantum dynamics after a quench in the 
one-phase region cannot be an equilibrium state}. 

\item[\bf (2)] 
For a critical quench, we need to distinguish the different cases I-III listed 
in Table~\ref{tab:eff}. Since the relaxational time scale $\tau_{\rm r}$ which 
characterises the sub-critical quench diverges when criticality is approached, 
we now have an algebraic long-time behaviour as in Eq.~\eqref{gl:g-crit}, with a 
multiplicative prefactor which we denote by $g_c$ and an exponent $\digamma$ 
which is the same as for the effective dynamics. In order to calculate the 
remaining exponents $\lambda_{C,R}$, $a$, and $b$, from Eqs.~\eqref{eq:lambdas} 
and \eqref{gl:vieux} we discuss below the correlation and response functions. 
The equal-time correlation function $C_{\vec{k}}(t)$ can be decomposed into two 
contributions,
\begin{subequations}\label{gl:corr-tempsegaux} 
\begin{equation}
 C_{\vec{k}}(t)  =  C_{\vec{k}}^{\rm (ic)}(t) +  C_{\vec{k}}^{\rm (n)}(t) 
\label{gl:corr-tempsegaux-a}, 
\end{equation}
the first being determined by the  correlations in the initial state (ic) and 
the second by the quantum noise (n). In turn, as we shall show in 
Sec.~\ref{sec:observables}, these quantities admit the following scaling 
behaviours in the scaling limit~(\ref{eq:spatiotemp}):
\begin{align}
  C_{\vec{k}}^{\rm (ic)}(t) &\simeq  \frac{c_\alpha}{g_c} (k^2 t)^{\-\digamma} 
e^{-2 k^2 t /\gamma} k^{\digamma + \alpha/2}, \label{gl:corr-tempsegaux-b}  \\[.25cm]
  C_{\vec{k}}^{\rm (n)}(t)  &\simeq -\frac{4\hbar}{\pi \gamma} \left[ {\rm C}_E 
+\ln \left(\frac{\tto k^2}{\gamma}\right)\right] -\frac{2\hbar}{\pi \gamma} 
\left(\frac{k^2t}{\gamma}\right)^{-\digamma} \Psi\left( \frac{k^2 t}{\gamma}, 
\frac{k^2 t}{\gamma}\right), \label{gl:corr-tempsegaux-c}
\end{align}
\end{subequations}
with a universal scaling function $\Psi(\rho,\rho)$ whose form -- reported in 
Eq.~\eqref{gl:Psi-scal-funk} -- depends only on $\digamma$ and $c_\alpha$ and is 
given by the initial conditions in Eq.~(\ref{eq:C0}). From the integral 
representation (\ref{gl:Psi-scal-funk}) derived in Sec.~\ref{sec:solution} 
below, it follows that  $\Psi(\rho,\rho) \stackrel{\rho\gg1}{\sim} 
\rho^{\digamma-1}$ for large arguments $\rho$. By itself, this scaling form 
fixes the dynamical exponent $z=2$, because the relevant variable turns out to 
be $k^2t$ and therefore $t \sim k^{-2}$. In case I, both terms in 
Eq.~(\ref{gl:corr-tempsegaux-a}) equally contribute to the final expression and 
their relative importance is fixed by the non-universal amplitude 
$c_{\alpha}/g_c$. In cases II and III, instead, the initial term  in 
Eq.~(\ref{gl:corr-tempsegaux-b}) dominates over the quantum noise contribution 
in the scaling limit. The two-time auto-response and autocorrelation functions 
are obtained from a careful asymptotic analysis of the double integral involving 
the quantum noise memory kernel. In the scaling limit, see 
Eq.~(\ref{eq:scale2}), we find 
\begin{align}\label{eq:RCc}
 R(t,s) &= R_{(0)} s^{-d/2} \left(\frac{t}{s}\right)^{-\digamma/2} 
\left(\frac{t}{s}-1\right)^{-d/2}, \quad
 C(t,s) \sim s^{-d/2} \frac{(t/s)^{-\digamma/2}}{(1+t/s)^{(d+\alpha)/2}} \ .
\end{align}
Using Eq.~(\ref{eq:lambdas}), we read off the autoresponse and 
autocorrelation exponents
\begin{equation}
 \lambda_R = d+\digamma,\quad\mbox{and}\quad \lambda_C = d +\alpha +\digamma,
\end{equation}
respectively. These results are identical to those of effective dynamics 
reported in Table~\ref{tab:eff}.
%

 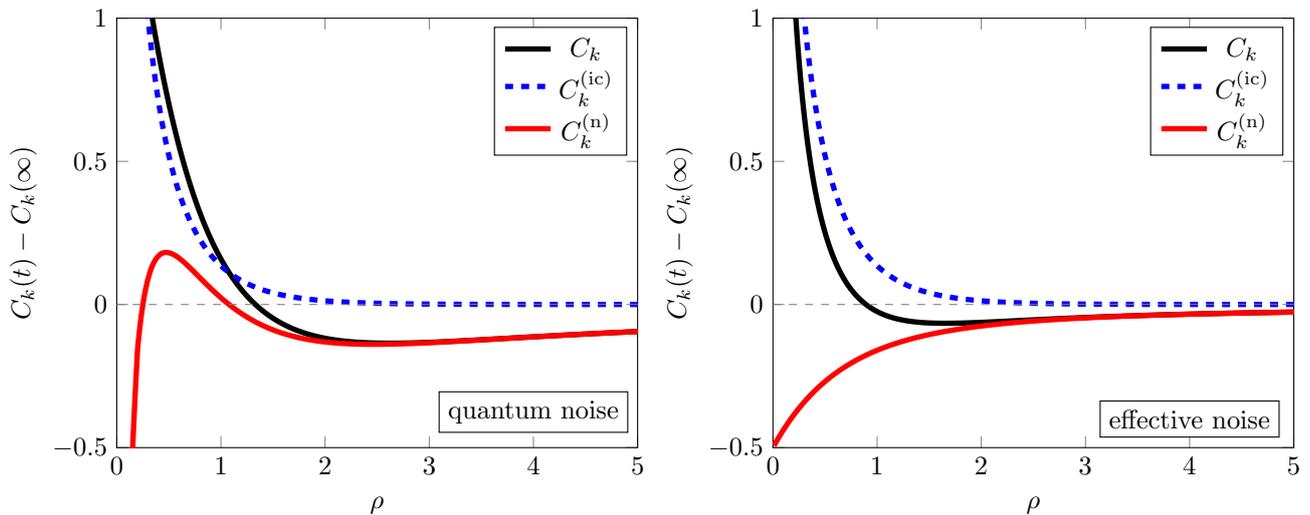
\begin{figure}[t]
\begin{subfigure}{.2\textwidth}
\begin{tikzpicture}
\begin{axis}[xmin=0,xmax=5,ymin=-0.5,ymax=1,
    xlabel={$\rho$},
    ylabel={$C_k(t)-C_k(\infty)$}]

        \node[draw,align=left] at (40,70) {quantum noise};
     \addplot[smooth, line width = 2pt, black]
     table [y=t, x=rho]{dat_C_t_QN.dat};
     \addlegendentry{$C_k$};
     
          \addplot[ smooth, line width = 2pt, blue,dashed] table [y=ic, x=rho]{dat_C_t_QN.dat};
     \addlegendentry{$C_k^{\rm (ic)}$}

    \addplot[smooth, line width = 2pt, red] table [y=n, x=rho]{dat_C_t_QN.dat};
     \addlegendentry{$C_k^{\rm (n)}$}
     
        \addplot[domain = 0:5,samples = 100, dashed, gray] {0};

    \end{axis}
      
\end{tikzpicture}

\end{subfigure}
\hspace*{5cm}
\begin{subfigure}{.2\textwidth}
\begin{tikzpicture}
\begin{axis}[xmin=0,xmax=5,ymin=-0.5,ymax=1,
    xlabel={$\rho$},
    ylabel={$C_k(t)-C_k(\infty)$}]

     \addplot[smooth, line width = 2pt, black]
     table [y=t, x=rho]{tmp1.dat};
     \addlegendentry{$C_k$};
     
          \addplot[ smooth, line width = 2pt, blue,dashed] table [y=ic, x=rho]{tmp1.dat};
     \addlegendentry{$C_k^{\rm (ic)}$}

    \addplot[smooth, line width = 2pt, red] table [y={n}, x=rho]{tmp1.dat};
     \addlegendentry{$C_k^{\rm (n)}$}
     
        \addplot[domain = 0:5,samples = 100, dashed, gray] {0};

     \node[draw,align=left] at (40,10) {effective noise};

    \end{axis}
\end{tikzpicture}

\end{subfigure}

 \caption[Equal-time correlator]{
Equal-time correlation function $C_k(t) - C_k(\infty)$ for $\gamma = 1$ 
as a function of the 
dimensionless parameter $\rho$ (see Eq.~(\ref{eq:spatiotemp}) for 
the case I of a quench to the critical point,  in the presence of either the 
quantum noise in Eq.~(\ref{gl:corr-tempsegaux}) (left panel) or the effective 
dynamics in Eq.~(\ref{gl:effektiv_Ct}) (right panel). 
The decay of the initial correlations in (\ref{eq:C0}) has been chosen with with 
$\alpha=-1$. 
 The contributions $C_{\vec{k}}^{\rm (ic)}(t)$ coming from the initial noise 
correlators and $C_{\vec{k}}^{\rm (n)}(t)$ coming from the bath noise 
correlators are indicated separately.}
\label{fig:2tempsegaux_corr}
\end{figure}
%
%
In Fig.~\ref{fig:2tempsegaux_corr} we plot the equal-time correlators for the 
critical quench in case I, for both the quantum noise (left panel) and the 
effective dynamics (right panel) with the time-independent term 
$C_{\vec{k}}(\infty)$ in Eqs.~\eqref{gl:effektiv_Ct} and 
\eqref{gl:corr-tempsegaux-c} being subtracted. The qualitative behaviour is very 
similar. In both cases, the behaviour of the correlations for small values of 
the scaling variable $\rho$  (see Eq.~\eqref{eq:spatiotemp}) is dominated by the 
contribution of the initial noise $C_{\vec{k}}^{\rm (ic)}(t)$ which depends on a 
non-universal amplitude $c_\alpha/g_c$, whereas for larger values of $\rho$, the 
corresponding universal bath noise term $C_{\vec{k}}^{\rm (n)}(t)$ dominates. 
Qualitatively, these bath noise terms are distinct, see the red lines in 
Fig.~\ref{fig:2tempsegaux_corr}.  While for $\rho\gg 1$, they are both 
anticorrelated and decay to zero as $\sim \rho^{-1}$ (note that anticorrelations 
are absent for the classical dynamics), their difference emerges for finite 
values of $\rho$. For the effective noise, $C_{\vec{k}}^{\rm (n)}(t)$ grows 
monotonically as a function of $\rho$ while for the quantum noise the shape of 
the scaling function is not monotonic  and for $\rho\simeq 1$ the contribution 
is positively correlated and has a maximum.  Although  the effective noise 
(right panel in Fig.~\ref{fig:2tempsegaux_corr}) does faithfully reproduce the 
qualitative scaling behaviour of the two-time observables and provides exactly the same 
exponents as the actual quantum noise (left panel), it is not adequate for a 
precise quantitative description of single-time observables. 

\item[\bf (3)]
For a quench across the critical point, that is for $r < r_0^c$, the 
self-consistent function $g(t)$ still shows an algebraic behaviour that 
coincides with the one of  effective dynamics discussed above and even with what 
happens in the presence of classical white noise~\cite{Pico02}. 
Equation~\eqref{gl:g-crit} holds with a multiplicative prefactor $g_d$ and the 
exponent $\digamma = - (d+\alpha)/2$. This is actually not surprising since the 
quantum noise is sub-dominant compared to the classical white noise and the 
white noise itself is, in turn, sub-dominant compared to the propagation of the 
initial correlations for a deep quench. Using again 
Eq.~(\ref{gl:corr-tempsegaux}), the equal-time correlation function in the 
scaling limit ($k\to0$, $t\to\infty$ and $\rho =\text{cst.}$) becomes 
\begin{align} \label{gl:corr-tempsegaux-ord} 
 C_{\vec{k}}(t) \simeq \frac{c_\alpha }{g_d} k^{-d}\rho^{(d+\alpha)/2} e^{-2\rho}, 
\end{align}
in terms of the scaling variable $\rho$ in Eq.~\eqref{eq:spatiotemp}, which also 
implies $z=2$. For any correlated initial state with $\alpha<0$, the propagation 
of the initial correlation dominates over the one of the contribution due to the 
quantum noise. Only for a fully uncorrelated initial state, corresponding to 
$\alpha=0$, both initial and quantum noise correlations would contribute 
similarly to the scaling function. 

\vspace{.25cm}
\textcolor{black}{Another commonly used prescription for quantum dynamics is 
a master equation of the Lindblad form. It is instructive to compare Eq.~(\ref{gl:corr-tempsegaux-ord}) with the corresponding prediction when 
the quantum Langevin dynamics considered here 
is replaced by a Lindblad master equation (Li). 
The Lindblad dynamics of the quantum spherical model has been analysed 
in Ref.~\cite{Wald18}:}
the lower critical dimension turns out to be 
$d_l^{\rm (Li)}=1$, the upper critical dimension $d_u^{\rm (Li)}=3$, and the 
critical point of the dynamics is the same as in equilibrium. In addition, at 
the leading non-trivial order of the semi-classical expansion, the 
fluctuation-dissipation theorem applies in the single-phase region and the 
long-time behaviour of the single-time correlator becomes independent of the 
dissipative dynamical coupling $\gamma$. In $d=2$ dimensions, one finds a 
scaling form $C_{\vec{k}}(t) = k^{-2} \Psi( |\vec{k}|t)$, hence the dynamical 
exponent is $z=1$, as expected for a closed quantum system with unitary dynamics 
\cite{Cala16,Delf18}. For $d>2$ dimensions, instead, logarithmic corrections to 
scaling are found.\footnote{Indeed, if $z=2$ a dimensionless scaling variable 
can only be of the form $k^2 t/\gamma$, assuming the microscopic velocity $v$ is set to $v = 1$. For $z=1$, a scaling variable $k 
t$ is dimensionless and no further time scale $1/\gamma$ is needed.}
\textcolor{black}{This explicit example shows that quantum Langevin dynamics and 
Lindblad dynamics lead to completely different results.}
\vspace{.25cm}

Returning to the quantum Langevin equation, for
the sub-critical quench we are considering here, the two-time autoresponse 
and autocorrelation function read
\begin{align}
R(t,s) &= R_{(0)} s^{-d/2} \left(\frac{t}{s}\right)^{(d+\alpha)/4} 
\left(\frac{t}{s}-1\right)^{-d/2},\quad  C(t,s) \sim 
\frac{(t/s)^{{(d+\alpha)}/{4}}}{[1+(t/s)]^{(d+\alpha)/2}},
\end{align}
with the same $R_{(0)}$ as in Eq.~(\ref{eq:RCc}). The corresponding exponents 
can be determined by comparing these expressions with Eqs.~\eqref{gl:vieux} and 
\eqref{eq:lambdas} as
\begin{equation}
 \lambda_R = \frac{d-\alpha}{2}, \qquad \lambda_C = \frac{d+\alpha}{2}.
\end{equation}
All these predictions are identical to those of the case of the effective 
dynamics, reported in Tab.~\ref{tab:eff}. 
\end{enumerate}
In summary, our explicit calculations have shown that:
\begin{enumerate}
\item
As far as the occurrence of a phase transition is concerned, both the effective 
dynamics and the quantum dynamics lead to a dimensional shift $d\mapsto d-2$ 
with respect to the classical dynamics.
\item 
The stationary state at temperature $T=0$ reached after a quench cannot be an 
equilibrium state. This is due to the fact that the overdamped limit (see 
Eq.~(\ref{eq:over})) prevents the emergence of any kind of equipartition.
\item 
The effective dynamics, characterised by a rescaled Markovian noise in 
Eq.~\eqref{eq:eff_qn_anticommu} instead of the actual quantum noise in 
Eq.~\eqref{eq:qudyn}, turns out to be sufficient in order to reproduce correctly 
all the universal exponents which describe the ageing after a quantum quench. 
The additional long-time memory effects in the quantum noise are not strong 
enough to yield further modifications. The entries of Table~\ref{tab:eff} 
are therefore the same for both the effective and the quantum dynamics.
\item 
The scaling of the equal-time correlator, which describes the coarsening 
occurring after a quantum quench, is sensibly determined by the long-time memory 
effects of the quantum noise. In particular, this influence is manifest above 
the critical point (see, e.g., Fig.~\ref{fig:1phase_corr}) and in region I (see 
Fig.~\ref{fig:dalpha}) at criticality. In the critical regions II, III (see 
Fig.~\ref{fig:dalpha}) and for sub-critical quenches, the quantum noise merely 
leads to corrections to the leading scaling behaviour. 
\end{enumerate}

Finally, let us reconsider the role of the cutoff scale $\tto$, introduced into 
the quantum noise correlator in Eq.~(\ref{eq:qn_anticommu}) at temperature 
$T=0$. In those cases and observables for which the quantum noise turned out to 
be irrelevant, the actual value of $\tto$ has no influence on the long-time 
behaviour of the dynamics.   The cases in which the quantum noise is important 
concern the equal-time correlator $C_{\vec{k}}(t)$, either for region I of a 
critical quench or  for a quench to the disordered phase. For case I of the 
critical quench, according to the classification of Tab.~\ref{tab:eff}, 
Eq.~(\ref{gl:corr-tempsegaux-c}) shows that $\tto$ merely enters a simple 
additive and time-independent term. This term describes the logarithmic 
singularity which arises if one attempts to take the limit $\tto\to 0$.  
However, the other aspects of the dynamics do not depend explicitly on $\tto$. 
For a quench to the disordered phase, Eq.~(\ref{gl:correl-stat-disor}) shows 
that $\tto$ is merely setting a scale for the stationary correlator.  Here it is 
adequate to recall the heuristic argument from \ref{app:commu} that this scale 
is set at $\tto \sim \gamma^{-1}$. In any case, the value of $\tto$ does not 
influence the long-time quantum dynamics of the quantum spherical model. 

A detailed discussion and comparison with the results of classical dynamics is 
provided in Sec.~\ref{sec:conclusion}.

\section{Non-Markovian equation: details of the solution}
\label{sec:solution}

We analyse here the effects of the quantum spherical constraint on the dynamics 
of the model. We shall first show that the function $g(t)$ defined in Eq.~ 
(\ref{gl:g}) obeys a non-linear integro-differential equation. Focusing on the 
non-Markovian quantum noise correlations (\ref{eq:qn_anticommu_cut}), we then 
show how this equation can be embedded into a larger set of linear Volterra 
integro-differential equations for a function $G(t,s)$ of two variables, such 
that $g(t)=G(t,t)$. The asymptotic solution of this linear integro-differential 
equation renders the sought long-time behaviour in Eqs.~(\ref{gl:qn-regime1}), 
(\ref{gl:qn-regime2}), and (\ref{gl:qn-regime3}) of the function $g(t)$. This 
piece of information is necessary for the calculation of physical observables, 
which we discuss in Sec.~\ref{sec:observables}.

\subsection{Formulation of the quantum spherical constraint}

The solution of the non-equilibrium dynamics of either the spherical model or 
the $O(n)$-model at large $n$ relies on the fact that the field 
$\phi_{\vec{k}}(t)$ basically evolves according to a (time-dependent) linear 
equation, which is essential for finding the general formal solution in 
Eqs.~(\ref{eq:sol}) and (\ref{gl:RC-formels}). All aspects of the many-body 
interaction are self-consistently introduced through the spherical constraints 
of these models, represented by $g(t)$, which, in turn, is determined by the 
equal-time correlator $C(t,t)$.  In order to determine the evolution of the 
constraint, we begin by rewriting Eq.~(\ref{gl:C-formel}) in the following 
compact form
\begin{align}
C(t,t) = \frac{1}{g(t)}\bigg[A(t)+\left(g_2\dast F\right)(t,t)\bigg],
\label{eq:C}
\end{align}
where we introduced the two-variable function 
\begin{equation}\label{eq:g2}
 g_2(t,s): = \sqrt{g(t)g(s)} \ , 
\end{equation}
two auxiliary functions $A(t)$ and $F(t,s)$ and the double convolution  denoted 
by $\dast$, specified further below in Eq.~(\ref{eq:dcon}). The functions $A(t)$ 
and $F(t,s)$ describe the propagation of initial and quantum noise correlation  
and are respectively given by
\begin{align}
 A(t)& :=\int_{\vec{k}} \exp\left( {-2\frac{k^2}{\gamma}t}\right) c_\alpha 
k^\alpha, 
 \quad 
F(t,s) := \gamma^{-2}\int_{\vec{k}} \exp\left({-\frac{k^2}{\gamma}(t+s)}\right) 
m_{\vec{k}}(t-s),
\label{eq:AandF}
\end{align}
with the memory-kernel $m_{\vec{k}}$ of the quantum noise in 
Eq.~\eqref{eq:def-k-m}. Herein, we do not yet specify whether the momentum 
integrals $\int_{\vec{k}}$ require a cutoff at some large momentum 
$|\vec{k}|=\Lambda$, or not, see below. The double convolution $\dast$ in 
Eq.~\eqref{eq:C} is defined for two functions $h_1,h_2: 
\mathbb{R}_+^2\to\mathbb{C}$,  as (see also  \ref{app:prop_lap})
\begin{equation}\label{eq:dcon}
  (h_1\dast h_2)(t,s) = \int_0^t \!\D x \int_0^s \!\D y \:  h_1(x,y) h_2(t-x,s-y) .
\end{equation}
The spherical constraints Eqs.~(\ref{eq:Hon}) and (\ref{eq:Hsm}) applied to 
Eq.~(\ref{eq:C}) now produce integro-differential equations for $g(t)$. For the 
spherical model, the constraint $C(t,t)=1/\lambda$ leads to
\begin{subequations} \label{eq:selfconv} 
\begin{align}\label{eq:selfconvoSM}
 \frac{1}{\lambda}g(t) = A(t) + (g_2\dast F)(t,t),
\end{align}
while for the $O(n)$-model at large $n$, the constraint $r(t)=r_0+ u C(t,t)/12$ 
becomes 
\begin{align}\label{eq:selfconvo}
 \frac{6{\gamma}}{u} \frac{\D  g(t)}{\D t} -\frac{12 {r}_0}{u} g(t) = A(t) + 
(g_2\dast F)(t,t).
\end{align}
\end{subequations}
For the effective or classical dynamics, where the noise $m_{\vec{k}}(\tau)$ is 
$\delta$-correlated in the time $\tau$, Eqs.~\eqref{eq:selfconv} become  
(generalised) linear Volterra integral (or integro-differential) equations for 
$g(t)$. For completeness and comparison, this case is discussed and solved 
in~\ref{app:eff}.

\subsection{Equivalence of spherical and $O(n)$-models, at large $n$}

As anticipated above, determining the dynamics of the various models in the 
presence of the different type of noises discussed in the previous sections is 
essentially reduced to solving one of Eqs.~(\ref{eq:selfconv}) for the 
constraint, depending on which one of the specific models introduced in 
Sec.~\ref{sec:equilibrium} one is focusing on. 
Eqs.~(\ref{eq:selfconvoSM}) and~(\ref{eq:selfconvo}) do not coincide, as they 
refer to two \emph{a priori} different models. However, the formal limit 
${\gamma} \to 0$ together with the identification $1/\lambda = - 12 
{r}_0/u$ transforms Eq.~(\ref{eq:selfconvo}) into Eq.~(\ref{eq:selfconvoSM}). 
Accordingly, the spherical model dynamics can always be obtained from the 
solution of Eq.~(\ref{eq:selfconvo}) via this limit. We shall see later in 
Sec.~\ref{ssec:formal} that the extra terms contained in 
Eq.~(\ref{eq:selfconvo}) do not modify the universal features of the long-time 
behaviour. In addition, asymptotic solutions of the form $g(t)\sim 
e^{t/\tau_{\rm r}}$ or $g(t)\sim \mbox{\rm cst.}$ occur for certain parameter 
ranges in both equations.  

Because of the equivalence of the long-time behaviour of the two models, we 
shall focus henceforth on the $O(n)$-model in the large-$n$ limit and therefore 
concentrate on solving Eq.~(\ref{eq:selfconvo}).

\subsection{Formal solution of the spherical constraint}
\label{ssec:formal}
For the quantum noise, such as that in Eq.~(\ref{eq:qn_anticommu_cut}), which is 
not $\delta$-correlated in time, the constraint in Eq.~(\ref{eq:selfconvo}) is a 
\emph{non-linear} integro-differential equation for the function $g(t)$, without 
obvious explicit solutions. The non-linearity can be formally avoided via the 
introduction of the function $g_2(t,s)=\sqrt{g(t) g(s)\,}$, according to 
Eq.~(\ref{eq:g2}), depending on two variables $t$ and $s$, combined with a 
double convolution. In principle, the knowledge of either function $g$ 
or $g_2$ determines the other one via Eq.~(\ref{eq:g2}). Accordingly, it might 
appear promising to try to determine the symmetric function $g_2(t,s)=g_2(s,t)$ 
directly.

However, instead of solving directly the non-linear equation 
(\ref{eq:selfconvo}), we  rather consider a \emph{different} 
integro-differential equation for a symmetric function $G(t,s)=G(s,t)$. The 
function $g=g(t)$, whose long-time behaviour is sought, is recovered in the 
equal-time limit $g(t)=G(t,t)$. The equation reads 
\begin{align}\label{eq:selfconvo2d}
 \frac{3{\gamma}}{u}\left[\frac{\partial G(t,s)}{\partial t} +\frac{\partial 
G(t,s)}{\partial s}\right]-\frac{12 {r}_0}{u} G(t,s) = 
A\left(\frac{t+s}{2}\right) + (G\dast F)(t,s)
\end{align}
and has two desirable properties: (a) it is \emph{linear} in terms of the 
function $G(t,s)$ and (b) for $s\to t$ it reduces to Eq.~(\ref{eq:selfconvo}) as 
$\lim_{s\to t}G(t,s)=G(t,t)=g(t)$, although, generically, $G(t,s) \neq 
g_2(t,s)$. For solving Eq.~\eqref{eq:selfconvo2d}, we impose the initial 
condition $G(0,0)=g(0)=1$ implied by Eq.~(\ref{gl:g}). In addition, a boundary 
condition for $G(t,0)=G(0,t)$ will be required.\footnote{This requirement is 
actually unnecessary in the $\gamma\to 0$ limit.}


For given initial and boundary conditions, Eq.~(\ref{eq:selfconvo2d}) has a 
unique solution $G(t,s)=G(s,t)$ and furnishes $g(t)=G(t,t)$. In addition 
$g_2(t,s)=g_2(s,t)$ is another solution of either Eq.~(\ref{eq:selfconvo2d}) or 
(\ref{eq:selfconvo}), but {\it a priori} with different boundary conditions. It 
will turn out that one finds in the long-time limit (i) either a leading 
exponential behaviour $g(t)\sim e^{t/\tau}$ or (ii) a leading algebraic 
behaviour $g(t)\sim t^{\digamma}$. As we shall show below, the boundary 
conditions do not modify the value of $\digamma$. In addition, the boundary 
conditions do not affect which of the above two possibility occurs. Analogously, 
setting $\gamma=0$ in Eq.~(\ref{eq:selfconvo2d}) and substituting $\lambda=- 
u/12 r_0$, the solution of Eq.~(\ref{eq:selfconvo2d}) will produce the leading 
long-time exponent $\digamma$ of Eq.~(\ref{eq:selfconvoSM}). 
Accordingly, we now show that:

{\it The leading long-time behaviour of $g(t)=G(t,t)$ derived from 
Eq.~(\ref{eq:selfconvo2d}) is independent of the boundary conditions on 
$G(t,0)=G(0,t)$. It therefore agrees with the leading long-time behaviour of 
$g_2(t,t)$ found from Eq.~(\ref{eq:selfconvo}) for $\gamma\ne 0$ or 
Eq.~(\ref{eq:selfconvoSM}) in the $\gamma\to 0$ limit.}

\vspace{.2cm}

In order to prove these statements, we begin with the formal solution of 
Eq.~(\ref{eq:selfconvo2d}). This is readily obtained, via a two-dimensional 
Laplace transform \cite{Voel50,Ditk62,Debnath2016} according to the definition
\begin{equation}\label{eq:doubleL}
 \dlap{f}(p,q) := \int_0^\infty\!\D t \int_0^{\infty} \!\D s\:  f(t,s) 
e^{-pt-qs} .
\end{equation}
Using the properties listed in~\ref{app:prop_lap}, we straightforwardly obtain 
the Laplace transform $\dlap{G}(p,q)$ of the solution $G(t,s)$ of 
Eq.~(\ref{eq:selfconvo2d}), as
\begin{align}
\label{eq:solution0}
\dlap{G}(p,q) = \frac{\dlap{A}(p,q) + (3{\gamma}/u)\left[\,\lap{G}_0(q)+ 
\lap{G}_0(p)\right]}{(3\gamma/u) \left(p+q\right) - (12 r_0/u) -\dlap{F}(p,q)},
\end{align}
with the following boundary terms (recall that $G(t,0)=G(0,t)$)
\begin{equation}\label{G0}
 \lap{G}_0(p) := \int_0^\infty \!\D t\: e^{-p t} G(t,0)  = \int_0^\infty \!\D 
s\: e^{-p s} G(0,s).
\end{equation}
These boundary terms require a specific analysis, using the symmetry 
$G(t,s)=G(s,t)$ and the requirement that $G(t,s)$ is partially differentiable 
with respect to $t$ and $s$. It will turn out to be convenient to set
\begin{align}
G(t,s) =: H\left(\frac{t+s}{2}, \frac{t-s}{2}\right) = H(\alpha,\beta) \mbox{\rm 
~~ with ~~} \alpha := \frac{t+s}{2} \quad \text{and} \quad  \beta := 
\frac{t-s}{2}.
\end{align}
The symmetry condition under the exchange of $t$ and $s$ implies 
\begin{align} \label{eq:Hsymm}
H(\alpha,\beta) = H\left(\frac{t+s}{2}, \frac{t-s}{2}\right) 
=H\left(\frac{t+s}{2}, \frac{s-t}{2}\right) = H(\alpha,-\beta).
\end{align}
Since $G(t,s)$ is partially differentiable, apply the symmetry relation 
(\ref{eq:Hsymm}) to find 
\begin{align}
\demi \bigl( \partial_{\alpha} + \partial_{\beta}\bigr) H(\alpha,\beta) = 
\partial_t G(t,s) = \partial_t G(s,t) = \demi \bigl( \partial_{\alpha} - 
\partial_{\beta}\bigr) H(\alpha,-\beta) ,
\end{align}
or equivalently 
\begin{align}
\partial_{\beta} \bigl[ H(\alpha,\beta) +H(\alpha,-\beta) \bigr] = 
\partial_{\alpha} \bigl[ H(\alpha,-\beta) -H(\alpha,\beta) \bigr] = 0,
\end{align}
where we used the symmetry~(\ref{eq:Hsymm}). This implies $\partial_{\beta} 
H(\alpha,\beta) =0$. Accordingly,
\begin{align}
H = H(\alpha) = H\left(\frac{t+s}{2}\right),
\end{align}
i.e., $H$ actually depends only on  a single variable. After these 
preliminaries, we consider the function $G_0(t):=G(t,0)=H\bigl( t/2\bigr)$. 
Returning to Eq.~(\ref{eq:selfconvo2d}), the double convolution term will vanish 
when the limit $s\to 0$ is considered and $\bigl(\partial_t + 
\partial_s\bigr)=\partial_{\alpha}$. This leads to
\begin{align}
\lim_{s\to 0} \frac{3\gamma}{u} \partial_{\alpha} H\left(\frac{t+s}{2}\right) 
= \frac{3\gamma}{u} \partial_{\alpha} H(\alpha) = \frac{12 r_0}{u} H(\alpha) + A(\alpha) \;\; 
\mbox{where}\;\; \alpha = \frac{t}{2}.
\end{align}
Laplace-transforming this equation immediately produces $-\frac{3\gamma}{u} H(0) 
+ p \lap{H}(p) = \frac{12 r_0}{u} \lap{H}(p) + \lap{A}(p)$, hence with 
$H(0)=G(0,0)=1$, one finds
\begin{align} \label{eq:bord}
\lap{H}(p) = \frac{(3\gamma/u) + \lap{A}(p)}{p - 12 r_0/u}.
\end{align}
Since we had $G_0(t)=G(t,0)=H\bigl(t/2\bigr)$, it follows that $\lap{G}_0(p)=2 
\lap{H}(2p)$. At long last, the formal solution (\ref{eq:solution0}) of 
Eq.~(\ref{eq:selfconvo2d}) explicitly becomes in double Laplace space 
\begin{align}
\label{eq:solutionf}
\dlap{G}(p,q) = \frac{\dlap{A}(p,q) + ({6\gamma}/{u})\left[\,\lap{H}(2q)+ 
\lap{H}(2p)\right]}{(3\gamma/u) \left(p+q\right) - (12 r_0/u) -\dlap{F}(p,q)},
\end{align}
In Eqs.~(\ref{eq:solutionf}) $\dlap{A}(p,q)$ is the double Laplace transform of 
the function $A((t+s)/2)$, which can be derived from that of $A(t)$, see 
Eq.~(\ref{eq:lapA2}) below.

Well-known Tauberian theorems for the Laplace transform of a single variable 
\cite{Fell71} can be generalised to the present case of a double Laplace 
transforms.
For homogeneous functions, this generalisation is formulated and proven in 
\ref{app:Lap} and, accordingly, the long-time behaviour of $G(t,s)$ for large 
$t$ and $s$ we are interested in is related to the behaviour of $\dlap{G}(p,q)$ 
for small $p$ and $q$. In order to determine it, the asymptotic expansions of 
the functions $\dlap{A}(p,q)$,  $\dlap{F}(p,q)$ and $\lap{H}(p)$ for small $p$ 
and $q$ are needed.  

First, the long-time asymptotics of the function $A(t)= c_{\alpha} A_{\alpha}(t)$, 
and hence also 
$\lap{A}(p)$ for $p\ll 1$, is derived in~\ref{app:eff}, 
for the effective dynamics. There, it is shown that
\begin{equation} \label{eq:lapA}
  \lap{A}_\alpha(p) \simeq a_\alpha p^{\frac{d+\alpha}{2}-1}+
 \sum_{n=0}^{\lfloor\frac{d+\alpha}{2}-1\rfloor} (-1)^n A_n^{(\alpha)} p^n,
\end{equation}
with explicit expressions for the constants $a_{\alpha}$ and $A_n^{(\alpha)}$. 
Since we have chosen in Eq.~\eqref{eq:selfconvo2d} 
the two-time form as $A(t,s) = A((t+s)/2)$, 
we can use an identity proven in~\ref{app:prop_lap}, which states that 
\begin{equation} \label{eq:lapA2}
 \dlap{A}(p,q) = \frac{\lap{A}(p/2) - \lap{A}(q/2)}{(p-q)/2},
\end{equation}
in order to determine the small-$p$ and $q$ behaviour of $\dlap{A}(p,q)$ in 
terms of that of $\lap{A}$ given in Eq.~(\ref{eq:lapA}). This allows us to 
conclusively examine the relevance of the terms $\sim\lap{H}(p)$ in 
Eq.~(\ref{eq:solutionf}) coming from the boundary conditions. The general 
results from appendix~\ref{app:Lap} on inverting double Laplace transformation 
state that the leading long-time behaviour can be read from the small-$p$ 
scaling of the generic form $p^{\mathfrak{a}} f\bigl(\frac{q}{p}\bigr)$. The 
leading small-$p$ contributions of the numerator in Eq.~(\ref{eq:solutionf}) are 
proportional to either $\dlap{A}(p,q)\sim p^{\frac{d+\alpha}{2}-2}$, using 
(\ref{eq:lapA},\ref{eq:lapA2}), or to $\lap{H}(2p)+\lap{H}(2q)\sim 
p^{\frac{d+\alpha}{2}-1}$, using (\ref{eq:bord},\ref{eq:lapA}), times scaling 
functions in $q /p$. Clearly, the contribution of the boundary term merely 
generates a correction to the leading scaling behaviour.  Thus, in the 
subsequent analysis the boundary terms need not be included.

For the function $\dlap{F}(p,q)$, the non-Markovian character of the noise makes 
the analysis of the function considerably more difficult than that of 
$\dlap{A}$. We outline here only the main points of this analysis and refer 
to~\ref{app:F} for further details and explicit calculations. The double Laplace 
transform of $F(t,s)$ is explicitly obtained as
\begin{align}
 \dlap{F}(p,q) = \frac{2\hbar}{\pi\gamma}\int_{\vec{k},(\Lambda)} 
 \int_0^\infty \D t \int_0^\infty \D s \: e^{-pt - qs}  e^{-\frac{k^2}{\gamma} (t+s)} 
 \frac{\tto^2 - (t-s)^2}{\left[\tto^2 +( t-s)^2 \right]^2},
\end{align}
where we now also indicate the need of a cutoff at large momentum 
$|\vec{k}|=\Lambda$.  Using the symmetry of the integrand in $t$ and $s$ for 
$p=q=0$, the integration domain in $(t,s)$ can be reduced to a wedge domain with 
$s<t$. Then, we can introduce diagonal coordinates $(x,y)  = (t-s,t+s)$ such 
that the double Laplace transform reduces to a single integral of the form
\begin{equation} \label{eq:int}
 \mathcal{J} = \int_0^\infty \D x\, f(x)\frac{\tto^2 - x^2}{\left(\tto^2 + 
x^2\right)^2} 
\end{equation}
with an explicitly known function $f(x)$. In~\ref{app:F} we present a general 
approach for evaluating such an integral, based on interpreting the quantum 
noise correlator as a generalised function acting on the test function $f(x)$. 
For a generic function $f$, we shall see in \ref{app:F} that it is useful to 
introduce the formal series\footnote{These constants $a_j$ must not be confused 
with the constants $a_{\alpha}$ in Eq.~(\ref{eq:lapA}).} in $x$
\begin{equation}
 f(x) = {\sum_j}' a_j x^{\alpha_j} + \sum_{n=0}^\infty b_{2n+1} x^{2n+1} \ ,
\end{equation}
where the first sum contains all even powers and all non-analytic contributions 
in $x$ and the second sum contains all odd powers in $x$.
We further show that the integral in Eq.~(\ref{eq:int}) may be evaluated as
\begin{align}
\mathcal{J} = &-\frac{\pi}{2} {\sum_{j}}' \frac{\tto^{\alpha_j-1} a_j  \alpha_j 
}{\cos\left(\pi \alpha_j/2\right)} +\sum_{n=0}^\infty 
\left(-\tto^2\right)^n\bigg[ (1+(2n+1)\ln\tto)b_{2n+1} - (2n+1)B_n \bigg] 
  \label{eq:int_final}
\end{align}
with the constants
\begin{equation}
 B_n = \lim_{z \to -(2n+1)} \frac{\D}{\D z}\left[ \left( z+2n+1\right) \int_0^\infty \D y \,y^{z-1} f(y) \right].
\end{equation}
For the particular case of an exponential function $f(x)=e^{-\nu x}$, we show 
in~\ref{app:F} how to evaluate explicitly the series in Eq.~\eqref{eq:int_final} 
in terms of the auxiliary function $g_{\rm AS}$ defined in Eq.~(\ref{eq:aux}). 
Exploiting the known asymptotics of $g_{\rm AS}$ (see Eq.~\eqref{eq:aux}) one 
can determine the leading terms  for $p,q\to 0$ of the double Laplace transform 
as
\begin{subequations} \label{eq:dlapF-final}
\begin{align}
\label{eq:dlapF-final-a}
\dlap{F}(p,q)  &\simeq-\frac{2\hbar}{\pi\gamma}\frac{\Omega_d}{(2\pi)^d} 
\int_0^\Lambda \D k \: k^{d-1} \left[ \mathscr{F}(p,q) + 
\mathscr{F}(q,p)\right],
\end{align}
where
\begin{align}
\mathscr{F}(p,q) &:=  \frac{ \left( k^2/\gamma+p\right) \left[{\rm C}_E 
+\ln\left(\tto\left(k^2/\gamma+p\right)\right)\right] }{k^2/\gamma+(p+q)/2}.
\end{align}
\end{subequations}
As further shown in \ref{app:F}, the  integral in Eq.~\eqref{eq:dlapF-final-a} 
decomposes into a non-universal, regular (i.e. analytic) part and a universal, 
irregular (i.e., non-analytic) part
%
\begin{equation} \label{eq:dlapF-final2}
 \dlap{F}(p,q)  = \dlap{F}_{\rm reg}(p,q) + \dlap{F}_{\rm irr}(p,q) \ .
\end{equation}
%
These can be evaluated explicitly, to their respective lowest order in $\tto$
\begin{subequations} \label{eq:dlapF-regirr}
\begin{align}
  \dlap{F}_{\rm reg}(p,q) &\simeq 
-\frac{4\hbar}{\pi\gamma}\frac{\Omega_d}{(2\pi)^d} \left\{ \frac{\Lambda ^d}{d} 
\left[\ln \left(\Lambda^2\frac{\tto }{\gamma }\right)+ {\rm C}_E  
-\frac{2}{d}\right] + \demi\frac{\gamma  \Lambda ^{d-2}}{d-2}(p+q) + {\rm 
O}(\tto) \right\} ,\label{eq:dlapF-regirr-a}\\[.25cm] \dlap{F}_{\rm irr}(p,q) 
&\simeq -\frac{4\hbar}{\pi\gamma}\frac{\Omega_d}{(2\pi)^d} (\gamma 
p)^{\frac{d}{2}} \mathbb{F}(q/p) + {\rm O}(p,q), 
  \label{eq:dlapF-regirr-b}
\end{align}
with the scaling function
\begin{align}
  \mathbb{F}(z) &=  \int_0^{\infty}\D x\, x^{d-1} \frac{\left(x^2+z\right)
  \ln \left(1+z/x^2\right)+\left(x^2+1\right) \ln \left(1+1/x^2\right)}{
  \left(x^2+z\right)+\left(x^2+1\right)}.
\end{align}
\end{subequations}
Equations \eqref{eq:dlapF-final2} and \eqref{eq:dlapF-regirr},  provide the 
asymptotic expansion of $\dlap{F}(p,q)$ we are interested in. In particular, by 
using Eq.~\eqref{eq:solution0}, it turns out that the critical point of the 
$O(n)$-model is $r_0^c = -(u/12)\dlap{F}(0,0)$, the value of which follows from  
Eq.~\eqref{eq:dlapF-regirr-a}, resulting in Eq.~\eqref{gl:r0c-qu} of 
Sec.~\ref{sec:results}.

We can now analyse the relative importance of the various contributions to the 
exact formal solution in Eq.~(\ref{eq:solutionf}). Considering, first, the 
denominator in Eq.~(\ref{eq:solutionf}), we see from the leading terms of the 
expansions in Eqs.~(\ref{eq:dlapF-final2}) and (\ref{eq:dlapF-regirr}) that 
there are two distinct non-constant terms. There is a term $\sim p^{d/2}$ which 
comes exclusively from the quantum noise and there is a term $\sim p$ which has 
contributions from several different sources. Hence, for $d<2$ the leading 
behaviour will be determined by the quantum noise, but for $d>2$, the leading 
exponent is independent of it. However, the associated amplitude may receive a 
non-vanishing contribution from the quantum noise.

Finally, we briefly return to the relationship between the spherical model and the 
$O(n)$-model for $n\to \infty$. The difference between them comes from the terms 
in Eq.~(\ref{eq:solutionf}), which contain a factor $\gamma$.\footnote{In the 
limit $\gamma\to 0$,  Eq.~(\ref{eq:selfconvo2d}) reduces to a linear Volterra 
integral equation in two variables.} Clearly, these terms can at most 
renormalise certain amplitudes. Accordingly, one can draw the conclusion that 
{\it the leading universal properties of the spherical and of the $O(n)$-model 
for $n\to\infty$ are   identical also in these non-equilibrium conditions.}

\subsection{Asymptotic behaviour of the self-consistent solution for $d<2$}
\label{ssec:g}

For $d<2$, given that $\alpha\leq 0$, one also has $d+\alpha<2$. Using 
Eqs.~(\ref{eq:lapA}) and (\ref{eq:lapA2}), the function $\dlap{A}(p,q)$ may then 
be written in terms of a scaling function $\mathbb{A}(z)$ according to
\begin{equation}
 \dlap{A}(p,q) \simeq c_\alpha a_\alpha 2^{2-\frac{d+\alpha}{2}} 
p^{\frac{d+\alpha}{2}-2}\mathbb{A}(q/p), \quad {\rm with} \quad \mathbb{A}(z) = 
\frac{1-z^{(d+\alpha)/2-1}}{1-z}.
\end{equation}
Equation (\ref{eq:solutionf}) can then be rewritten asymptotically as follows, 
using  Eqs.~(\ref{eq:dlapF-final2}) and (\ref{eq:dlapF-regirr}),
\begin{align}
\label{eq:solution_scale}
\dlap{G}(p,q) \simeq \frac{ c_A  p^{\frac{d+\alpha}{2}-2} \mathbb{A}(q/p)}{M^2  
+ c_F p^{d/2} \mathbb{F}(q/p)},
\end{align}
where $M^2 = {r}_0^c - {r}_0 $ quantifies the distance from the critical point 
${r}_0^c = -(u/12)\dlap{F}(0,0)$, already quoted above in Eq.~(\ref{gl:r0c-qu}). 
In order to simplify the notation, in the equation above we introduced the 
constants $c_A$ and $c_F$ which account for all numerical factors in the 
corresponding functions, following from \sw{Eq.~\eqref{eq:solutionf}}. We now 
consider iteratively the different quench protocols schematically illustrated 
in Fig.~\ref{fig:quenches}.
\begin{enumerate}
 \item[\bf (A)]
 
 For a {\bf quench above criticality}, i.e., with $r_0 > r_0^c$, the constant 
$M^2$ introduced in Eq.~\eqref{eq:solution_scale} is formally negative, i.e., 
$M^2<0$. Accordingly, the denominator in Eq.~(\ref{eq:solution_scale}) has a 
line of poles in the $(p,q)$ plane. Each of these poles corresponds to a mode of 
relaxation. To be explicit, consider a quench close to the critical point: the 
denominator vanishes along a curve $\Gamma$ of points $(p_0,q_0)$ which yields a 
curve of poles for the solution $\dlap{G}(p,q)$. Sufficiently close to 
criticality, $\Gamma$ is defined by the asymptotic equation for $p,q\to 0$, i.e., 
\begin{equation}\label{eq:poles}
 M^2+c_F \left[p_0^{d/2}+q_0^{d/2}\right] = 0 .
\end{equation}
The function $g(t)$ is obtained by formally inverting the double Laplace 
transform, i.e., 
\begin{align}
 g(t)= G(t,t) =  \int_{c_1-\II\infty}^{c_1 + \II\infty} \!\frac{\D p}{2\pi 
\II}\: \int_{c_2-\II\infty}^{c_2 + \II\infty} \!\frac{\D q}{2\pi 
\II}\dlap{G}(p,q) e^{(p+q) t} ,
\end{align}
and these two contour integrals in the complex plane can be evaluated with the 
poles on the curve $\Gamma$. Thus it is apparent that many relaxation times
\begin{equation} \label{eq:taur2d}
 1/\tau_{\text{r}} \simeq p_0  + q_0
\end{equation}
with $(p_0,q_0)\in \Gamma$ contribute to the integral. Generally the system will 
choose the slowest relaxational time scale as each of them contributes 
exponentially. Anyhow, all time scales satisfy
\begin{equation}\label{eq:relax}
 1/\tau_{\text{r}} \simeq \left( -M^2/c_F\right)^{2/d} 
\end{equation}
and consequently, $g(t)\sim e^{t/\tau_{\rm r}}$ is of exponential 
form.\footnote{A comment on stationary critical exponents: since $\tau_{\rm 
r}\sim \left|r_0-r_0^c\right|^{-\nu z}$, Eq.~(\ref{eq:relax}) implies that $\nu 
z = 2/d$.}

\item[\bf (B)]
For {\bf quenches onto or across criticality}, i.e., with $r_0 \le r_0^c$ and 
therefore either $M^2=0$ or $M^2>0$, the long-time behaviour of $g(t)$ becomes 
algebraic. This can be immediately seen from Eq.~(\ref{eq:relax}), since the 
relaxational time-scale diverges as criticality is approached, i.e.,
\begin{equation}
\tau_{\rm r} \to\infty \quad {\rm for} \quad {r}_0 \to {r}_0^c.
\end{equation}
Since $d<2$, the leading singularity of $\dlap{G}$ in Eq.~(\ref{eq:solutionf}) 
is no longer a pole but we have the small-$p$ behaviour 
\begin{align} \label{gl:qn-regime1-pq}
 \dlap{G}(p,q) \simeq 
 \begin{dcases}
   \frac{c_A}{c_F} p^{\frac{\alpha}{2}-2} \frac{ \mathbb{A}(q/p)}{ 
\mathbb{F}(q/p) } &\quad \mbox{for} \quad M^2 = 0,\\[.25cm]
   \frac{ c_A}{M^2}   p^{\frac{d+\alpha}{2}-2}\mathbb{A}(q/p) &\quad \mbox{for} 
\quad M^2 > 0.  
 \end{dcases}
\end{align}
Thus, the function $ \dlap{G}(p,q)=p^{\mathfrak{g}-2}\mathbb{G}(q/p)$ assumes a 
scaling form with an algebraic pre-factor and a certain exponent $\mathfrak{g}$. 
In~\ref{app:Lap} we show that this is equivalent to a scaling form 
$G(t,s)=s^{-\mathfrak{g}}\mathscr{G}(t/s)$. In particular it follows that 
$g(t)=G(t,t)=t^{-\mathfrak{g}}\mathscr{G}(1)$.  The actual value of the exponent 
$\mathfrak{g}$ can be directly identified from Eq.~(\ref{gl:qn-regime1-pq}). 
Hence, we conclude that
\begin{align}  \label{gl:qn-regime1}
 g(t) \simeq 
		  \begin{dcases}
                  g_c t^{-\alpha/2} & \quad \mbox{for} \quad M^2 = 0, \\[.25cm]
                  g_d t^{-(d+\alpha)/2} & \quad \mbox{for} \quad M^2 > 0,
		  \end{dcases}
\end{align}
where $g_c$ and $g_d$ denote, respectively, the non-universal pre-factors for a 
critical and a deep quench which can be obtained by evaluating the 
Laplace-inverted scaling function $\mathscr{G}(1)$. 
However, their explicit values will never be required in our analysis.
\end{enumerate}

\noindent
We finish with a comment on the form of the effective parameter $r(t)$, see 
Eq.~(\ref{eq:Hon}), which is implied by these results. In general, the 
asymptotic long-time behaviour $g(t) = g_0 t^{\digamma} + g_1 t^{-\kappa}$ is 
expected, where we included the leading correction $g_1 t^{-\kappa}$ 
with the yet unknown exponent $\kappa$ and we admit $\digamma\geq0$ 
and $\kappa>0$. Using Eq.~(\ref{gl:g}), we find for large times the 
``effective mass''
\begin{align} \label{gl:rgg}
r(t) = \frac{\gamma}{2} \frac{g'(t)}{g(t)} \simeq \frac{\gamma}{2}\left[ \frac{\digamma}{t} -\frac{g_1}{g_0}\frac{\digamma+\kappa}{t^{1+\digamma+\kappa}} + \ldots\right].
\end{align}
As long as $\digamma\ne 0$, this produces the asymptotic scaling $r(t)\simeq 
\frac{\gamma}{2}\frac{\digamma}{t}+\ldots$\ . This estimate suggests the very 
natural {\em light-cone ansatz} $r(t)\sim t^{-1}$ which is usually 
introduced on the basis of dimensional analysis and which simplifies calculations 
considerably, also for the field-theoretical Keldysh formalism 
\cite{Gagel14,Gagel15}, see, e.g., Eq. (70) in Ref.~\cite{Gagel15}. Now, 
consider a quench onto the critical point, which for $d<2$ corresponds to regime 
I and also focus on the limit of short-ranged spatial initial correlations 
(studied throughout in Ref.~\cite{Gagel14,Gagel15})  which corresponds to 
$\alpha=0$~\cite{Pico02}. Equation~(\ref{gl:qn-regime1}) then implies 
$\digamma=-\frac{\alpha}{2}\to 0$. Thus, Eq.~(\ref{gl:rgg}) rather yields 
asymptotically $r(t)\simeq-\frac{\gamma}{2}\frac{g_1}{g_0} \kappa\, 
t^{-1-\kappa}$ and the ansatz mentioned above for $r(t)$ no longer applies.

\subsection{Asymptotic behaviour of the self-consistent solution for $d > 2$}

In more than two spatial dimensions, the leading quantum noise contribution 
is regular and the self-consistency function $\dlap{G}(p,q)$
can be written as
\begin{align}
\dlap{G}(p,q) \simeq \frac{ \dlap{A}(q,p) + 12 {\gamma}/u}{M^2  + \tilde{c}_F p 
\left( 1+ q/p\right) }.
\end{align}
As seen in~\ref{app:eff} for the effective noise, the result of the subsequent 
analysis depends on the initial condition through the value of $d+\alpha$. For a 
quench towards the disordered phase, the generic behaviour derived for $d<2$ 
carries over to $d>2$ and only the value of the relaxational time-scale 
$\tau_{\rm r}$ changes. Accordingly, all qualitative behaviours emerging for 
$d<2$ extend to $d>2$. 

We now look again at the {\bf quenches onto and across the critical point} but we 
need to distinguish regions II and III in Fig.~\ref{fig:dalpha}.

\begin{enumerate}
 \item[\bf (B1)] 
We start by focusing on the case $d+\alpha <2$, which corresponds to spatially 
long-ranged correlated initial conditions. The scaling forms for the function 
$\dlap{G}(p,q)$ are then written as
\begin{align}
\dlap{G}(p,q) \simeq 
\begin{dcases}
\frac{c_A}{\tilde{c}_F} p^{\frac{d+\alpha}{2}-3}  \frac{\mathbb{A}(q/p)}{1+q/p} 
& \quad \mbox{for}\quad M^2 = 0, \\[.25cm]
\frac{c_A}{M^2} p^{\frac{d+\alpha}{2}-2} \mathbb{A}(q/p) & \quad \mbox{for}\quad 
M^2 > 0.
\end{dcases}
\end{align}
As before, the techniques of \ref{app:Lap} allow the formal inversion of these 
scaling forms and yield the following leading long-time behaviour 
\begin{align}  \label{gl:qn-regime2}
 g(t) \simeq 
		  \begin{dcases}
                  g_c t^{1-\frac{d+\alpha}{2}}  &  \quad \mbox{for}\quad
                  M^2 = 0, \\[.25cm]
                  g_d t^{-\frac{d+\alpha}{2}}   & \quad \mbox{for}\quad
                  M^2 > 0.
		  \end{dcases}
\end{align}

\item[\bf (B2)]
In the case $2<d+\alpha$, we must reconsider the scaling form of 
$\dlap{A}(p,q)$. Now,  $A(t,s)$ is integrable and the value of $\dlap{A}(0,0)$ 
is finite. On the other hand,
since the boundary conditions are given by $\lap{G}_0(p) = \lap{G}(p,0)$, see 
Eq.~(\ref{G0}), and these are only defined for $p\geq 0$, we can analytically 
continue $G_0 = G_0(t) = G_0(-t)$ to an even function.
Going back to Eq.~(\ref{eq:solution0}) and expanding, we find  
\begin{align}
\label{eq:solution_scale2}
\dlap{G}(p,q) \simeq \begin{dcases}
  \frac{\dlap{A}(0,0) + (6{\gamma}/u)\left[ \lap{G}_0(p) + \lap{G}_0(q)\right] 
}{\tilde{c}_F}\frac{1}{p+q} & \mbox{for}\quad M^2 = 0,\\[.25cm]
  \frac{\dlap{A}(0,0) + (6{\gamma}/u)\left[ \lap{G}_0(p) + \lap{G}_0(q)\right] 
}{M^2} +\frac{ c_A }{M^2}p^{\frac{d+\alpha}{2}-2} \mathbb{A}(q/p) &  
\mbox{for}\quad M^2 > 0.                    \end{dcases}
\end{align}
We first invert  this in the critical case $M^2=0$. We define a new even 
function $\mathscr{R}(t)$ such that 
$\lap{\mathscr{R}}(p)+\lap{\mathscr{R}}(q)=\dlap{A}(0,0)+(6\gamma/u)\left[\lap{G}_0(p)+\lap{G}_0(q)\right]$. Using the identity (\ref{gl:C:hksymm}), we obtain 
$G(t,s) = {\tilde{c}_F}^{-1}\mathscr{R}(t-s)$. Because of the boundary condition 
$G(0,0)= {\tilde{c}_F}^{-1} \mathscr{R}(0)=1$, we finally have $g(t)=G(t,t)= 
{\tilde{c}_F}^{-1} \mathscr{R}(0)=1$, i.e., $g$ does not depend on time. 

Next, 
for the case $M^2>0$, the first term in Eq.~(\ref{eq:solution_scale2}) will 
merely lead to terms concentrated around $t=0$ or $s=0$, which we neglect in our 
study of the long-time behaviour. The inversion of the last term  in 
Eq.~(\ref{eq:solution_scale2})  proceeds as before, so that we have the leading 
long-time behaviour 
\begin{align} \label{gl:qn-regime3}
 g(t) \simeq 
		  \begin{dcases}
                  g_c t^{0} & \mbox{for}\quad M^2 = 0,\\[.25cm]
                  g_d t^{-(d+\alpha)/2} & \mbox{for}\quad M^2 >0. 
		  \end{dcases}
\end{align}
Writing the long-time behaviour as in Eq.~\eqref{gl:g-crit}, we see that the 
results in Eqs.~(\ref{gl:qn-regime1}), (\ref{gl:qn-regime2}), and 
(\ref{gl:qn-regime3}) for the exponent $\digamma$ are identical to those found 
in Sec.~\ref{sec:dynamics} for the effective Markovian dynamics either in the 
critical regions I, II, III or for $r_0<r_0^c$. Hence the results listed in 
Tab.~\ref{tab:eff} hold true for both  the actual quantum dynamics and the 
effective dynamics. 
\end{enumerate}

\section{Physical observables - the quantum noise case}
\label{sec:observables}

Having solved the spherical constraint in the previous section, we can now 
determine the long-time behaviour of the physical correlation and response 
functions defined in Sec.~\ref{sec:results}. We shall begin with the one-time 
correlation function before we move on to the two-time quantities.

\subsection{One-time correlation function}

The equal-time correlation function is found from the formal solution 
Eqs.~(\ref{eq:sol}) and~(\ref{gl:C-formel}) as
\begin{align}\label{eq:Ccrit}
 C_{\vec{k}}(t) &= \frac{e^{-2k^2 t /\gamma}}{g(t)}\left[ c_\alpha k^\alpha 
+\frac{2 \hbar }{\pi \gamma} \int_0^t \!\D t_1 \int_0^t \!\D t_2 \: g_2(t_1, 
t_2)\, e^{k^2 (t_1+t_2) /\gamma} \frac{\tto^2 - (t_1-t_2)^2}{\left(\tto^2 + 
(t_1-t_2)^2\right)^2} \right] 
 =:  C_{\vec{k}}^{(\rm ic)}(t)+ C_{\vec{k}}^{\rm (n)}(t).
\end{align}
Here, we decompose the total correlator into a contribution  $C_{\vec{k}}^{(\rm 
ic)}(t)$ due to the initial correlator and a contribution $C_{\vec{k}}^{(\rm 
n)}(t)$ of the quantum noise. In what follows, we shall analyse the relative 
importance of these terms. We emphasise that it is not the function $G(t,s)$, 
but rather the function $g_2(t,s) = \sqrt{g(t)g(s)}$, defined in 
Eq.~(\ref{eq:g2}), which arises in the quantum noise integral contribution 
$C_{\vec{k}}^{(\rm n)}(t)$.  

We proceed by considering the quenches above, onto and below the critical point 
$r_0^c$. 
\begin{enumerate}
 \item[\bf (A)]
For a {\bf quench above the critical point}, in the previous section we had 
derived the following exponential long-time behaviour
\begin{equation} \label{eq:g2-asy-exp}
 g_2(t,s)  \sim \exp\left(\frac{t+s}{2\tau_{\rm r}}\right).
\end{equation}
First, it follows that the initial correlations  $C_{\vec{k}}^{(\rm ic)}(t)$ are 
exponentially suppressed at large times and it suffices to treat the quantum 
noise correlations $C_{\vec{k}}^{\rm (n)}(t)$. Second, the main contribution to 
that integral in Eq.~(\ref{eq:Ccrit}) comes from the upper integration bound and 
we may replace $g_2(t,s)$ by its leading asymptotic form. Third, since $g_2$ 
depends only on the sum $t+s$, we can reduce the integration domain in 
Eq.~(\ref{eq:Ccrit}) to the triangle $0\leq t_2\leq t_1\leq t$ and then change 
coordinates to diagonal coordinates (see also \ref{app:F}). This yields  
\begin{equation}
 C^{\rm (n)}_{\vec{k}}(t) = e^{-(1/\tau_{\rm r}+ 2k^2/\gamma )t} \frac{2 \hbar 
}{\pi \gamma} \int_0^t \!\D u \int_u^{2t-u} \!\D v \:  e^{ \left[(2\tau_{\rm 
r})^{-1}+ k^2/\gamma\right] v } \frac{\tto^2 - u^2}{\left(\tto^2 + u^2\right)^2}.
\end{equation}
Calculating the $v$-integral we find that the quantum noise acts as a 
distribution on two exponential test functions
\begin{equation}
 C^{(n)}_{\vec{k}}(t) = \frac{2 \hbar }{\pi \gamma} \int_0^t \!\D u\: 
\frac{e^{-\left[(2\tau_{\rm r})^{-1}+k^2/\gamma\right] u} -e^{-\left[(2\tau_{\rm 
r})^{-1}+k^2/\gamma\right](2t-u)}}{(2\tau_{\rm r})^{-1}+k^2/\gamma} \frac{\tto^2 
- u^2}{\left(\tto^2 + u^2\right)^2}.
\end{equation}
The first term can be evaluated, in the limit $t\to \infty$, as discussed 
in~\ref{app:F}, in particular Eq.~(\ref{eq:Dexp}), while the second term is 
exponentially suppressed. We eventually arrive at the stationary correlator 
already quoted in Eq.~(\ref{gl:correl-stat-disor}) 
\begin{equation}
 C_{\vec{k}}(\infty) = \frac{2 \hbar }{\pi \gamma} g_{\rm AS }(\tto( (2\tau_{\rm 
r})^{-1} + k^2/\gamma)),
\end{equation}
with the function $g_{\rm AS}$ defined in Eq.~(\ref{eq:aux}).

 \item[\bf (B)] For a {\bf critical quench}, we shall study the scaling limit 
defined in Eq.~(\ref{eq:spatiotemp}) of the equal-time correlator. We derived in 
Sec.~\ref{sec:solution} the leading long-time behaviour $ g(t) \simeq g_c 
t^\digamma$ with $\digamma\geq0$, see also Tab.~\ref{tab:eff}. The contribution 
from the initial correlations shows a scaling behaviour
\begin{equation}
 C_{\vec{k}}^{\rm (ic)} (t) \simeq \frac{c_\alpha}{g_c} (k^2 t)^{-\digamma} 
e^{-2 k^2 t /\gamma} k^{\digamma + \alpha/2}.
\end{equation}
The noise contribution reads
\begin{align}\nonumber
 C_{\vec{k}}^{(n)}(t) &\simeq \frac{2\hbar}{\pi \gamma} \frac{1}{g(t)} 
\int_{[0,t]^2}\! \!\D t_1 \D t_2 \: e^{-k^2 (2t-t_1-t_2)/\gamma} g_2(t_1,t_2) 
\frac{\tto^2 - (t_1-t_2)^2}{\left(\tto^2 + (t_1-t_2)^2\right)^2}\\[.25cm]
&=  \frac{2\hbar}{\pi \gamma} \frac{(g_2 \ast \ast h)(t,t)}{g(t)} ,
\end{align}
and is expressed via a double convolution, with the auxiliary function
\begin{equation}
 h(t_1,t_2) = e^{-\frac{k^2}{\gamma}(t_1+t_2)}  \frac{\tto^2 - 
(t_1-t_2)^2}{\left(\tto^2 + (t_1-t_2)^2\right)^2}.
\end{equation}
In order to analyse this double convolution, we study first the two-time double 
convolution with two distinct arguments $(g_2 \ast \ast h)(t,s)$ and  then set 
them equal at the end. We emphasise that this procedure does {\it not} correspond 
to studying a two-time correlation function. The double Laplace transform of $h$ 
is evaluated using~\ref{app:F}, see Eq.~(\ref{eq:Dexp}). For the sought 
long-time scaling limit, one should fix $\bar{p} = p \gamma/k^2$ and $\bar{q} = 
q \gamma/k^2$ and then expand for $p,q,k^2$ small. We find 
\begin{align} 
 \dlap{h}(p,q) 
  &= \frac{\frac{k^2}{\gamma}+p}{\frac{k^2}{\gamma}+\frac{p+q}{2}} 
  g_{\rm AS}\left(\tto\left( \frac{k^2}{\gamma}+p\right)\right)
 +\frac{\frac{k^2}{\gamma}+q}{\frac{k^2}{\gamma}+\frac{p+q}{2}} 
 g_{\rm AS}\left(\tto\left(\frac{k^2}{\gamma}+q\right)\right) \nonumber \\[.25cm] 
&\simeq -2 \left[ {\rm C}_E +\ln \left(\frac{\tto k^2}{\gamma}\right)\right]
%
%
-\frac{\left(1+\bar{p}\right) \ln \left( 1+\bar{p}\right) + \left(1+\bar{q}\right) 
\ln \left( 1+\bar{q}\right)}{1+\left(\bar{p}+\bar{q}\right)/2}.
\label{eq:88}
\end{align}
Then, by using the form of $\dlap{h}$ in Eq.~(\ref{eq:88}) and the identities 
from \ref{app:prop_lap}, we find
\begin{align}
 \frac{(g_2\dast h)(t,t)}{g(t)} &= \frac{1}{g(t)} \mathcal{L}_2^{-1} \left( 
\dlap{g_2}(p,q) \dlap{h}(p,q) \right)(t,t) \nonumber \\
&= -2 \left[ {\rm C}_E +\ln \left(\frac{\tto k^2}{\gamma}\right)\right] -  
\frac{1}{g(t)}\mathcal{L}_2^{-1} \left( \dlap{g_2}(p,q) 
\frac{(1+\bar{p})\ln(1+\bar{p})+(1+\bar{q})\ln(1+\bar{q})}{1+\demi\left( 
\bar{p}+\bar{q}\right)} \right)(t,t)  \nonumber \\
&= -2 \left[ {\rm C}_E +\ln \left(\frac{\tto k^2}{\gamma}\right)\right]  - 
\left(\frac{k^2}{\gamma}\right)^{-2}\Gamma^2\left(1+\frac{\digamma}{2}\right) 
\left(\frac{k^2 t}{\gamma}\right)^{-\digamma} \times \nonumber \\
& ~~~ \times \mathcal{L}_2^{-1} \left( 
\left(\bar{p}\bar{q}\right)^{-1-\digamma/2} 
\frac{(1+\bar{p})\ln(1+\bar{p})+(1+\bar{q})\ln(1+\bar{q})}{1+\demi\left( 
\bar{p}+\bar{q}\right)} \right)(t,t)  .
 \label{eq:89}
\end{align}
In \ref{app:Lap}, we show that the scaling function  
$\Phi(\bar{p},\bar{q})=\Phi(p\gamma/k^2,{q}\gamma/k^2)$  is the double Laplace 
transform of the  scaling function $(k^2/\gamma)^{2} \phi(k^2 t/\gamma, k^2 
s/\gamma)$, provided $\Phi(u,v)=\dlap{\phi}(u,v)$. Applying this to the last 
line of Eq.~(\ref{eq:89}), we find the scaling form of the noise contribution 
\begin{align}
 C_{\vec{k}}^{\rm (n)}(t) \simeq -\frac{4\hbar}{\pi \gamma} \left[ {\rm C}_E 
+\ln \left(\frac{\tto k^2}{\gamma}\right)\right] -\frac{2\hbar}{\pi \gamma} 
(k^2t/\gamma)^{-\digamma} \Psi(k^2 t/\gamma, k^2 t/\gamma),
\end{align}
with a universal scaling function $\Psi(t,t')$ which can be obtained by inverting
\begin{equation}
\dlap{ \Psi}(p,q) = \Gamma^2\left(1+\frac{\digamma}{2}\right) 
(pq)^{-1-{\digamma}/{2}}\, \frac{ (1+p)\ln(1+p) + (1+q)\ln(1+q)}{1+(p+q)/2}  .
\end{equation}
This inversion can be  carried out by using Eq.~(44) at p.~186 of 
Ref~\cite{Voel50} or Eq.~(\ref{gl:C:tt}), followed by  Eq.~(2.5.2.1) of 
Ref.~\cite{Prud5}. This leads to the integral representation
\begin{subequations} \label{gl:Psi-scal-funk}
\begin{align} \nonumber
 \Psi(x,x) &= 4\int_0^x \D \xi\ e^{-2(x-\xi)} \xi^{\digamma+1}\bigg\{ \frac{\ 
_{2}F_2\left(1,1;2,2+\frac{\digamma}{2}; -\xi\right)}{1+\frac{\digamma}{2}} 
+\frac{1}{\xi}\left[\psi\left(1+\frac{\digamma}{2}\right) - \ln \xi  \right]
 \\
 &~~~+\frac{1}{\xi} \ _{2}F_2\left(1,1;2,1+\frac{\digamma}{2}; -\xi\right) 
+\frac{\digamma}{2\xi^2}\left[\psi\left(\frac{\digamma}{2}\right) - \ln \xi  
\right] \bigg\}\\[.75cm]
 &\sim 
 \begin{dcases}
  x^{\digamma} \ln x  & \mbox{\rm ~~~ for $x\ll 1$}, \\
  x^{\digamma-1} & \mbox{\rm ~~~ for $x\gg 1$} ,
 \end{dcases}
 \end{align}
 \end{subequations}
where $\psi(x)$ is the digamma function \cite{Abra65}. The asymptotics for small 
arguments is obtained by expanding the integrand above. The asymptotics of 
$\Psi$ for large scaling arguments is obtained, instead, from a Laplace 
approximation~\cite{Cop65}, or by using the results presented in~\ref{app:Lap}. 

Finally, we can compare the contributions due to initial correlations to those 
due to noise correlations. We make use of the values of $\digamma$ listed in 
table~\ref{tab:eff} to do this. In region I (see Fig.~\ref{fig:dalpha}), both 
terms equally contribute and we need to consider the full equal-time 
correlator~(\ref{eq:Ccrit}). In regions II and III instead, the initial term 
dominates for small momenta, because of the prefactors $k^{1-d/2}$ or 
$k^{\alpha/2}$, respectively.

 \item[{\bf (C)}] For a {\bf quench across criticality} the noise contribution 
is still given by Eq.~(\ref{eq:Ccrit}) upon the replacement $\alpha\mapsto 
d+\alpha$. The initial correlations are nevertheless more relevant in this case 
since their scaling form reads
\begin{equation}
 C_{\vec{k}}^{\rm (ic)}(t) = \frac{c_\alpha}{g_d} k^{-d} (k^2 t)^{(d+\alpha)/2} 
e^{-2 k^2 t/\gamma}.
\end{equation}
Because of the factor $k^{-d}$ in this expression, 
the initial correlations dominate in the scaling limit.
\end{enumerate}

\subsection{Two-time response and correlation functions}
We consider first the two-time response function. Since the underlying equations 
of motion are linear, it is clear from Eq.~(\ref{eq:R}) that the response 
function remains unaffected by the noise structure, be it classical, effective 
or quantum. It follows that in all these cases
\begin{equation} \label{eq:93}
 R_{\vec{k}}(t,s) = \gamma^{-1} \sqrt{\frac{g(s)}{g(t)}} e^{-k^2(t-s)/\gamma} 
\Theta(t-s).
\end{equation}
Using $g(t)$ as derived in Sec.~\ref{ssec:g} yields the results discussed in 
Sec.~\ref{ssec:qresults}.

In contrast, the two-time correlation function  
\begin{equation}\label{eq:c2}
 C_{\vec{k}}(t,s) = \frac{e^{-\frac{k^2}{\gamma} (t+s)}}{g_2(t,s)}\left[
 c_\alpha k^\alpha + \frac{2\hbar}{\pi \gamma} \int_0^t \!\D t' \int_0^s \!\D s'\: e^{\frac{k^2}{\gamma}(t'+s')} g_2(t',s') 
 \frac{\tto^2 - (t'-s')^2}{\left[\tto^2 + (t'-s')^2\right]^2} 
 \right] ,
\end{equation}
does depend on the noise correlator. We need to study carefully this 
non-Markovian integral in order to analyse its relevance with respect to the 
contribution of the correlations in the initial state, given by the first term 
in the brackets. We shall use the decomposition $C_{\vec{k}}(t,s) = 
C_{\vec{k}}^{(\rm ic)}(t,s) + C_{\vec{k}}^{\rm (n)}(t,s)$ to refer to the first 
contribution from the initial correlations (ic) and the second contribution 
coming from the noise correlations (n), corresponding to the first and second 
term in brackets respectively. 

\begin{enumerate}
 \item[\bf (A)]
For a {\bf quench to the disordered region} the self-consistent function shows 
an exponential behaviour. We shall evaluate the two-time correlator in the 
asymptotic limit with $s\to\infty$ but fixed $\tau=t-s$. Later on, we shall also 
consider the case in which $\tau$ becomes large. Since we are generally 
interested in $t>s$, we may separate the integration  in Eq.~(\ref{eq:c2}) into 
two terms $\int_0^t \!\D t' \int_0^s \!\D s' = \int_0^s \!\D t' \int_0^s \!\D s' 
+  \int_s^t \!\D t' \int_0^s \!\D s' $. The first term will contribute to the 
equal-time correlation function $C_{\vec{k}}^{(n)}(s)$, see 
Eq.~(\ref{eq:Ccrit}). The second integral retains only the dependence on $\tau$ 
in the asymptotic limit. In fact, this can be seen as follows. First, we write
\begin{align}\nonumber
C_{\vec{k}}^{(n)}(s+\tau,s) &\simeq e^{-\left[k^2/\gamma+(2\tau_{\rm 
r})^{-1}\right]\tau} C_{\vec{k}}^{(n)}(s)\\
& ~~~+  \frac{2\hbar}{\pi\gamma}  \int_s^{s+\tau}\!\!\!\D t'\int_0^s \!\D s'\: 
e^{(k^2/\gamma+(2\tau_{\rm r})^{-1})(t'+s'-2s-\tau)} \frac{\tto^2 - 
(t'-s')^2}{\left[ \tto^2 + (t'-s')^2\right]^2} .
\end{align}
We identify in the first term the part $C_{\vec{k}}^{(n)}(s)$ of the equal-time 
correlator, using Eq.~(\ref{eq:Ccrit}). The second term is dominated by the 
contributions near to the upper limits of integration, so that we immediately 
substituted the exponential form~(\ref{eq:g2-asy-exp}) for $g_2$. Now, we can 
express the first term as a response function by using Eq.~(\ref{eq:93}) and 
again Eq.~(\ref{eq:g2-asy-exp}), in the asymptotic limit. In the the second 
term, the new variables $u=t'-s-\tau$ and $v=s-s'$ are introduced, yielding
\begin{align}
C_{\vec{k}}^{(n)}(s+\tau,s) &\simeq \gamma R_{\vec{k}}(\tau) 
C_{\vec{k}}^{(n)}(s) 
+ \frac{2\hbar}{\pi\gamma}  \int_{-\tau}^{0}\!\!\D u\int_0^s \!\D v\: 
e^{(k^2/\gamma+(2\tau_{\rm r})^{-1})(u-v)} \frac{\tto^2 - (\tau+u+v)^2}{\left[\tto^2 
+ (\tau+u+v)^2\right]^2}
\end{align}
Next, we rescale $u$ and $v$ such that 
the limit $\tto\to 0$ can be taken (we also let $s\to\infty$). This yields
\begin{align}
C_{\vec{k}}^{(n)}(s+\tau,s) &\simeq \gamma R_{\vec{k}}(\tau) 
C_{\vec{k}}^{(n)}(\infty) - \frac{2\hbar}{\pi\gamma}  \int_0^{\infty} \!\D v 
\int_{-1}^{0}\!\D u\: \frac{e^{\tau(k^2/\gamma+(2\tau_{\rm 
r})^{-1})(u-v)}}{(1+u+v)^2} .
\end{align}
Next, the $u$-integration above is calculated,
for $\tau$ large, by using the Laplace approximation \cite{Cop65} and finally, the 
remaining $v$-integral is estimated, again for $\tau$ large. We eventually find
\begin{align}
C_{\vec{k}}^{(n)}(s+\tau,s) &\simeq \gamma R_{\vec{k}}(\tau)  
C_{\vec{k}}^{(n)}(\infty) -\frac{2\hbar}{\pi\gamma} \frac{1}{\left[(2\tau_{\rm 
r})^{-1} + k^2/\gamma\right]^2} \frac{1}{\tau^2}.
\end{align}
The first term in $C_{\vec{k}}^{(n)}(s+\tau,s)$ decays exponentially upon 
increasing $\tau$, see Eq.~(\ref{eq:93}), while the second term depends 
algebraically on $\tau$. For large $\tau$, we conclude that 
\begin{equation}
C_{\vec{k}}(s+\tau,s)\simeq -\frac{2\hbar}{\pi\gamma}\frac{1}{\left[(2\tau_{\rm 
r})^{-1}+ k^2/\gamma\right]^2}\frac{1}{\tau^2},
\end{equation}
since the initial contribution $C_{\vec{k}}^{\rm (ic)}(t,s)$ is exponentially 
small as $\tau$ increases. This proves Eq.~(\ref{eq:eff-C-single}) in 
Sec.~\ref{sec:results}.

 \item[\bf (B)]  For a {\bf quench to the critical point} we need to reconsider 
the noise contribution. Since we are mainly interested in the universal 
exponents defined in Eq.~(\ref{eq:lambdas}) we shall immediately work with the autocorrelation function  $C(t,s) = 
\int_{\vec{k}} C_{\vec{k}}(t,s)$ from which they can be readily obtained. The noise contribution then reads
\begin{equation}
 C^{\rm (n)}(t,s) = \frac{2\hbar}{\pi \gamma} \frac{\Omega_d}{(2\pi)^d} (t 
s)^{-\digamma/2} \int_0^t \!\D t' \int_0^s \!\D s'\: \frac{(t's')^{\digamma/2} 
}{(t+s-t'-s')^{d/2}} \frac{\tto^2 - (t'-s')^2}{\left(\tto^2 + 
(t'-s')^2\right)^2}.
\end{equation}
This can be written via a weighted convolution, defined 
in~\ref{app:prop_lap}, as
\begin{equation} \label{eq:99}
 C^{\rm (n)}(t,s) = \frac{2\hbar}{\pi \gamma} \frac{\Omega_d}{(2\pi)^d}
 (t s)^{-\digamma/2} \left( h_1 \dast_w h_2 \right)(t,s)
\end{equation}
with the identifications
\begin{equation}
 h_1(t,s) = (ts)^{\digamma/2}, \quad h_2(t,s) = (t+s)^{-d/2} \quad \text{and}
 \quad
 w(t-s) = \frac{\tto^2-(t-s)^2}{\left[ \tto^2 +(t-s)^2\right]^2}.
\end{equation}
Using Eq.~(\ref{eq:Lw}), we need to evaluate the double Laplace transforms of 
$h_2(t,s)$ and of $h_1(t,s) \cdot w(t-s)$, in order to factorise the weighted 
convolution (\ref{eq:99}) and study its asymptotics, followed by the application 
of a Tauberian theorem. While the former is straightforwardly evaluated as
\begin{equation} \label{eq:102}
\dlap{h_2}(p,q) = \Gamma \left(1-\frac{d}{2}\right) \frac{ p^{\frac{d}{2}-1}- 
q^{d/2-1}}{q-p},
\end{equation}
finding the latter is a non-trivial task. Starting from the formal definition of 
the double Laplace transform, using the techniques from \ref{app:F} in order to 
go over to diagonal coordinates yields
\begin{align}
 \dlap{\bigl(h_1 w\bigr)}(p,q) &= \int_0^\infty\!\D t\int_0^\infty\! \D s\:(t 
s)^{\digamma/2} \frac{\tto^2-(t-s)^2}{\left[\tto^2 +(t-s)^2\right]^2} e^{-pt-qs} 
\nonumber
\\[.2cm]
 &= \int_0^\infty \!\D t\int_0^t \!\D s\: (t s)^{\digamma/2} 
\frac{\tto^2-(t-s)^2}{\left[\tto^2 
+(t-s)^2\right]^2}\left(e^{-pt-qs}+e^{-ps-qt}\right)
= \int_0^\infty \D u \frac{\tto^2-u^2}{\left( \tto^2 +u^2\right)^2} H(u),
\label{eq:CDL}
\end{align}
with the auxiliary function $H(u)$. This is evaluated by using Eq. (2.3.6.10) of 
Ref.~\cite{Prud1} and where $K_{\nu}$ is a modified Bessel function 
\cite{Abra65}. Calculating the integral and then expanding in $u$, we find  
\begin{align}
 H(u) &= 2^{-\digamma}\cosh\left(\frac{u}{2}(p-q)\right) 
e^{-\frac{u}{2}(p+q)}\int_0^\infty  \left(v^2+2 uv\right)^{\digamma/2} 
e^{-\frac{v}{2}(p+q)} \ \D v \nonumber \\[.2cm]
&=\frac{\digamma\,\Gamma\left(\frac{\digamma}{2}\right)}{ \sqrt{\pi}} 
\left(\frac{u}{p+q}\right)^{\frac{1}{2}+\frac{\digamma }{2}} 
\cosh\left(\frac{u}{2}(p-q)\right) K_{\frac{1}{2}+\frac{\digamma 
}{2}}\left(\frac{p+q}{2}  u\right) \nonumber \\[.2cm]
&\stackrel{u\ll 1}{\simeq} 
\frac{\digamma\,\Gamma\left(\frac{\digamma}{2}\right)}{ \sqrt{\pi}} 
\left[2^{\digamma }  \Gamma \left(\frac{1+\digamma }{2}\right) 
(p+q)^{-\digamma-1} +2^{-\digamma -2} \Gamma \left(\frac{-\digamma -1}{2}\right) 
\, u^{ 1+\digamma } \right]. 
 \label{eq:104}
\end{align}
We are now able to evaluate the double Laplace transform using the small-$u$ 
expansion of $H(u)$ in Eq.~(\ref{eq:104}), according to the analysis presented 
in~\ref{app:F}. First, the $u$-independent constant term does not contribute to 
the integral for $\dlap{h_1 w}$, see Eq.~(\ref{eq:integral-potenz}) with $s=0$.  
Next, the lowest-order correction in $u$ is  independent of $p$ and $q$. Thus,  
using Eqs.~(\ref{eq:102}) and (\ref{gl:C:hk}), we conclude that to leading order 
the weighted convolution behaves as
\begin{equation}
 (h_1 \dast_w h_2)(t,s) \sim (t+s)^{-d/2},
\end{equation}
which implies that $C^{\rm (n)}(t,s)$ is less relevant than the initial 
correlations $C^{\rm (ic)}(t,s)$, if $\alpha<0$. We should remark, however, that 
this argument does not hold if $\digamma = 0$, e.g., in region I with $\alpha 
=0$ (see Fig.~\ref{fig:dalpha}). Then the term we calculated above is constant 
as well and the next-to-leading contribution in $u$ must be worked out  by 
expanding $H(u)$ in Eq.~(\ref{eq:104}) to the next order in $u$ and using again 
\ref{app:F}. Then the scaling of the  weighted convolution behaves as 
\begin{equation}
 (h_1 \dast_w h_2)(t,s) \sim (t+s)^{-d/2-1}
\end{equation}
for $t$ and $s$ large. Accordingly, the initial correlations always dominate the 
asymptotic behaviour of the two-time correlation function. The two-time 
autocorrelation function thus reads
\begin{equation}
 C(t,s) \simeq C^{\rm (ic)}(t,s) \simeq s^{-{d}/{2}} f_C(t/s) \quad 
\mbox{with}\quad f_C(x) = \frac{c_{\alpha}}{g_c} 
\frac{\Omega_d}{(2\pi)^d}\frac{x^{-{\digamma}/{2}}}{(1+x)^{(d+\alpha)/2}}.
\end{equation}

 \item[\bf (C)] 
 For a {\bf quench across criticality} all the steps presented above for the 
analysis of the noise contribution of the critical auto-correlator apply upon 
replacing $\alpha \mapsto \alpha +d$. In particular, the initial correlations 
remain the dominating contribution such that finally 
\begin{equation}
 C(t,s) \simeq s^{0} f_C(t/s) \quad \mbox{with}\quad f_C(x) = 
\frac{c_{\alpha}}{g_d} 
\frac{\Omega_d}{(2\pi)^d}\frac{x^{{(d+\alpha)}/{4}}}{(1+x)^{(d+\alpha)/2}}.
\end{equation}
Summarising, for quantum quenches onto or across the critical point, the ageing 
scaling behaviour of the two-time auto-responses and auto-correlators is the 
same as the one derived for the effective dynamics in Sec.~\ref{sec:results}. 
\end{enumerate}

\section{Conclusions}
\label{sec:conclusion}

We have presented a detailed study of the relaxational dynamics and of the 
ageing phenomena in a many-body quantum system in contact with an external bath 
at temperature $T=0$. In comparison with classical dynamics, we have considered 
two distinct types of noise correlators, characterised by the corresponding 
correlators:
\begin{enumerate}[label={\bf (\alph*)}]
\item 
Non-Markovian, {\it quantum noise} the correlators of which are given in 
Eqs.~(\ref{eq:qn_anticommu_T}) and~(\ref{eq:qn_commu}) and are derived from the 
system-interaction-bath method which is known to reproduce all physically 
desirable properties of the system, including the quantum 
fluctuation-dissipation theorem for temperatures $T>0$ 
\cite{Cal81,Ford65,Ford87,Arau19}. Setting $T=0$, a regularisation, such as that 
in Eq.~(\ref{eq:qn_anticommu}), is necessary \cite{Gard04}. 
\item
A Markovian {\it effective noise}, with correlators (\ref{eq:eff_qn_anticommu}) 
which resembles a classical white noise, but with a momentum-dependent effective 
temperature $T_{\rm eff}=\frac{\mu}{2} |\vec{k}|^2$. This was chosen such as to 
reproduce the scaling dimensions of the actual quantum noise.
\end{enumerate} 
Comparing these two noises allows us to study the non-Markovian memory effects, 
which are present in the quantum noise correlators, but not in the Markovian 
effective noise. In addition, the scaling of these two noises is different from 
the one of the classical white noise. We chose to investigate two paradigmatic 
and exactly solvable models of statistical mechanics, namely the quantum 
spherical model and the quantum $O(n)$-model, with $n\to\infty$. For simplicity, 
only ``ferromagnetic'' nearest-neighbour interactions were considered, in $d$ 
spatial dimensions, but generalisations should not be very difficult. Accounting 
for the effects of a ``paramagnetic'' initial state with spatially long-ranged 
correlations  as in Eq.~(\ref{eq:C0}) \cite{Pico02} turned out to be an 
important for understanding the ensuing dynamics. 

In analogy with what was already known at $T>0$ from the quantum equilibrium 
state and the classical dynamics of these two models, {\it the long-time 
relaxational behaviour, at $T=0$, of the quantum spherical and the  $O(n)$-model 
for $n\to\infty$ belongs to the same universality class}, for both cases of 
noise considered. These predictions were not obtained for the full quantum 
Langevin equation, but rather for its over-damped limit (\ref{eq:qle}) which has 
been derived in a particular scaling limit~\cite{Arau19}. That these models are 
solvable is due to the exact reduction of the many-body dynamics to a single 
integro-differential equation for a function $g(t)$. Since that equation is 
strongly non-linear, it is solved by embedding it into a system of linear 
equations for a function $G(t,s)$ of two variables, for which $G(t,t)=g(t)$ 
holds. Then the long-time behaviour of $g(t)$ has been derived via Mellin 
transform methods and Tauberian theorems, discussed in detail in the appendices. 
From this, the long-time behaviour of the correlation and response functions of 
the order parameter can be obtained. 

The meaning of our results becomes clearer from a comparison with those of 
classical dynamics, which are summarised in Table~\ref{tab:class}. 
\begin{table}[t]
 \captionsetup{width=.8\textwidth}
\setlength{\tabcolsep}{10pt} 
\renewcommand{\arraystretch}{1.5} 
\centering
\caption{Non-equilibrium exponents for classical dissipative dynamics, taken 
from Ref.~\cite{Pico02}.  For a critical quench to $T=T_c$, one should 
distinguish a number of cases, denoted by I$_c$ - V$_c$. For a sub-critical 
quench to $T<T_c$ such a distinction is not necessary.} 
{\small\begin{tabular}{|lr||c||c|c|c|c|c|}\hline
\multicolumn{2}{|c||}{region}     & $\digamma$             & $\lambda_C$ 
& $\lambda_R$      	     & $a$ 		       & $b$                 \\ \hline 
I$_c$   & $2<d<4$, $0<d+\alpha<2$ & $-1-\frac{\alpha}{2}$  & $d+\frac{\alpha}
{2}-1$ & $d-\frac{\alpha}{2}-1$ & $\frac{d}{2}-1$ & $\frac{d}{2}-1$     \\ \hline
II$_c$  & $4<d$, $0<d+\alpha<2$   & $1-\frac{d+\alpha}{2}$ & $1+\frac{d+\alpha}{2}
$ & $\frac{d-\alpha}{2}+1$ & $\frac{d}{2}-1$ & $1$ 		           \\ \hline
III$_c$ & $2<d<4$, $d+\alpha>2$   & $\frac{d}{2}-2$        & $\frac{3}{2}d-2$  	    
& $\frac{3}{2}d-2$       & $\frac{d}{2}-1$ & $\frac{d}{2}-1$      \\  \hline
IV$_c$  & $4<d$, $d+\alpha>2$, $\alpha>-2$ & 0             & $d$       	            
& $d$                    & $\frac{d}{2}-1$ & $\frac{d}{2}-1$      \\  \hline
V$_c$   & $4<d$, $d+\alpha>2$, $\alpha<-2$ & 0             & $d+\alpha$       	    
& $d$                    & $\frac{d}{2}-1$ & $\frac{d+\alpha}{2}$ \\ \hline \hline 
\scriptsize{$T<T_c$} & $2<d$      & $-\frac{d+\alpha}{2}$  & $\frac{d+\alpha}{2}$   
& $\frac{d-\alpha}{2}$   & $\frac{d}{2}-1$ & $ 0$		 	  \\ \hline
\end{tabular}}
\label{tab:class}
\end{table}
\setlength{\tabcolsep}{6pt}
\renewcommand{\arraystretch}{1.0}

The main  results of our analysis can be stated as follows.
\begin{enumerate}[label={\bf (\arabic*)}]
\item {\it The stationary state for a quench in the one-phase region
at temperature $T=0$ is not an 
equilibrium state.} This holds for both the quantum and the effective dynamics. 
This conclusion is based on the following evidence:
\begin{enumerate} 
\item For the existence of a critical point, between classical and quantum 
dynamics there is a dimensional shift $d+2\mapsto d$. We find a dynamical 
quantum phase transition, with a finite critical coupling $r_0^c$, for any 
dimension $d>0$. However, an {\it equilibrium} quantum phase transition only 
exists for $d>1$. 
\item For quenches into the single-phase region, both responses and correlators 
rapidly become stationary and do not depend on the initial conditions. The 
two-time response $R_{\vec{k}}(s+\tau,s)$ is equal to the classical one and 
decays exponentially upon increasing $\tau$. For the effective dynamics, this is 
also the case for the two-time correlator $C_{\vec{k}}(s+\tau,s)$ and the {\it 
classical} fluctuation-dissipation theorem with the effective temperature 
$T_{\rm eff}(\vec{k})\ne 0$ is satisfied. 

For the quantum noise, instead, the two-time correlator $C_{\vec{k}}(s+\tau,s)$ 
decays algebraically upon increasing $\tau$. This mismatch of functional forms 
makes it impossible to satisfy the quantum fluctuation-dissipation theorem.  
\end{enumerate}
\item ``Quantum ageing'' may be  characterised via the scaling behaviour in 
Eq.~(\ref{gl:vieux}) of the two-time response and correlation functions for 
quantum quenches onto or across the quantum phase transition. The universal 
exponents which describe ageing turn out to be the same for the effective and 
the quantum dynamics and they are listed in Table~\ref{tab:eff}.  Accordingly, 
{\it quantum memory effects do not appear to be relevant for the ageing 
dynamics.}
\item The scaling properties observed during  ``quantum ageing'' are subtly 
different from those of classical dynamics, as it can be inferred from comparing 
the relevant characteristic exponents reported in Tables~\ref{tab:eff} 
and~\ref{tab:class}. \begin{enumerate} \item For a critical quench, 
Table~\ref{tab:eff} distinguishes the cases I - III (see Fig.~\ref{fig:dalpha}) 
of ``quantum ageing'', and Table~\ref{tab:class} those I$_c$-V$_c$ of classical 
ageing. The classical cases III$_c$ and IV$_c$, which correspond to the 
paradigmatic instance of fully uncorrelated initial conditions 
\cite{Ronc78,Godr00}, do not have an analogue in ``quantum ageing''. %
\item We 
observe the correspondences $\mbox{\rm I} \leftrightarrow \mbox{\rm I}_c$, 
$\mbox{\rm II} \leftrightarrow \mbox{\rm II}_c$ and $\mbox{\rm III} 
\leftrightarrow \mbox{\rm V}_c$. The exponents $\lambda_{C,R}$ in the cases 
I$_c$ and I are related by the dimensional shift $d-1\mapsto d$.  This 
relationship however does not extend to the exponents $a$ and $b$. 

In the remaining two correspondences, the respective exponents are all 
identical. The difference between classical and the quantum ageing which
we observe here, merely comes from the dimensional shift $d+2\mapsto d$.
\end{enumerate}
\item For critical quenches, both the effective and the quantum noises appear to 
be weaker than classical white noise. This comes about since those classical 
cases III$_c$ and IV$_c$, where the bath noise dominates the contributions of 
the initial correlations, do not have a quantum analogue.  Only in the region I 
(and analogously, I$_c$ for classical noise) are their corresponding 
contributions of the same order. Accordingly, {\it critical quantum systems 
appear to be more sensitive than classical ones to spatially long-ranged initial 
correlations.} 
\item For a critical quantum quench, spatially long-ranged initial correlations 
with $\alpha<0$ are necessary for a long-time scaling behaviour distinct from 
that predicted by mean-field theory. This is not the case for classical 
dynamics. 
\item Quantum memory effects are apparent for the equal-time structure factor. 
\begin{enumerate}
\item For a critical quench in region I, Fig.~\ref{fig:2tempsegaux_corr} shows 
that although the time-dependence of $C_{\vec{k}}(t)$ for the quantum noise is 
qualitatively very similar to the one of the effective dynamics, there are also quantitative 
differences, notably for smaller values of the scaling variable $\rho = k^2 
t/\gamma$.
\item The stationary structure factor $C_{\vec{k}}(\infty)$ resulting from the 
quantum noise is well approximated by an Ornstein-Zernicke form in the 
single-phase region, see Fig.~\ref{fig:1phase_corr}. The different form for the 
effective dynamics comes from the momentum-dependence of $T_{\rm eff}(\vec{k})$. 
 
\end{enumerate}
\item 
In the single-phase region, for large waiting times $s$, the two-time correlator 
$C_{\vec{k}}(s+\tau,s)$, becomes independent of $s$ and decays upon increasing $\tau$ algebraically for quantum 
noise and exponentially for the effective dynamics. This evidence for dynamical 
scaling above the quantum critical point is surprising. Its origin may require 
further investigations in the future. 
\item
The dynamical exponent $z=2$ of the dissipative dynamics of open quantum systems 
is distinct from the value $z=1$ of the unitary dynamics of closed quantum 
systems \cite{Cala16,Delf18} (or of Markovian approximations of quantum dynamics 
via Lindblad equations \cite{Wald18}).  If one considers
a dynamics where $T\to 0$ such that the stationary state 
is an equilibrium state, dissipative quantum dynamics still leads to $z=2$ but 
different values of the exponents of ageing are obtained~\cite{Gagel14,Gagel15}. 
\item For quantum quenches across criticality, the long-time dynamics is 
dominated by the initial correlations and is identical to classical dynamics, 
see Tables~\ref{tab:eff} and~\ref{tab:class} This is somewhat expected, since 
the quantum noise can be considered to be weaker than the classical white noise. 
\end{enumerate} It would be interesting to see which of the above conclusions 
are valid also for different non-equilibrium universality classes. 

In summary, it appears that the main differences between ``quantum'' and 
classical ageing come from the different scaling of the noises. Our exact 
results for the quantum spherical model at $T=0$ suggest that it should be often 
sufficient to replace the scaling properties  of the quantum noise correlators 
(\ref{eq:qudyn}) -- with its temporal non-locality -- by a well-chosen effective 
Markovian noise (\ref{eq:eff_qn_anticommu}), with a spatial non-locality. Both 
of them are distinct from the classical white noise (\ref{eq:classical}).  
Non-Markovian long-term memory effects generically appear as relatively minor 
quantitative details, notably for the structure factor, and hardly ever lead to 
qualitative changes of the long-time behaviour. It would be interesting to 
investigate if similar phenomenological prescriptions could be formulated beyond 
the model-specific context of the present work. 

The coarsening dynamics and defect formation after quenches across critical 
points can also be described by the {\it Kibble-Zurek mechanism} 
\cite{Kibble76,Kibble80,Zurek85,Zurek96}. Because of the divergence of length- 
and time-scales at criticality (known as the critical slowing-down), even for 
``slow'' quenches across criticality the system's adiabatic dynamics can no longer 
equilibrate. Then the dynamics is analogous to the one of a rapid  quench such 
that our predictions might be seen as a reliable benchmark for Kibble-Zurek 
studies of transitions at zero temperature, such as the one recently carried out 
for the classical spherical model in Ref.~\cite{Scopa18}.

A different question for future work concerns the possible consequences of 
different long-time memory effects in stochastic complex systems, for which some 
of the new mathematical tools developed here might become useful. 

\medskip
\textbf{Acknowledgments}:
We are grateful to 
A. Chiocchetta,
M. Hase,
M. Heyl,
P. McClarty,
S. de Nicola and
J. Schmalian
for useful discussions. 
MH and SW are grateful to the MPIPKS for warm hospitality where part of 
this work was done. SW is grateful to SISSA where this project was initiated.


\appendix

\setcounter{footnote}{0}   

\newpage

\section{\textcolor{black}{Overdamping as long-time scaling limit}}
\label{app:over}
\textcolor{black}{
We illustrate the main steps connected to the scaling limit indicated
in Eq.~(\ref{eq:over}) that yields the overdamped Langevin 
equation~(\ref{eq:qle}), following the steps outlined in 
Ref.~\cite{Arau19}. For clarity, we focus on a single degree of freedom 
which suffices for  the analysis of the spherical model and the 
$O(n)$ model for $n\to \infty$ as the equations of motion decouple in 
Fourier space. The corresponding quantum Langevin equations read}
\begin{subequations}
\begin{align}
\partial_t \phi(t) &= \lambda \pi(t) + \eta^{(\phi)}(t),\\
\partial_t \pi(t) &= -\frac{1}{\lambda} r(t) \phi(t)
-\gamma\pi(t)+\eta^{(\pi)}(t),
\end{align}
\end{subequations}
\textcolor{black}{which yield a single second-order
quantum Langevin equation,}
\begin{align}
\label{eq:qletrafo}
\partial_t^2 \phi(t) &= - r(t) \phi(t)
-\gamma\lambda \pi(t)+\lambda \eta^{(\pi)}(t) + 
\partial_t\eta^{(\phi)}(t).
\end{align}
\textcolor{black}{We apply the following scaling transformation
to Eq.~(\ref{eq:qletrafo})}
\begin{align}
 \wit{t} = \lambda t , \quad 
\eta^{(\phi)}(t) = \lambda^{0} \wit{\eta}^{(\phi)}(\wit{t}), \quad
\eta^{(\pi)}(t) = \lambda^{0}\wit{\eta}^{(\pi)}(\wit{t}), \quad 
\phi(t) = \lambda^{1}\wit{\phi}(\wit{t}),
\end{align}
\textcolor{black}{such that the quantum Langevin equation takes the 
form}
\begin{align}
 \lambda^2 \partial_{\wit{t}}^2 \wit{\phi}(\wit{t})
= - r(\wit{t})\wit{\phi}(\wit{t}) - \wit{\gamma} \partial_{\wit{t}}\wit{\phi}(\wit{t})  
+ \wit{\xi}(\wit{t})
\label{eq:qle-scale}
\end{align}
\textcolor{black}{with $\wit{\gamma} = \gamma \lambda$ and the composite 
noise
$\wit{\xi}(\wit{t}) =  \wit{\eta}^{(\pi)}(\wit{t}) + \wit{\gamma}\lambda^{-2} \wit{\eta}^{(\phi)}(\wit{t})+  \partial_{\wit{t}}\wit{\eta}^{(\phi)}(\wit{t})$. In the scaling
limit~(\ref{eq:over}) 
one lets $\lambda\to0$ such
that the second-order time derivative on the left-hand-side of 
Eq.~(\ref{eq:qle-scale}) is suppressed and 
we obtain the overdamped quantum Langevin equation}
\begin{align}
  \wit{\gamma} \partial_{\wit{t}}\wit{\phi}(\wit{t}) 
= - r(\wit{t})\wit{\phi}(\wit{t})  + \wit{\xi}(\wit{t}).
\end{align}
\textcolor{black}{
It remains to analyse the effects of the overdamped limit on the noise 
correlations. To this end we report the complete noise (anti-)
commutators~\cite{Arau19}}
\begin{align}
\left\langle\bigl\{ \eta^{(\phi)}(t) ,\eta^{(\pi)}(t') \bigr\}\right\rangle 
= \frac{\hbar\gamma}{2\pi} J\left(\frac{\hbar}{2T}, t-t'\right), \quad
\left\langle\bigl[ \eta^{(\phi)}(t) , \eta^{(\pi)}(t') \bigr]\right\rangle 
=\II\hbar\gamma \delta(t-t'),
\end{align}
\textcolor{black}{with the function 
$J(a,\tau) = -\II \int_{\mathbb{R}}\D \nu \coth(a\nu)e^{\II\nu\tau}$.
For the noise correlators to be well-defined, we observe that the temperature $T$ needs to be rescaled in the overdamped as $\wit{T} = T/ \lambda$. A careful 
analysis reveals that the (anti-) commutation relation of the composite noise 
in the overdamped Langevin equation is given by}
\begin{align}
\left\langle\bigl\{  \wit{\xi}(\wit{t}),
 \wit{\xi}(\wit{t}')\bigr\}\right\rangle =
\frac{\hbar\wit{\gamma}}{\pi} I\left(\frac{\hbar}{2\wit{T}},\wit{t}-\wit{t}'\right),
\quad 
\left\langle\bigl[  \wit{\xi}(\wit{t}), \wit{\xi}(\wit{t}')\bigr]\right\rangle &=
2\II\hbar\wit{\gamma} \delta'(\wit{t}-\wit{t}'),
\end{align}
\textcolor{black}{with $I(a,\tau)=\partial_{\tau}J(a,\tau)$. Two limiting cases 
are of special interest here, namely 
\begin{enumerate}
\item[a)] $\wit{T}\to\infty$, which reproduces the
classical white noise. 
\item[b)] $\wit{T}\to 0$, in which the zero-temperature noise 
correlation function reads
\begin{align}
\left\langle \bigl\{ \wit{\xi}(\wit{t}), \wit{\xi}(\wit{t}')
\bigr\}\right\rangle 
= \frac{\wit{\gamma} \hbar}{\pi} \int_{-\infty}^\infty |\omega| e^{\II \omega(
\wit{t} - \wit{t}')} \D \omega.
\end{align}
\end{enumerate}}
\noindent
\textcolor{black}{
In order to simplify the notation, 
in the main text we drop all tildes from the variables
$\wit{t},\ \wit{\gamma}$ and $\tilde{T}$. 
By considering now the case with many degrees of freedom discussed 
in the main text, one can reproduce the argument above for the 
Fourier modes of the field $\phi_{\vec{k}}(t)$.
We find the quantum Langevin equation~(\ref{eq:qle})
}
\begin{align}
 \gamma \partial_t \phi_{\vec{k}}(t) + \left(r(t)+k^2\right) \phi_{\vec{k}}(t) = 
\xi_{\vec{k}}(t),
\end{align}
\textcolor{black}{with the quantum noise correlation function in Eq.~(\ref{eq:qn_anticommu_T})}
\begin{align}
\left\langle \left\{ \xi_{\vec{k}}(t), \xi_{\vec{k}'}(t') \right\} 
\right\rangle &= \frac{2\gamma\hbar}{\pi} \int_0^\infty \!\D\omega\: 
\omega\operatorname{coth} \left( \frac{\hbar \omega}{T} \right) 
\cos(\omega(t-t')) \: \delta(\vec{k}+\vec{k}').
\end{align}

\section{Regularised quantum noises in the over-damped limit \label{app:commu}}

We focus on the following quantum Langevin equation for the harmonic oscillator 
with position operator $x$ and friction coefficient $\gamma$, i.e., 
\begin{equation}
 \epsilon \ddot{x} + \gamma\dot{x} + \Omega^2 x = \xi,
\end{equation}
where the ``mass'' $\epsilon$ allows us to keep track of the impact of the 
inertial term $\epsilon\ddot{x}$, while $\Omega$ quantifies the strength of the 
harmonic potential. We now proceed to study the equal-time commutation relation 
of $x$ and the canonically conjugate variable $p = \epsilon \dot{x}$. This is 
useful as we shall see that the inertia term acts as a regulator to guarantee 
that the canonical commutation relation is satisfied.

Assuming that the initial conditions of the dynamics are in the very remote 
past, and that they relax in time due to dissipation~\cite{Gard04}, the solution 
of the homogeneous equation vanishes and one is left only with the contribution 
generated by the noise. This can be readily determined by using a Fourier 
transform in time according to 
$\wit{x}(\omega)=\frac{1}{\sqrt{2\pi\,}}\int_{\mathbb{R}}\!\D t\, e^{-\II\omega 
t} x(t)$  \cite{Gard04}, which yields
\begin{align}
 \wit{x}(\omega) &= \frac{\wit{\xi}(\omega)}{\Omega^2 - \epsilon \omega^2 + \II \gamma \omega} \quad \mbox{and}\quad
 \wit{p}(\omega) = \frac{\II \omega \epsilon\, \wit{\xi}(\omega)}{\Omega^2 - \epsilon \omega^2 + \II \gamma \omega},
\end{align}
where $\wit{p}$ is defined in analogy with $\wit{x}$.
Using the quantum noise correlator in Eq.~\eqref{eq:qn_commu}, it is readily 
checked that the equal-time commutator does not depend on $\epsilon$, i.e., 
\begin{align}
\label{eqApp:CCR}
\left\langle [x(t),p(t)]\right\rangle &= 
 \int_{\mathbb{R}^2}\frac{\D\omega \D\omega'}{2\pi}\:\left\langle [\wit{x}(\omega),\wit{p}(\omega')]\right\rangle  
 e^{\II(\omega+\omega')t} 
 =\frac{\II\hbar\gamma}{\pi} \int_{-\infty}^{+\infty}\!\!\!\D\omega\: 
\frac{\omega^2}{\Omega^2 - \epsilon \omega^2 +\II\gamma \omega}
\frac{\epsilon}{\Omega^2 - \epsilon \omega^2 -\II\gamma \omega}
 =  \II \hbar.
\end{align}
We thus see this choice of the noise correlator guarantees that the canonical 
commutation relation is satisfied at all times. Note that this noise correlation 
diverges at short times and this is where the inertia term is relevant to ensure 
the convergence of the integral for large values of $|\omega|$, i.e., at short 
times. Heuristically, however, the late-time dynamics of the system we are 
interested in is expected to be effectively dominated by the dissipative terms 
and should be rather insensitive to what happens at short times, especially as 
far as the emergence of collective behaviours is concerned. This suggests that 
the limit $\epsilon\to 0$ could be taken from the outset in 
Eq.~\eqref{eqApp:CCR} if one introduces a suitable regularising function  
$\mathscr{R}(\omega)$  acting on the integrand evaluated for $\epsilon = 0$, 
i.e., 
\begin{align}
\left\langle [x(t),p(t)] \right\rangle &=\II\hbar\frac{\gamma}{\pi} 
\int_{\mathbb{R}}\!\D\omega \: \frac{\omega^2 \mathscr{R}(\omega)}{\Omega^4 + 
(\gamma \omega)^2}.
\end{align}
The choice $\mathscr{R(}\omega)= e^{-\tto|\omega|}$ (with $\tto$ thought to be a small quantity) gives
\begin{align}
\left\langle [x(t),p(t)]\right\rangle &=\II\hbar 
\frac{\gamma}{\pi}\int_{-\infty}^{\infty} \!\D\omega\: 
\frac{\omega^2e^{-\tto|\omega|}}{\Omega^4 + (\gamma \omega)^2} \simeq\II\hbar  
\frac{2}{\pi \tto\gamma} + \mbox{\rm O}(\tto),
\end{align}
which suggests the natural choice $\tto= 2/(\gamma \pi)\sim \gamma^{-1}$ for the 
scale of the cut-off in order to preserve the canonical commutation relations. 
Although we tested this practical prescription on the equal-time commutator, it 
works also for various other quantities. For example,  all the other relevant 
two-point functions yield the same conclusion concerning the cutoff as can be 
easily verified by similar calculations.

\section{Effective dynamics: details of the analysis}
\label{app:eff}

We outline here the calculations for the effective dynamics, with the Markovian 
noise correlator in Eq.~(\ref{eq:eff_qn_anticommu}). Using the formal solution 
(\ref{gl:C-formel}), the equal-time autocorrelator 
$C(t,t)=\int_{\vec{k},(\Lambda)} C_{\vec{k}}(t,t)$ is given by
\begin{align}
C(t,t) = \frac{1}{g(t)} \int_{\vec{k},(\Lambda)} e^{-2k^2 t/\gamma} 
C_{\vec{k}}(0) + \frac{1}{g(t)}\frac{\mu}{\gamma^2} \int_0^t \!\D s\: g(s) 
\int_{\vec{k},(\Lambda)} k^2 e^{-2 k^2(t-s)/\gamma}.
\end{align}
With the initial condition (\ref{eq:C0}), the definition 
$A_{\alpha}(t) := \int_{\vec{k},(\Lambda)} k^{\alpha} e^{-2k^2 t/\gamma}$ 
and  the convolution from~\ref{app:prop_lap}, this gives
\begin{align}
C(t,t) = \frac{1}{g(t)} \left[ c_{\alpha} A_{\alpha}(t) + \frac{\mu}{\gamma^2} 
(g\ast A_2)(t) \right].
\end{align}
We  rewrite the spherical constraint as a linear integral or 
integro-differential equation for the function $g(t)$, as follows for the two 
models considered here.

For the spherical model, the spherical constraint in Eq.~(\ref{eq:Hsm}) reads 
$C(t,t)=1/\lambda$. This directly produces, along with the formal exact solution 
in Laplace space,
\begin{subequations} \label{gl:B:g}
\begin{align} \label{gl:B:g-sm}
\frac{1}{\lambda} g(t) &= c_{\alpha} A_{\alpha}(t) + \frac{\mu}{\gamma^2} (g\ast 
A_2)(t) \quad \Rightarrow \quad \lap{g}(p) = \frac{ c_{\alpha} 
\lap{A_{\alpha}}(p)}{1/\lambda -( \mu/\gamma^2) \lap{A_2}(p)}.
\end{align}

For the $O(n)$-model with $n\to\infty$, the spherical constraint (\ref{eq:Hon}) 
is $r(t)= r_0 + \frac{u}{12} C(t,t)$. {}From the definition (\ref{gl:g}), one 
has $\frac{g'(t)}{g(t)}=\frac{2}{\gamma} r(t)$. This gives, again together with 
the formal solution and $g(0) = 1$
\begin{align}
 & ~~ \frac{6{\gamma}}{u} g'(t) - \frac{12{r}_0}{u} g(t) = c_{\alpha} 
A_\alpha(t) + \frac{\mu}{\gamma^2} (g\ast A_2)(t) \quad \Rightarrow \quad 
\lap{g}(p) = \frac{c_{\alpha}\lap{A_\alpha}(p) + {6\gamma}/{u}}{{6\gamma}p/u 
-12{r}_0/u - (\mu/\gamma^2)\lap{A_2}(p)} .
 \label{gl:B:g-on}
\end{align}
\end{subequations}
In both cases, the late-time behaviour of $g(t)$ is related, via Tauberian 
theorems \cite{Fell71}, to the small-$p$ behaviour of $\lap{g}(p)$.  In turn, in 
order to determine this, we need to know the small-$p$ expansion of 
$\lap{A_{\alpha}}(p)$.  Since the computation of this expansion is standard, 
see, e.g., Refs.~\cite{Godr00,Pico02}, we simply cite the results. The final 
expansion contains at least one $\Lambda$-independent term which is in general not an entire function 
of $p$ and, in 
addition, a sum of terms with integer powers of $p$ taking the form
\begin{equation}\label{eq:Aas}
 \lap{A_\alpha}(p) \simeq a_\alpha p^{(d+\alpha)/2-1}+ 
\sum_{n=0}^{\lfloor\frac{d+\alpha}{2}-1\rfloor} (-1)^n A_n^{(\alpha)} p^n,
\end{equation}
where $\lfloor x\rfloor$ is the largest integer $< x$ and the sum above is 
understood to be zero if its upper limit is negative. In addition, if $d+\alpha 
= 2m\in\mathbb{N}$ is a positive even integer, extra logarithmic factors arise 
which we neglect here. The constants in Eq.~(\ref{eq:Aas}) read explicitly 
\begin{equation}
  a_{\alpha} = \frac{\pi}{2}\frac{\Omega_d }{(2\pi)^d} 
\frac{\left(\gamma/2\right)^{(d+\alpha)/2}}{ \sin 
\left(\frac{\pi}{2}(d+\alpha)\right)} \;\;,\;\; \quad 
A_n^{(\alpha)} = \frac{\Omega_d }{(2\pi)^d} \left(\frac{\gamma}{2}\right)^{n+1} 
\int_0^\Lambda \!\D k\: k^{d+\alpha-3-2n},
\end{equation}
where $\Omega_d = 2\pi^{d/2}/\Gamma(d/2)$ is the surface of the unit hypersphere 
$S^d$ in $d$ dimensions. Clearly, the first term in Eq.~(\ref{eq:Aas}) is 
universal, while the other terms, if they occur, depend explicitly on the 
momentum cutoff $\Lambda$ and cannot be universal. 

Given the expansion (\ref{eq:Aas}), we can now compare the leading behaviour of 
$\lap{g}(p)$ for the two solutions of Eq.~(\ref{gl:B:g}). First, for $0<d<2$, 
the leading non-constant term in both denominators comes from $\lap{A_2}(p)\sim 
p^{d/2}$, such that the term $(6\gamma/u)p$ present in Eq.~(\ref{gl:B:g-on}) 
merely provides a correction to scaling. Since $\alpha\leq 0$, it follows that 
$d+\alpha<2$, thus the leading term in $\lap{A_{\alpha}}(p)\sim 
p^{(d+\alpha)/2-1}$ in both numerators will dominate over an eventual constant 
present in the numerator of Eq.~(\ref{gl:B:g-on}). Accordingly, the leading 
long-time behaviour of both the spherical and the $O(n)$-model is the same. 
Second, let $2<d$. Then one has schematically the leading structure 
$\lap{A_2}(p)\sim  p^0 + p^1 + p^{d/2}$, where we omitted to indicate the 
various constants. The extra terms in the denominator of Eq.~(\ref{gl:B:g-on}) 
can be absorbed into these, up to re-defining certain non-universal constants. 
For the numerators, if $d+\alpha<2$, then the leading terms comes from 
$\lap{A_{\alpha}}(p)\sim p^{(d+\alpha)/2-1}$ and the constant term present in  
the numerator of Eq.~(\ref{gl:B:g-on}) merely creates a finite-time correction. 
If, on the other hand, $d+\alpha>2$, then one has the structure 
$\lap{A_{\alpha}}(p) \sim p^0 + p^{(d+\alpha)/2-1}$ and the extra constant term 
in the numerator of Eq.~(\ref{gl:B:g-on}) can be absorbed, up to a redefinition 
of a non-universal constant. Again, we conclude that {\it the leading long-time 
behaviour of the spherical and $O(n)$ models is the same, for all $d>0$ and all 
initial conditions}. Although the leading exponents are the same, the 
corresponding amplitudes can be different, especially for $d>2$ and/or 
$d+\alpha>2$. 

We now determine the critical point from the formal solutions 
Eq.~(\ref{gl:B:g}).  If the denominator vanishes for some $p_c>0$, then the 
function $\lap{g}(p)$ has a simple pole at $p=p_c$ and it follows that 
asymptotically $g(t) \sim \exp(t/\tau_{\text{r}})$ which defines the relaxation 
time scale $\tau_{\text{r}}$. On the other hand, if $p_c\to 0$, then the 
behaviour of $\lap{g}(p)$ will change to $g(t)\sim t^{\digamma}$ becoming 
algebraic. The condition $p_c=0$ fixes the critical point. 
Expanding for $p\to 0$ and keeping the bath control parameter $\mu$ fixed, gives 
$1/\lambda_c = (\mu/\gamma^2) \lap{A_2}(0)$ for the spherical model and
$(12/u) r_0^c = - (\mu/\gamma^2) \lap{A_2}(0)$ for the $O(n)$-model. Specifically,
the critical values of the control parameters are given by:
\begin{subequations}
\begin{align}
  \frac{1}{\lambda_c} &= \frac{\mu }{\gamma^2}A_0^{(2)} \hspace{0.82truecm} = 
\frac{\mu}{\gamma} \frac{\Omega_d}{(2\pi)^d}  \int_0^\Lambda \!\D k\: k^{d-1} 
\hspace{1.8truecm}\mbox{\rm ~ for the spherical model} \\
  {r}_0^c &= - \frac{u}{12} \frac{\mu }{\gamma^2}A_0^{(2)} = - \frac{u}{12} 
\frac{\mu}{\gamma} \frac{\Omega_d}{(2\pi)^d}  \int_0^\Lambda \!\D k\: k^{d-1} 
\hspace{0.9truecm} \mbox{\rm ~~~~ for the $O(n)$-model} .
\end{align}
\end{subequations}
The rest of the analysis required for determining the leading relaxation time 
$\tau_{\rm r}$ as well as the exponents of the leading algebraic behaviours 
follows closely the approach used for  classical dynamics 
\cite{Godr00,Pico02,Henk10} and produces the results quoted in the main text of 
Sec.~\ref{ssec:eff}.

\section{Properties of double Laplace transforms}
\label{app:prop_lap}

We summarise here some useful properties of double convolutions, related to the 
double Laplace transform. First, we recall the definition of the simple Laplace 
transform of a function $h: \mathbb{R}_+\to\mathbb{C}$ of a single variable, 
namely $\lap{h}(p) = \mathcal{L}(h)(p) := \int_0^{\infty} \!\D t\: e^{-pt} 
h(t)$. The convolution of two functions $h_1$, $h_2$ of a single variable is 
defined as $(h_1\ast h_2)(t) := \int_0^{t} \!\D t'\, h_1(t') h_2(t-t')$. An 
important property is the factorisation identity $\lap{(h_1\ast h_2)}(p) = 
\lap{h_1}(p) \lap{h_2}(p)$, see, e.g., Refs.~\cite{Abra65,Doet74,Doet76}. 

The {\it double Laplace transform} of a function $h: \mathbb{R}_+^2 \to 
\mathbb{C}$ of two variables is defined as \cite{Voel50}
\begin{equation}
 \dlap{h}(p,q) = \mathcal{L}_2(h)(p,q) := \int_0^\infty \!\D t \int_0^\infty \!\D s \: e^{-pt-qs}\, h(t,s).
\end{equation}
We refer to the literature \cite{Voel50,Ditk62} for detailed discussions of the 
conditions under which these Laplace transforms exist and we rather concentrate 
here on formal identities for explicit calculations. First, if the function $h$ 
depends only on the sum of its two arguments, namely  $h(t,s) = k(t+s)$, the 
double Laplace transform $\dlap{h}$  is related to the simple Laplace transform 
$\lap{k}$ of $k$ via \cite{Voel50,Ditk62,Debnath2016}
\begin{equation} \label{gl:C:hk}
 \dlap{h}(p,q) = \frac{\lap{k}(p) - \lap{k}(q)}{p-q} .
\end{equation}
On the other hand, if  $h(t,s)=k(|t-s|)$, one has \cite{Ditk62}
\begin{equation} \label{gl:C:hksymm}
 \dlap{h}(p,q) = \frac{\lap{k}(p) + \lap{k}(q)}{p+q}
\end{equation}
This latter identity also holds if $h(t,s)=k(t-s)$, provided $k(\tau)=k(-\tau)$ 
is even \cite{Debnath2016}. Second, if $h(t,s)=h(s,t)$ is symmetric, it follows 
that $\dlap{h}(p,q)=\dlap{h}(q,p)$, i.e., $\dlap{h}$ is also symmetric. 
Equations~(\ref{gl:C:hk}) and (\ref{gl:C:hksymm}) provide some examples. Third, 
if $h(t,s) = h_1(t) h_2(s) $ then  $\dlap{h}(p,q) = \lap{h}_1(p) \lap{h}_2(q)$. 
Fourth, we note (see, e.g., Eq.~(44) at p. 186 of Ref.~\cite{Voel50})
\begin{equation} \label{gl:C:tt} 
\mathcal{L}_2^{-1} \left( \frac{\dlap{h}(p,q)}{c+p+q} \right)(t,t) = \int_0^{t} 
\!\D t'\: e^{-c t'}\, h(t-t',t-t') .
\end{equation}
The {\it double convolution} of two functions $h_{1,2}$ of two variables is 
defined as 
\begin{align}
(h_1 \ast \ast h_2)(t,s) := \int_0^t \!\D t' \int_0^s \!\D s'\: h_1(t',s') 
h_2(t-t',s-s').
\end{align}
The factorisation identity for the simple convolution via Laplace transform 
\cite{Abra65,Doet76} naturally carries over to the double Laplace transform 
\cite{Voel50,Ditk62}
\begin{equation} \label{C:lap-conv}
 \dlap{ \bigl( h_1 \dast h_2\bigr) }(p,q) = \mathcal{L}_2\bigl(h_1 \dast 
h_2\bigr)(p,q) = \dlap{h}_1(p,q)\, \dlap{h}_2(p,q)  .
\end{equation}
This property allows us to solve linear Volterra integral equations in two 
variables, as shown in the main text. We introduce a weighted convolution, 
defined as 
\begin{equation}
 (h_1 \dast_w h_2)(t,s) : = \int_0^t \!\D t' \int_0^s\! \D s'\: h_1(t',s') 
h_2(t-t',s-s') w(t'-s') ,
\end{equation}
with the weight function $w=w(t)$. Its double Laplace transformation factorises as  
\begin{align}\label{eq:Lw}
\mathcal{L}_2 (h_1 \dast_w h_2)(p,q) = \dlap{h}_2(p,q)\, 
\mathcal{L}_2\bigl(h_1(t,s)w(t-s)\bigr)(p,q) = \dlap{h}_2(p,q)\, 
\dlap{\bigl({h_1 w}\bigr)}(p,q) 
\end{align}
The proof of Eq.~(\ref{C:lap-conv}) is given in Refs.~\cite{Voel50,Ditk62} and 
merely uses Fubini's theorem. The proof of the new identity in Eq.~(\ref{eq:Lw}) 
is similar.

\section{Asymptotics of the quantum noise integrals}
\label{app:F}
Consider the double Laplace transform of the quantum noise correlation function, 
c.f. Eqs.~(\ref{eq:AandF}) and~(\ref{eq:qn_anticommu_cut}),
\begin{align} \label{D1}
\dlap{F}(p,q) =\frac{2\hbar}{\pi\gamma} \int_0^\infty \!\D t \int_0^\infty \!\D 
t' \int_{\vec{k},(\Lambda)} e^{-\frac{k^2}{\gamma}(t+t')} 
\frac{\tto^2-(t-t')^2}{\left[\tto^2+(t-t')^2\right]^2} e^{-p t-q t'}.
\end{align}
Since the original function $F(t,t')=F(t',t)$ is symmetric, this also holds for 
the double Laplace transform, $\dlap{F}(p,q)=\dlap{F}(q,p)$, see 
\ref{app:prop_lap}.  

\begin{figure}[t]
 \centering
 \includegraphics[width=.5\textwidth]{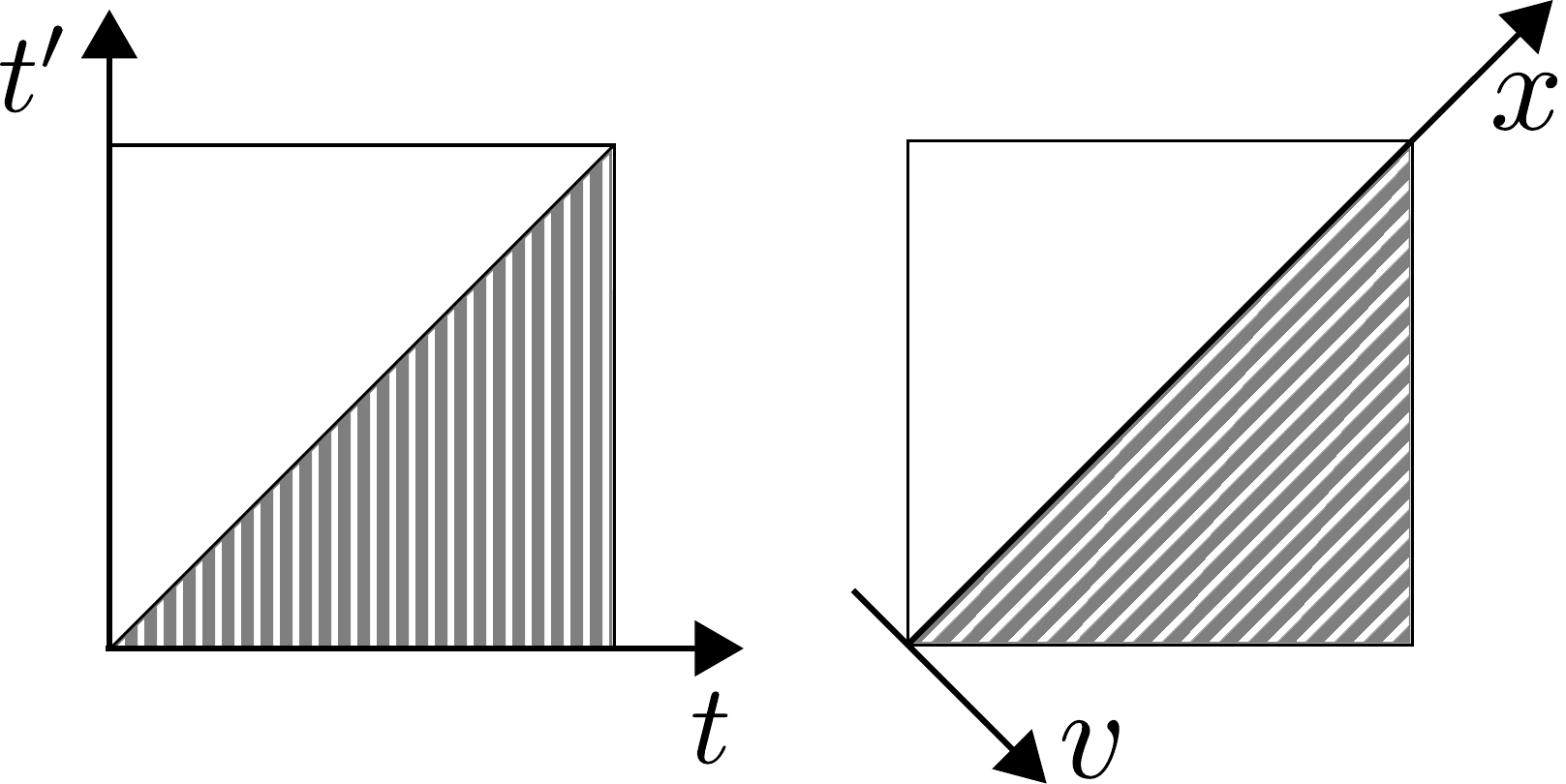}
 \caption[Change of variables]{Illustration of the change of variables employed 
in Eq.~(\ref{D2}) in order to to isolate the action of the quantum noise 
function.}
 \label{fig:sketch_int}
\end{figure}

We now reduce Eq.~(\ref{D1}) to a form for which the asymptotic behaviour, 
especially for $\tto\to 0$, can be easily determined. We decompose the square 
integration domain into two triangles, as in Fig.~\ref{fig:sketch_int}. The 
integration over the upper triangle indicated by the white domain in 
Fig.~\ref{fig:sketch_int} can be reduced to an integration over the lower 
triangle, denoted by the shaded domain in Fig.~\ref{fig:sketch_int}, by Fubini's 
theorem, and we also use the symmetry of $F(t,t')$. This leads to
\begin{align}
\dlap{F}(p,q) &=  \int_0^\infty \!\D t \int_0^t \!\D t'\: F(t,t')  e^{-p t-q t'} 
+  \int_0^\infty \!\D t \int_t^\infty \!\D t'\: F(t,t')  e^{-p t-q t'} \nonumber\\
&=  \int_0^\infty \!\D t \int_0^t \!\D t'\: F(t,t')  e^{-p t-q t'} 
+  \int_0^\infty \!\D t' \int_0^{t'} \!\D t\: F(t,t')  e^{-p t-q t'} \nonumber \\
&=  \int_0^\infty \!\D t \int_0^t \!\D t'\: F(t,t')  e^{-p t-q t'} 
+  \int_0^\infty \!\D t \int_0^{t} \!\D t'\: F(t',t)  e^{-p t'-q t} \nonumber \\
&=  \int_0^\infty \!\D t \int_0^t \!\D t'\: F(t,t') \left(  e^{-p t-q t'} +   e^{-p t'-q t}\right)  .
\label{D2}
\end{align}
Next, we change the integration variables according to $x=t+t'$, $v=t-t'$, such 
that the shaded domain of integration in figure~\ref{fig:sketch_int} is 
rewritten as $\int_0^{\infty}\!\D t \int_0^{t} \!\D t' = \demi \int_0^{\infty} 
\!\D v\int_v^{\infty} \!\D x$. 
Because of the identity
\begin{equation}
e^{-pt-qt'} + e^{-pt' -qt} = 2\, e^{-x(p+q)/2} \cosh\left(\frac{v}{2}(q-p)\right)
\end{equation}
the above change of variables casts the double integral (\ref{D2}) into a form 
where the quantum noise correlation acts as a distribution on a test function 
$f(v)$,  namely
\begin{align}\label{eq:mellinintegral}
\dlap{F}(p,q)&= \frac{2\hbar}{\pi\gamma} \int_0^\infty\D v\,  f(v)
\frac{\tto^2-v^2}{\left(\tto^2+v^2\right)^2} 
\end{align}
where the variables $p$ and $q$ are implicit in the test function $f$. The 
integrals~(\ref{D1}) and~(\ref{D2}) lead to the following integral 
representation of this test function 
\begin{align}
 f(v)=\int_v^\infty \D x \int_{\vec{k},(\Lambda)}  e^{-\frac{k^2}{\gamma}x} 
e^{-x\frac{p+q}{2}} \cosh\left(v \frac{q-p}{2} \right),
\end{align}
which is clearly invariant upon exchanging $p$ and $q$. This appendix analyses 
general integrals of the form (\ref{eq:mellinintegral}) in the limit $\tto\to0$. 
Note that setting $\tto=0$ from the outset would in general  lead to a divergent 
integral.
\subsection{The quantum noise memory kernel as a generalised function}
\begin{figure}[t]
 \centering
 \includegraphics[width=.7\textwidth]{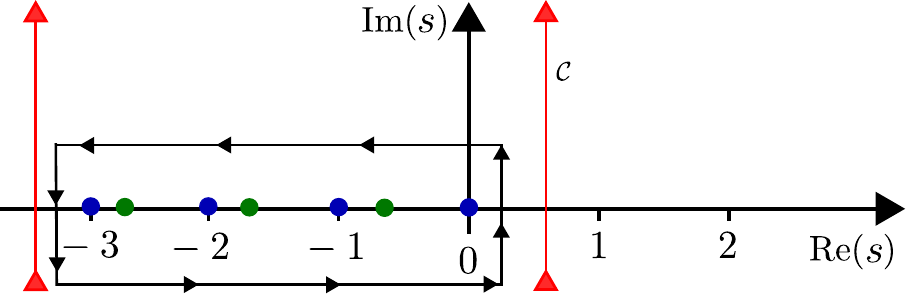}
 \caption[Contour]{Integration contours for carrying out the inverse Mellin transform: moving the integration contour $\mathcal{C}$ to the left yields
 contributions from the poles that are inside the black contour.}
 \label{fig:sketch_invMell}
\end{figure}

In classical dynamics, one may write the noise correlation as a generalized 
function by modeling a Markovian noise through a delta function, see 
Eq.~(\ref{eq:classical}). We are interested in interpreting the quantum noise 
correlation in a similar way. Consider the integral
\begin{equation}\label{eq:integral}
 \int_0^\infty\D x\, f(x) \frac{\tto^2-x^2}{\left( \tto^2 + x^2 \right)^2}.
\end{equation}
For certain choices of $f(x)$, this kind of integral can be calculated from the 
residue theorem. For example, with $-1<s<1$ 
\begin{equation}\label{eq:integral-potenz}
 \int_0^\infty \!\D x\: x^{-s} \frac{\tto^2-x^2}{\left( \tto^2 + x^2 \right)^2} 
 = \frac{\pi}{2} \frac{s}{\cos \bigl( \pi s/2\bigr)} \tto^{-1-s} .
\end{equation}
In order to study systematically the dependence of the integral 
(\ref{eq:integral}) on the cut-off parameter $\tto$ we use the {\it Mellin 
transform} which is defined as \cite{Flaj95} 
\begin{align}\label{eq:mellin}
 \mel{f}(s) = \mathscr{M}(f)(s) := \int_0^\infty \D x\, x^{s-1} f(x) \;\;,\;\; 
\quad f(x) = \frac{1}{2\pi \II}\int_{c-\II\infty}^{c+ \II \infty} \D s\, x^{-s} 
\mel{f}(s) 
\end{align}
where the real constant $c$ is chosen freely in the fundamental strip of the 
respective transform, as illustrated in Fig.~\ref{fig:sketch_invMell} by the 
right integration path. 
This {\it fundamental strip} is defined through the convergence of the integral 
and thus is set by the asymptotic behaviour of $f(x)$. For example, if 
$f(x)\stackrel{x\to 0}{\sim} x^{-\alpha}$ and $f(x)\stackrel{x\to\infty}{\sim} 
x^{-\beta}$ with $\alpha<\beta$, then the fundamental strip is a superset of the 
strip $\left\{ s = \sigma +\II \tau \in\mathbb{C} \bigl|  \tau\in\mathbb{R} 
\mbox{\rm  ~and~} \alpha<\sigma<\beta\bigr.\right\}$. On the fundamental strip, 
$\mel{f}(s)$ exists and is holomorphic \cite{Flaj95}. A monomial $f(x)=x^a$ does 
not admit a Mellin transform. 

For the moment, we do not specify $c$ in Eq.~(\ref{eq:mellin}) but we shall come 
back to this point, once we have correctly identified the necessary assumptions 
on the function $f(x)$. With Eq.~(\ref{eq:integral-potenz}), and in the 
fundamental strip $-1<c<1$, the integral~(\ref{eq:integral}) is rewritten as
\begin{align}
 \int_0^\infty \D x\,f(x) \frac{\tto^2-x^2}{\left( \tto^2 + x^2 \right)^2} 
  &= \frac{\pi}{2}\frac{1}{2\pi \II}\int_{c-\II\infty}^{c+ \II \infty} \!\D s\: \mel{f}(s) 
  \frac{ s \, \tto^{-1-s}}{\cos\left(\frac{\pi}{2} s\right)}.
  \label{eq:integral-mellin}
\end{align}
The remaining analysis depends on the function $\mel{f}(s)$. From its 
definition~(\ref{eq:mellin}) it is clear that the convergence of the integral 
for $x\to\infty$ as well as for $x\to0$ has to be guaranteed. We assume here 
that the function $f(x)$ does not cause any problem at infinity for some subset 
of $c \in (-1,1)$. Furthermore we assume that $f(x)$ has some formal series 
expansion (which does not necessarily represent an analytic function, but we 
assume $0< \alpha_0 < \alpha_1 < \ldots$) 
\begin{equation} \label{D8}
 f(x) = \sum_j a_j x^{\alpha_j}, \quad \text{for} \ x\to0 .
\end{equation}
According to the Direct Mapping Theorem \cite{Flaj95}, the exponents $\alpha_j$ 
in the expansion (\ref{D8}) correspond to poles $s_0$ of the Mellin transform, 
while the coefficients $a_j$ are the residues
\begin{equation}
 \operatorname{Res}_{s=-\alpha_j}\left( \mel{f}(s)\right) = a_j .
\end{equation}
Under these conditions, the fundamental strip of the Mellin transform is the 
segment $(\operatorname{max}\{-\alpha_0,-1\},1)$. Assume that 
$|\mel{f}(\sigma+\II\tau)|< \mathfrak{f}_0(\sigma) e^{\mathfrak{f}_1 |\tau|}$  
for $\tau\to\pm\infty$ such that $\mathfrak{f}_1<\frac{\pi}{2}$ and 
$\mathfrak{f}_0(\sigma)$ remains bounded for all $\sigma\in[-\infty,0]$.  Then, 
we can shift the integration contour in Eq.~(\ref{eq:integral-mellin}) to 
$c\to-\infty$ and write the integral as a sum over all residues to the left of 
the initial contour (figure~\ref{fig:sketch_invMell} shows the initial contour 
and the intermediate stage where one has already shifted $c\mapsto c-4$) 
\begin{equation}
 \int_0^\infty \D x f(x) \frac{\tto^2-x^2}{\left( \tto^2 + x^2 \right)^2} = 
\frac{\pi}{2}\sum_{s_0\leq -\min(\alpha_0,1)} 
\operatorname{Res}_{s=s_0}\left(\mel{f}(s) \frac{ s \tto^{-1-s} 
}{\cos\left(\frac{\pi}{2} s\right)}\right).
  \label{eq:integral-residues}
\end{equation}
Besides the simple poles of the Mellin transform $\mel{f}(s)$, the cosine 
function also generates simple poles, located at $s_0=-(2n+1)$ with 
$n\in\mathbb{N}_0$. Indeed, those poles $s_0$ of the Mellin transform which do 
{\it not} occur at a negative odd integer correspond to simple poles in 
Eq.~(\ref{eq:integral-residues}). Those poles $s_0$ which occur at a negative 
odd integer correspond, instead, to double poles in 
Eq.~(\ref{eq:integral-residues}). 
Accordingly, we decompose the formal expansion (\ref{D8}) of $f(x)$ according to
\begin{equation} \label{eq:serie-formelle}
 f(x) = {\sum_j}' a_j x^{\alpha_j} + \sum_{n=0}^\infty b_{2n+1} x^{2n+1},
\end{equation}
where the exponents $\alpha_j$ are ordered according to $-1<\alpha_0<\alpha_1 < 
\ldots$ and the $\alpha_j$ cannot be odd positive integers. Then the first 
formal series in Eq.~(\ref{eq:serie-formelle}) contains all even powers
and non-analytic terms in $x$ that generate first-order poles, 
while the second formal 
series in Eq.~(\ref{eq:serie-formelle}) contains all odd powers in $x$ that generate second-order poles. The integral can then 
be written as
\begin{align}
 \int_0^\infty \D x\ f(x) \frac{\tto^2-x^2}{\left( \tto^2 + x^2 \right)^2} = 
&-\frac{\pi}{2} {\sum_{j}}' a_j \frac{ \tto^{-1+\alpha_j}  \alpha_j 
}{\cos\left(\frac{\pi}{2} \alpha_j\right)}
+\frac{\pi}{2}\sum_{n=0}^\infty 
\operatorname{Res}_{s=-(2n+1)}\left(\mel{f}(s) \frac{s}{\cos\left(\frac{\pi}{2} 
s\right)} \tto^{-1-s} \right).
  \label{eq:D:13}
\end{align}
It remains to determine the residues at the second-order poles. These can be 
found in general as follows, see, e.g., Ref.~\cite{Henk10}. 
Consider two functions $h(z)$ and $g(z)$ such that, around $z\approx z_0$,
\begin{subequations} 
\begin{align}
 h(z) &= \frac{1}{z-z_0} \left[ P(z_0)+P'(z_0)(z-z_0)+\ldots \right], \\[.25cm]
 g(z) &= (z-z_0) \left[ Q(z_0)+Q'(z_0)(z-z_0)+\ldots \right],
\end{align}
\end{subequations}
with entire functions $P,Q$ and $Q(z_0)\ne 0$. 
The residue of the quotient function at $z=z_0$ is thus
\begin{align} \label{eq:2orderres}
\operatorname{Res}_{z=z_0}\left[ \frac{h(z)}{g(z)}\right] = \frac{P'(z_0)}{Q(z_0)} - \frac{P(z_0) Q'(z_0)}{Q(z_0)^2} .
\end{align}
In the case we are interested in, a second-order pole arises, from the second 
line in Eq.~(\ref{eq:D:13}), if and only if $\mel{f}(s)$ has a first-order pole 
at $s_0:=-(2n+1)$. Thus $(s+2n+1)\mel{f}(s)$ is well-defined and analytic in 
$s=s_0$. For $s\approx s_0$, we then have
\begin{align}
  \frac{\mel{f}(s) s \tto^{-1-s}}{\cos\left(\frac{\pi}{2} s\right)}
  &= \frac{1}{s-s_0} \frac{(s-s_0)\mel{f}(s) s \tto^{-1-s}}{\cos\left(\frac{\pi}{2} s\right)} \nonumber \\
  &\simeq
 \frac{2}{\pi}\frac{\left(-1\right)^n\tto^{-1-s_0}}{(s-s_0)^2}
 \bigg(\bigl[ b_{2n+1} +B_n (s-s_0) \bigr] 
 \bigl[1-\ln\left( \tto  \right)(s-s_0)\bigr]\bigl[  s_0 + (s-s_0) \bigr]
 \bigg) \nonumber \\ 
 &\simeq
\frac{2}{\pi}\frac{\left(-1\right)^n\tto^{-1-s_0}}{(s-s_0)^2}
\bigg[b_{2n+1} s_0 + \bigl(b_{2n+1} + B_n s_0 - b_{2n+1} s_0 \ln \tto\bigr) (s-s_0) + O((s-s_0)^2) \bigg],
\end{align}
with the constants
\begin{equation}
 b_{2n+1} = \lim_{s\to s_0}\left[(s-s_0) \mel{f}(s) \right],\;\; \quad
 B_n = \lim_{s\to s_0}\frac{\D}{\D s}\left[(s-s_0) \mel{f}(s) \right].
\end{equation}
The residue is read off from the pre-factor of the linear term in $s-s_0$, 
inside the brackets.

Collecting all results, the integral in Eq.~(\ref{eq:D:13}), already using the 
formal expansion in Eq.~(\ref{eq:serie-formelle}), can now be evaluated and 
gives
\begin{align} \label{eq:mellin_final}
\int_0^\infty \D x\, f(x) \frac{\tto^2-x^2}{\left( \tto^2 + x^2 \right)^2}  = 
&-\frac{\pi}{2\tto} {\sum_{j}}' \frac{\tto^{\alpha_j} a_j  \alpha_j 
}{\cos\left(\frac{\pi}{2} \alpha_j\right)}
+\sum_{n=0}^\infty \left(-1\right)^n\tto^{2n}\bigg[ 
\bigl(1+(2n+1)\ln\tto\bigr)b_{2n+1} - (2n+1)B_n \bigg] 
\end{align}
This equation is the central result of our approach, as it allows one to 
understand the behaviour as $\tto \to 0^+$.

We now investigate some specific examples, which we also checked numerically. 
First, we study the exponential function $f(x) = \exp(-\nu x)$, with $\nu>0$. 
Its Mellin transform is $\mel{f}(s)=\mathscr{M}(e^{-\nu x})(s) = 
\nu^{-s}\Gamma(s)$, involving the Gamma function $\Gamma(s)$~\cite{Abra65}. The 
power series $e^{-x}=\sum_{k=0}^{\infty} \frac{(-1)^k}{k!} x^k$ corresponds to 
the `singular expansion' $\Gamma(s) \asymp \sum_{k=0}^{\infty} \frac{(-1)^k}{k!} 
\frac{1}{s+k}$ \cite{Flaj95}. The decomposition according to 
Eq.~(\ref{eq:serie-formelle}) is achieved by writing $e^{-\nu x}=\cosh \nu x - 
\sinh \nu x$.  Now, both series in Eq.~(\ref{eq:mellin_final}) can be evaluated 
exactly in terms of sine and cosine integrals \cite{Abra65} which are themselves 
best written with the auxiliary function $g_{\rm AS}(x)$ defined in 
Eq.~(\ref{eq:aux})\footnote{The $B_n$ were evaluated using 
Eq.~(\href{functions.wolfram.com/06.05.06.0009.01}{06.05.056.0009.01}) of 
Ref.~\cite{Wolfram}.}
\begin{align}\label{eq:Dexp}
 \int_0^\infty \!\D x\: e^{-\nu x} \frac{\tto^2-x^2}{\left(\tto^2 + x^2 
\right)^2} &= -\nu \left[  \cos(\nu\tto) \operatorname{Ci}(\nu\tto) +\sin(\nu 
\tto) \operatorname{si}(\nu \tto) \right] = \nu g_{\rm AS}(\nu\tto) .
\end{align}
A straightforward generalisation for any $n\in \mathbb{N}_0$ is
\begin{align}\label{eq:Ddexp}
 \int_0^\infty \!\D x\: x^n e^{-\nu x} \frac{\tto^2-x^2}{\left(\tto^2 + x^2 
\right)^2} &= (-1)^n \frac{\D^n}{\D \nu^n}\bigg(\nu g_{\rm AS}(\nu\tto)\bigg)
\end{align}
Equations~(\ref{eq:Dexp}) and~(\ref{eq:Ddexp}) are also used in other appendices 
and in the main text. 

Since the exponential function is analytic everywhere, it is worthwhile to benchmark our 
method  as well with a function whose Taylor series has a finite radius of 
convergence. We choose the function $f(x) = (1+\sqrt{x}\,)^{-1}$. Evaluating the 
series in Eq.~(\ref{eq:mellin_final}) we find
\begin{align} 
 \int_0^\infty \!\D x\: \frac{1}{1+\sqrt{x}} \frac{\tto^2-x^2}{\left(\tto^2 + 
x^2 \right)^2}
&= \frac{\pi}{2\tto} \frac{4 \tto ^2+\sqrt{\tto } \sqrt{2}(\tto+1) 
\left(\tto^2-4 \tto +1\right)}{2 \left(\tto ^2+1\right)^2}  +\frac{1+\tto 
^2+(1-\tto ^2) \ln \tto}{\left(\tto ^2+1\right)^2}
\end{align}
%

\subsection{Asymptotic expansion}

The identity (\ref{eq:Dexp}), involving the exponential function, can now be used in order to evaluate the asymptotic 
behaviour of the quantum noise correlation function in Eq.~(\ref{D1}).
%
%
We see that the integrand is invariant under the exchange $t_1\leftrightarrow 
t_2$ apart from the $p$ and $q$ exponential contributions. For a general 
symmetric function $F(t_1,t_2) = F(t_2,t_1)$ we may write
\begin{align}
 \int_0^\infty \!\D t_1 \int_0^\infty \!\D t_2\: F(t_1,t_2) e^{-pt_1}e^{-qt_2}
 &=\int_0^\infty \!\D t_1 \int_0^{t_1} \!\D t_2\: F(t_1,t_2) \left( 
e^{-pt_1}e^{-qt_2} + e^{-pt_2}e^{-qt_1}\right).
\end{align}
Having explicitly symmetrised the above integral, we can now introduce the 
diagonal coordinates, as in Fig.~\ref{fig:sketch_int}. Using 
diagonal coordinates has the advantage that we can now isolate the distribution 
and re-use the formul{\ae} derived in the above examples, in particular in 
Eq.~(\ref{eq:Dexp}), to obtain 
\begin{align}\nonumber
\dlap{F}(p,q)&=
\frac{2\hbar}{\pi\gamma}
\int_0^\Lambda \D k\: k^{d-1}\int_0^\infty \D t_1\int_0^{t_1} \D t_2 \:
e^{-\frac{k^2}{\gamma}(t_1+t_2)}
\frac{\tto^2 - (t_1-t_2)^2}{\left(\tto^2 + (t_1-t_2)^2\right)^2}
\left[ e^{-pt_1}e^{-qt_2}
 + e^{-pt_2}e^{-qt_1}\right]
\\ \nonumber
&=\frac{2\hbar}{\pi\gamma}
\int_0^\Lambda \D k\: k^{d-1}\int_0^\infty \D v\int_v^{\infty} \D u\:  e^{-u
\frac{k^2}{\gamma}}
\frac{\tto^2 - v^2}{\left(\tto^2 + v^2\right)^2}
\left(e^{-\frac{p}{2}( u+v) - \frac{q}{2}(u-v)} + 
e^{-\frac{p}{2}(u-v) - \frac{q}{2}(u+v)}\right)\\ \nonumber
&=
\frac{2\hbar}{\pi\gamma}\int_{\vec{k},(\Lambda)} \int_0^\infty \!\D 
v\int_v^{\infty} \!\D u \: e^{-u\left( \frac{k^2}{\gamma}+\frac{p+q}{2}\right)} 
\frac{\tto^2 - v^2}{\left(\tto^2 + v^2\right)^2} \left(e^{ \frac{q-p}{2}v} + 
e^{- \frac{q-p}{2}v}\right)\\[.25cm] \nonumber
&=\frac{2\hbar}{\pi\gamma}\int_{\vec{k},(\Lambda)} 
\frac{1}{{k^2}/{\gamma}+(p+q)/{2}} \int_0^\infty \!\D v\:  \frac{\tto^2 - 
v^2}{\left(\tto^2 + v^2\right)^2} \left(e^{-v\left( 
\frac{k^2}{\gamma}+p\right)}+ e^{-v\left( 
\frac{k^2}{\gamma}+q\right)}\right)\\[.25cm]
&=\frac{2\hbar}{\pi\gamma}\int_{\vec{k},(\Lambda)} \frac{ \left( 
k^2/\gamma+p\right) g_{\rm AS}\left(\tto\left( k^2/\gamma+p\right)\right) + 
\left( k^2 /\gamma+q\right) g_{\rm AS}\left(\tto \left( 
k^2/\gamma+q\right)\right) }{k^2/\gamma+(p+q)/2}.
\label{eq:dlapFg}
\end{align}
We want to extract the leading scaling behaviour  of this integral 
representation, especially for $\tto \ll 1$ and for $p$ and $q$ 
small.\footnote{From \ref{app:commu}, we have $\tto\sim \gamma^{-1}$, and the 
equation of motion (\ref{eq:qle}) is in the over-damped limit $\gamma$ large.} 
Accordingly, we replace the auxiliary function $g_{\rm AS}$ by its 
small-argument asymptotics \cite{Abra65}
\begin{equation} \label{eq:D:g-expand}
 g_{\rm AS}(x) \simeq -\left(\ln x+{\rm C}_E\right) +\frac{\pi}{2} x + {\rm 
O}(x^2) ,
\end{equation}
which allows us to identify the leading behaviour of the quantum noise function, 
up to the order needed in the main text. In general, it turns out 
$\dlap{F}(p,q)$ has a non-universal regular part and an universal irregular 
part 
\begin{equation}
 \dlap{F}(p,q)  = \dlap{F}_{\rm reg}(p,q) + \dlap{F}_{\rm irr}(p,q)  .
\end{equation}
To linear order in $p$ and $q$, the regular part may be obtained by inserting the expansion (\ref{eq:D:g-expand}) 
into  Eq.~(\ref{eq:dlapFg}), with the result
\begin{align}
  \dlap{F}_{\rm reg}(p,q) &\simeq 
-\frac{4\hbar}{\pi\gamma}\frac{\Omega_d}{(2\pi)^d} \frac{\Lambda ^d}{d} \left\{ 
\ln \left(\frac{\tto\Lambda^2 }{\gamma }\right)+ {\rm C}_E  -\frac{2}{d} - 
\frac{\pi}{2}\frac{d}{d+2}\frac{\tto \Lambda^{2}}{\gamma} \right. \nonumber \\ 
& ~~~ \left. +\left[ -\frac{\pi}{4} 
+\frac{1}{2}\frac{d}{d-2}\frac{\gamma}{\tto\Lambda^{2}}\Theta(d-2)\right] 
\tto(p+q) + {\rm O}(\tto^2) + {\rm o}(p,q) \right\}.
\end{align}
We point out that the term of order zero in $p$ and $q$ exists for all $d>0$ and 
that certain contributions to the first-order term only exist for $d>2$. This is 
expressed above by the Heaviside function $\Theta$. Higher-orders terms arise 
for larger dimensions. In the main text, Eq.~(\ref{eq:dlapF-regirr-a}) neglects 
the terms of order O($\tto$). The presence of the cutoff parameters $\tto$ and 
$\Lambda$ signals that $\dlap{F}_{\rm reg}(p,q)$ depends on the details of the 
cutoff procedures, both temporal and in momentum space (indeed, they only arise 
through the scaling variable $\tto \Lambda^2/\gamma$), and they are therefore 
non-universal. 

The irregular part is obtained by subtracting $\dlap{F}(0,0)$ from 
Eq.~(\ref{eq:dlapFg}),  rescaling the integral according to $x = k/\sqrt{\gamma 
p}$ and finally taking the limit $p\to 0$. To lowest order, this procedure 
yields
\begin{equation}
  \dlap{F}_{\rm irr}(p,q) \simeq 
  -\frac{4\hbar}{\pi\gamma}\frac{\Omega_d}{(2\pi)^d}
  (\gamma p)^{\frac{d}{2}} \mathbb{F}(q/p), \quad \mbox{as}\ p,q\to 0, \quad {\rm with} \quad   \mathbb{F}(1) = \frac{\pi}{d}\frac{1}{\sin \left(\frac{\pi  d}{2}\right)} \ .
\end{equation}
Here, we introduced the scaling function
\begin{align}
  \mathbb{F}(z) &= \int_0^{\infty } \!\D x\:  x^{d-1} \frac{\left(x^2+z\right)
  \ln \left(1+z/x^2\right)+\left(x^2+1\right) \ln \left(1+1/x^2\right)}{
  \left(x^2+z\right)+\left(x^2+1\right)}  \ .
 \end{align}
Remarkably, this can be evaluated explicitly, in terms of hypergeometric  and incomplete Beta functions~\cite{Abra65}, i.e.,
\begin{align}\nonumber
  \mathbb{F}(z)
&= \frac{\pi  \csc \left(\frac{\pi  d}{2}\right)}{4}  \bigg\{\frac{2}{d}
      \left[\, _2F_1\left(1,-\frac{d}{2};1-\frac{d}{2};\frac{z+1}{2}\right)+\frac{4 z^{\frac{d}{2}+1}}{(d+2) (z+1)}\right]
      \\ \nonumber
&\quad-\frac{2}{d-2} \, _2F_1\left(1,1-\frac{d}{2};2-\frac{d}{2};\frac{z+1}{2}\right)
+2^{-\frac{d}{2}} (z+1)^{\frac{d}{2}-1} \bigg(\pi  (1-z) \cot \left(\frac{\pi  d}{2}\right)\\
&\quad +2 z B_{\frac{2 z}{z+1}}\left(\frac{d}{2}+1,0\right)-(z+1) B_{\frac{2 z}{z+1}}\left(\frac{d}{2}+2,0\right)\bigg)\bigg\}.
\end{align}
This is Eq.~(\ref{eq:dlapF-regirr-b}) in the main text. Since this scaling 
function does not contain the parameters $\tto$ and $\Lambda$ of the 
regularisations, it is universal. In addition, the damping parameter $\gamma$ 
only enters via the scaling variable $p\gamma$ and as a trivial scale factor. On 
the other hand, the form of $\dlap{F}_{\rm irr}(p,q)$ should depend on having 
assumed Ohmic damping.

In the special case $d=d_u=2$, logarithmic corrections to scaling are expected 
to be present, in analogy to classical dynamics. We do not present a detailed 
analysis of this case here, but it can be done as outlined above

\section{Homogeneity and double Laplace transforms}
\label{app:Lap}
Well-known Tauberian theorems for the Laplace transform, which go back to Hardy 
and Littlewood, and Karamata, and Feller, relate the asymptotics of a function 
$f(x)$ for $x\to\infty$ with the behaviour of its Laplace transform $\lap{f}(p)$ 
as $p\to 0$, see~ch. XIII.5 in~\cite{Fell71}. The non-local structure of 
the quantum noise correlations requires us to find an extension of these results 
for functions $f(x,y)$ of two variables and their double Laplace transform, see 
Eq.~(\ref{eq:doubleL}) and \ref{app:prop_lap}. In what follows, $f$ is assumed 
to be such that $\dlap{f}$ exists, see Refs.~\cite{Voel50,Ditk62} for sufficient 
conditions. We are interested in how the asymptotics of $f(x,y)$ for $x$ and $y$ 
both large is related to the properties of $\dlap{f}(p,q)$. {}From 
Ref.~\cite{Fell71}, the scaling limit $x,y\to \infty$ with fixed $x/y>1$ 
corresponds to the limit $p,q\to 0$ with fixed $q/p$. We are mainly interested 
in homogeneous functions and look for an explicit transformation formula for the 
scaling functions, in order to relate the respective asymptotics.\\

\noindent
{\bf Lemma 1:} {\it The double Laplace transform of a homogeneous function 
$f(x,y)=y^{-\alpha} \phi(x/y)$ where $\alpha<2$ and $\phi(0)$ is a finite 
constant and where $\phi(u)\simeq \phi_{\infty} u^{-\lambda}$, asymptotically 
for $u\to\infty$, also admits a scaling form}
\begin{align}
\dlap{f}(p,q) &=  p^{\alpha-2} \Phi(q/p),
\end{align}
{\it with the scaling function}
\begin{align}\label{eq:phipremellin}
 \Phi(u) &= \Gamma(2-\alpha)\, u^{\alpha-1} \int_0^{\infty} \!\D\xi\: \phi(\xi 
u)\, (\xi+1)^{\alpha-2} .
\end{align}
{\it In particular, for $0<\lambda<1$ and $\alpha<1+\lambda$, one has 
asymptotically for $u\to \infty$}
\begin{subequations}\label{Lemma1}
\begin{align} \label{Lemma3amp}
\Phi(u) &\simeq \Phi_{\infty} u^{\alpha-1-\lambda} \quad \mbox{with} \quad 
\Phi_{\infty} = \phi_{\infty} \Gamma(1-\lambda) \Gamma(1+\lambda-\alpha).
\end{align}
{\it For $1<\lambda<2$, one has asymptotically for $u\to\infty$}
\begin{align}\label{Lemma3amp-bis}
\Phi(u) &\simeq  \phi^{(1)} u^{\alpha-2} + \Phi_{\infty} u^{\alpha-1-\lambda} 
\quad \mbox{with} \quad \phi^{(n)} = 
(-1)^{n-1}\frac{\Gamma(n+1-\alpha)}{(n-1)!}\int_0^{\infty} \!\D u\: u^{n-1} 
\phi(u).
\end{align}
{\it More generally, for $n<\lambda<n+1$ with $n\in\mathbb{N}$, one has asymptotically}  
\begin{align}\label{Lemma3amp-ter}
\Phi(u)\simeq \phi^{(1)} u^{\alpha-2} + \ldots + \phi^{(n)} u^{\alpha-1-n} 
+\Phi_{\infty} u^{\alpha-1-\lambda}
\end{align}
\end{subequations}

\noindent
{\bf Proof:} 
The scaling assumption on $f(x,y)$ is equivalent to requiring homogeneity 
\begin{equation} \label{A2}
f(\ell x, \ell y) = \ell^{-\alpha} f(x,y), \nonumber
\end{equation}
with the index $\alpha$ and for all positive $\ell\in\mathbb{R}_+$. It follows 
that $\dlap{f}(p,q)$ is homogeneous with index $2-\alpha$, i.e.,
\begin{align}
\dlap{f}(\ell p,\ell q) &= 
\ell^{-(2-\alpha)} \dlap{f}(p,q).
\label{A3}
\end{align}
Choosing $\ell=1/p$ in Eq.~(\ref{A3}) gives the scaling form of the double Laplace transform
\BD \label{A4}
\dlap{f}(p,q) = p^{-2+\alpha} \dlap{f}(1,q/p) = p^{\alpha-2} \Phi\left( 
q/p\right),
\ED
with the scaling function
\begin{align}
\Phi(u) &= p^{2-\alpha} \dlap{f}(p,p u) 
=  u^{\alpha-1} \int_0^{\infty} \!\D x\int_0^{\infty} \!\D y\: y^{-\alpha} \phi\left(\frac{x}{y} u\right) e^{-x-y}
\nonumber \\
&=u^{\alpha-1} \int_0^{\infty} \!\D\xi\: \phi(\xi u) 
\int_0^{\infty} \!\D\eta\: \eta^{1-\alpha} e^{-(\xi+1)\eta} 
\:=\: \Gamma(2-\alpha)\, u^{\alpha-1} \int_0^{\infty} \!\D\xi\: \phi(\xi u)\, (\xi+1)^{\alpha-2},
\nonumber
\end{align}
 as anticipated in Eq.~(\ref{eq:phipremellin}).
We now derive the large-$u$ behaviour of $\Phi(u)$. We begin with a heuristic 
discussion. In general, one expects a decomposition into a regular part and an irregular part
\begin{align}
\Phi(u) &= \Gamma(2-\alpha) u^{\alpha-2} \bigl[ \Phi_{\rm reg}(u) + \Phi_{\rm 
irr}(u) \bigr] \nonumber \\[.25cm]
&= \Gamma(2-\alpha) u^{\alpha-2} \left[ \int_0^{\eta} \!\D \xi\: \phi(\xi)\left( 
1 + \frac{\xi}{u}\right)^{\alpha-2} + \int_{\eta}^{\infty} \!\D \xi\: 
\phi(\xi)\left( 1 + \frac{\xi}{u}\right)^{\alpha-2}\right],
\nonumber 
\end{align}
with a cut $\eta$. Expanding formally the regular part  in $u$ leads to
\BD
\Phi_{\rm reg}(u) = \sum_{n\geq 0} \left(\vekz{\alpha-2}{n}\right) 
\int_0^{\eta} \!\D\xi\: \phi(\xi)\left(\frac{\xi}{u}\right)^n 
\ED
and taking the limit $\eta\to\infty$, one only keeps those terms where the corresponding moment $\phi^{(n)}$ exists, which 
depends on the value of $\lambda$. These are the regular terms in (\ref{Lemma3amp-ter}). The irregular term
is estimated as follows, where for sufficiently large $\eta$ the asymptotic form of $\phi(u)$ is used 
%
%
\begin{align}
\Phi_{\rm irr}(u) 
= u \int_{\eta/u}^{\infty} \!\D \xi\: \phi(\xi u)\left( 1 + {\xi}\right)^{\alpha-2} 
\simeq u^{1-\lambda} \phi_{\infty} \int_{\eta/u}^{\infty} \!\D \xi\: \xi^{-\lambda}\left( 1 + {\xi}\right)^{\alpha-2}
\stackrel{u\to\infty}{=~}
 u^{1-\lambda} \phi_{\infty} \frac{\Gamma(\lambda+1-\alpha)\Gamma(1-\lambda)}{\Gamma(2-\alpha)} ~~~ \nonumber,
\end{align}
where, in the second step, we consider first the asymptotic limit $u\to \infty$ 
and then express the integral via a Beta function \cite{Abra65}. The final 
result is then independent of the cut $\eta$ and corresponds to the second part 
of Eq.~(\ref{Lemma3amp-ter}). 

Not all terms in this formal expansion really occur. For example, for 
$0<\lambda<1$ and also with $\alpha<1+\lambda$, we consider the regular part as 
taken from (\ref{eq:phipremellin}). The asymptotic approximation $\phi(u)\sim 
u^{-\lambda}$ should work as long as $\xi\gtrsim 1/u$ is sufficiently large. If 
on the other hand,  $\xi\lesssim 1/u$ and if $\phi(0)$ is a finite constant, 
that part of the integral contributes a term of order ${\rm O}(\phi(0)/u)$, 
compared to the contribution ${\rm O}(u^{-\lambda})$ from the main term. 
Accordingly, the small-$\xi$ contribution, for $\lambda<1$, will be a 
sub-dominant correction, see Eq.~(\ref{Lemma3amp}). 

We now turn to a more systematic method which does not rely on heuristics. It is convenient to re-write the scaling function as 
\BD
 \Phi(u) = \Gamma(2-\alpha) \int_0^{\infty} \D z\ \phi\left(1/z\right) z^{-2}\left(u+\frac{1}{z}\right)^{\alpha-2} .
\ED
The required asymptotics for large $u$, we are interested in, is obtained by 
first renaming $\phi(1/z) = f(z)$ and second expressing $f(z)$ through its 
Mellin transform, see Eq.~(\ref{eq:mellin}) in \ref{app:F}. Exchanging the order 
of integrations, we first calculate the $z$-integration and find 
\begin{align}
 \Phi(u)
%
%
%
&= \frac{1}{2\pi \II} \int_{c-\II\infty}^{c+\II\infty}\! \D s\:  \mel{f}(s)
u^{s+\alpha -1} \Gamma(s+1) \Gamma(1-s-\alpha) \nonumber 
= \sum_{s_0} \operatorname{Res}_{s=s_0} \left[ \mel{f}(s)
u^{s+\alpha -1} \Gamma(s+1) \Gamma(1-s-\alpha) \right],
\nonumber
\end{align}
with $c\in(0,1-\alpha)$. As explained in \ref{app:F}, we then shift the contour 
of integration towards having $c \to -\infty$ and express the integral as a sum 
over the set of enclosed poles $s_0$. Summing all relevant residues yields an 
ordered series in $u$, beginning with the most relevant contributions as $u\gg 
1$.  The integrand has three potentially singular contributions, i.e., for $s_0 
\in \{ -1-n \ | \ n \in \mathbb{N}_0 \}$, the poles of the Mellin transform 
itself and for  $s_0 \in \{ 1-\alpha+n \ | \ n \in \mathbb{N}_0 \}$. The last 
ones do not contribute to the asymptotic behaviour since they are located to the 
right of the original integration domain. We thus need to identify the poles of 
the Mellin transform. This is done by specifying the asymptotic behaviour of the 
function $\phi(u)$, e.g.,
\begin{align}
 \phi(u) &\simeq u^{-\lambda}  \left( A_0 + \frac{A_1}{u} + \frac{A_2}{u^2} + 
\ldots \right) + B_0 + \frac{B_1}{u}+ \frac{B_2}{u^2}+ \ldots \quad \text{for} 
\quad u\to\infty,
\nonumber
\end{align}
which translates into
\begin{align}
f(z) &\simeq z^{\lambda}  \left( A_0 + A_1 z + A_2z^2 + \ldots \right)
+ B_0 + B_1 z+ B_2 z^2+ \ldots \quad \text{for} \quad z\to0.
\nonumber
\end{align}
We also use the decomposition $f(z)=f_A(z) + f_B(z)$ if we want to consider 
these two series separately. The poles of the Mellin transform are located at 
$s_0 \in \{-\lambda -n \ | \ n\in\mathbb{N}_0  \} \cup \{ -n \ | \ 
n\in\mathbb{N}_0  \}$ \cite{Flaj95}. We assume $\lambda \not\in\mathbb{N}$ such 
that the first series has only simple poles. Evaluation of the residues leads to 
the following asymptotic series for the scaling function
\begin{align}\nonumber
 \Phi(u) &\simeq \sum_{n=0}^\infty \bigg\{ 
 A_n \Gamma(1+\lambda\textcolor{black}{-}\alpha+n)\Gamma(1-\lambda-n)  u^{-\lambda+\alpha-n-1} 
\bigg.\\
 &\quad \bigg. + \frac{\Gamma(2-\alpha)}{1-\alpha} B_0
 -\sum_{n=1}^\infty B_n (-1)^n n \frac{\Gamma(1+n-\alpha)}{\Gamma(n+1)}\psi(n) u^{\alpha-n-1} \bigg\}
\label{eq:scaleL}\\ 
&\quad +\textcolor{black}{\sum_{m\geq 1}\mel{f}_A(-m)(-1)^{m-1}\frac{\Gamma(m+1-\alpha)}{(m-1)!}u^{\alpha-m-1}} \nonumber  ,
\end{align}
where $\psi(n)$ is the digamma function \cite{Abra65} and the terms in the last 
line have to be included as long as $\mel{f}_A(-m)$ exists.  

In the special case, in which $A_0=\phi_{\infty}$ and $A_n=B_n=B_0=0$ for all 
$n\geq 1$, we recover Eq.~(\ref{Lemma3amp}) for $0<\lambda<1$. For 
$1<\lambda<2$, we formally have $\mel{f}_A(-1)=\int_0^{\infty} \!\D z\: z^{-2} 
f(z) = \int_0^{\infty} \!\D u\: \phi(u)$ and we obtain what we anticipated in 
Eq.~(\ref{Lemma3amp-bis}). Similarly, for different ranges of $\lambda$, the 
terms contained in Eq.~(\ref{Lemma3amp-ter}) are read off. This completes the 
proof. \hfill $\square$ \\

It follows that a derived asymptotic behaviour $\Phi(u) \sim u^{-\vartheta}$ must be interpreted carefully in order
to identify the exponent $\lambda$ in $\phi(u)\sim u^{-\lambda}$ correctly. If effectively $\vartheta>n-\alpha$ is found, 
the expansion must be carried up to terms O($u^{\alpha-n-1}$). \\

\noindent
{\bf Corollary:} {\it Consider a function $f(x,y)$ of two variables and such 
that its double Laplace transform $\dlap{f}(p,q)$ exists. Assume that 
$f(x,y)=y^{-\alpha} \bigl(\ln \frac{1}{y}\bigr)^{-\beta} \phi(x/y)$ admits a 
logarithmic scaling form, with $\alpha<2$ and $\phi(0)$ being a finite constant. 
 Then the double Laplace transform admits the scaling form}
\begin{align}
\dlap{f}(p,q) &=  p^{\alpha-2} \bigl( \ln p\bigr)^{-\beta} \Phi(q/p), 
\quad \mbox{with} \quad
\Phi(u)       =  \Gamma(2-\alpha)\, u^{\alpha-1} \int_0^{\infty} \!\D\xi\: \phi(\xi u)\, (\xi+1)^{\alpha-2} .
\end{align}

A different kind of scaling arises if there is a further auxiliary variable, labeled $k$ here. 
We can formulate the following elementary result. \\

\noindent
{\bf Lemma 2:} {\it Consider a function $f(x,y;k)$ of two variables $x,y$ and such that its double Laplace transform 
$\dlap{f}(p,q;k)$ with respect to these variables exists. 
Assume that $f$ admits the scaling form $f(x,y;k) = k^{\alpha z} \phi(k^z x, k^z y)$. 
Then the double Laplace transforms admits the scaling form}
\BEQ \label{E:8}
\dlap{f}(p,q;k) = k^{(\alpha-2)z} \Phi(p k^{-z}, q k^{-z})
\quad \mbox{with} \quad
\Phi(u,v) = \dlap{\phi}(u,v).
\EEQ

\noindent
{\bf Proof:} The scaling assumption on $f$ is equivalent to the homogeneity 
property
\BD
f(\ell x, \ell y; \ell^{-1/z} k) = \ell^{-\alpha} f(x,y;k) .
\ED
Laplace-transforming this with respect to $x$ and $y$  leads to the transformed homogeneity property
\BD
\dlap{f}\left( \frac{p}{\ell}, \frac{q}{\ell}; \ell^{-1/z} k\right) = \ell^{2-\alpha} \dlap{f}(p,q;k)
\ED
and setting $\ell=k^z$ gives the scaling form in Eq.~(\ref{E:8}). The scaling functions are identified as 
$\phi(x,y) := f(x,y;1)$ and $\Phi(p,q) := \dlap{f}(p,q;1)$. 
The relationship between these scaling functions, as stated in Eq.~(\ref{E:8}),  
readily follows from the definitions. \hfill $\square$ 


\section*{References}

\bibliographystyle{prr.bst}

\bibliography{library}


\end{document}